\documentclass[modern]{aastex63}

\usepackage{xspace}
\usepackage{amsmath}

\newcommand{\teff}{\ensuremath{T_{\mathrm{eff}}}\xspace}  
\newcommand{\logg}{\ensuremath{\log g}\xspace} 
\newcommand{\feh}{[Fe/H]\xspace}
 
\newcommand{\prot}{\ensuremath{P_\mathrm{rot}}\xspace}
\newcommand{\rvar}{\ensuremath{R_\mathrm{var}}\xspace}
\newcommand{\vtan}{\ensuremath{v_\mathrm{tan}}\xspace}
\newcommand{\vb}{\ensuremath{v_\mathrm{b}}\xspace}

\newcommand{\mstar}{$M_*$\xspace}
\newcommand{\rstar}{$R_*$\xspace}

\newcommand{\rearth}{$R_\oplus$\xspace}
\newcommand{\logage}{$\text{log(age)}$\xspace}
\newcommand{\rp}{$R_P$\xspace}
\newcommand{\rprstar}{$R_P/R_\star$\xspace}

\newcommand{\logrp}{$\log_{10}(R_P/R_\oplus)$\xspace}
\newcommand{\deltaage}{$\Delta \log_{10}(\mathrm{age~yr}^{-1})$\xspace}
\newcommand{\deltamstar}{$\Delta M_*$\xspace}
\newcommand{\deltateff}{$\Delta T_\mathrm{eff}$\xspace}

\newcommand{\galex}{\textit{GALEX}\xspace}
\newcommand{\gaia}{\textit{Gaia}\xspace}
\newcommand{\kepler}{\textit{Kepler}\xspace}

\newcommand{\stardate}{\texttt{stardate}\xspace}

\newcommand{\rtau}{$R_\tau$\xspace}
\newcommand{\npla}{156\xspace}
\newcommand{\nplb}{238\xspace}
\newcommand{\nplc}{124\xspace}
\newcommand{\npld}{190\xspace}

\newcommand{\rvshift}{0.07\xspace}
\newcommand{\avmax}{0.5\xspace}

\newcommand{\nplbase}{1443\xspace}
\newcommand{\nstbase}{871\xspace}
\newcommand{\nplfilt}{732\xspace}
\newcommand{\nstfilt}{466\xspace}

\newcommand{\samplea}{\texttt{isoc\_fgk\_1to2}\xspace}
\newcommand{\sampleb}{\texttt{isoc\_fgk\_lt2}\xspace}
\newcommand{\samplec}{\texttt{isoc\_fg\_lt2}\xspace}
\newcommand{\sampled}{\texttt{gyro\_gk\_lt3}\xspace}
\newcommand{\gold}{\texttt{gold}\xspace}
\newcommand{\goldyng}{\texttt{gold\_lt3}\xspace}
\newcommand{\goldold}{\texttt{gold\_gt3}\xspace}

\received{November 19, 2020}
\revised{February 22, 2021}
\submitjournal{AAS Journals}

\shorttitle{Evolution of the Exoplanet Radius Gap}
\shortauthors{David et al.}
\begin{document}

%\title{Small Planet Sizes Evolve Over Billions of Years}
\title{Evolution of the Exoplanet Size Distribution:\\Forming Large Super-Earths Over Billions of Years}

\correspondingauthor{Trevor J. David}
\email{tdavid@flatironinstitute.org}

\author[0000-0001-6534-6246]{Trevor J.\ David}
\affil{Center for Computational Astrophysics, Flatiron Institute, New York, NY 10010, USA}
\affil{Department of Astrophysics, American Museum of Natural History, Central Park West at 79th Street, New York, NY 10024, USA}

\author[0000-0002-3011-4784]{Gabriella Contardo}
\affil{Center for Computational Astrophysics, Flatiron Institute, New York, NY 10010, USA}

\author[0000-0003-1133-1027]{Angeli Sandoval}
\affiliation{Department of Physics and Astronomy, Hunter College, City University of New York, New York, NY 10065, USA}

\author[0000-0003-4540-5661]{Ruth Angus}
\affil{Department of Astrophysics, American Museum of Natural History, Central Park West at 79th Street, New York, NY 10024, USA}
\affil{Center for Computational Astrophysics, Flatiron Institute, New York, NY 10010, USA}
\affil{Department of Astronomy, Columbia University, 550 West 120th Street, New York, NY, USA}

\author[0000-0003-4769-3273]{Yuxi (Lucy) Lu}
\affil{Department of Astronomy, Columbia University, 550 West 120th Street, New York, NY, USA}
\affil{Department of Astrophysics, American Museum of Natural History, Central Park West at 79th Street, New York, NY 10024, USA}

\author[0000-0001-9907-7742]{Megan Bedell}
\affil{Center for Computational Astrophysics, Flatiron Institute, New York, NY 10010, USA}

\author[0000-0002-2792-134X]{Jason L.~Curtis}
\affil{Department of Astrophysics, American Museum of Natural History, Central Park West at 79th Street, New York, NY 10024, USA}

\author[0000-0002-9328-5652]{Daniel Foreman-Mackey}
\affil{Center for Computational Astrophysics, Flatiron Institute, New York, NY 10010, USA}

\author[0000-0003-3504-5316]{Benjamin J.\ Fulton}
\affil{California Institute of Technology, Pasadena, CA 91125, USA}
\affil{IPAC-NASA Exoplanet Science Institute Pasadena, CA 91125, USA}

\author[0000-0003-4976-9980]{Samuel K. Grunblatt}
\affil{Department of Astrophysics, American Museum of Natural History, Central Park West at 79th Street, New York, NY 10024, USA}
\affil{Center for Computational Astrophysics, Flatiron Institute, New York, NY 10010, USA}

\author[0000-0003-0967-2893]{Erik A.\ Petigura}
\affil{Department of Physics and Astronomy, University of California, Los Angeles, CA 90095, USA}




\begin{abstract}
The radius valley, a bifurcation in the size distribution of small, close-in exoplanets, is hypothesized to be a signature of planetary atmospheric loss. Such an evolutionary phenomenon should depend on the age of the star-planet system. In this work, we study the temporal evolution of the radius valley using two independent determinations of host star ages among the California--\kepler Survey (CKS) sample. We find evidence for a wide and nearly empty void of planets in the period-radius diagram at the youngest system ages ($\lesssim$2--3~Gyr) represented in the CKS sample. We show that the orbital period dependence of the radius valley among the younger CKS planets is consistent with that found among those planets with asteroseismically determined host star radii. Relative to previous studies of preferentially older planets, the radius valley determined among the younger planetary sample is shifted to smaller radii. This result is compatible with an atmospheric loss timescale on the order of gigayears for progenitors of the largest observed super-Earths. In support of this interpretation, we show that the planet sizes which appear to be unrepresented at ages $\lesssim$2--3~Gyr are likely to correspond to planets with rocky compositions. Our results suggest the size distribution of close-in exoplanets, and the precise location of the radius valley, evolves over gigayears.
\end{abstract}


\section{Introduction} \label{sec:intro}
By far the most intrinsically common planets known are small ($<$4~\rearth), close-in ($<$1~au) planets. NASA's \kepler mission \citep{Borucki2010} revealed the surprising abundance of these planets; some 30--60\% of Sun-like stars host a small, close-in planet, depending on assumptions about the intrinsic multiplicity and inclination dispersion within planetary systems \citep{Fressin2013, Petigura2013, Zhu2018, He2019}. An enduring mystery posed by small planets is how some accreted sizable atmospheres while others appear to have avoided runaway accretion altogether \citep[e.g.][]{Ikoma2012, Lee2014, Lee2016}. Often times \kepler multiplanet systems host both planets with and without atmospheres, in some cases separated from one another by only a hundredth of an astronomical unit \citep{Carter2012}.

Recent progress in understanding small planets has been fueled by improved precision in stellar and planetary parameters. Through homogeneous spectroscopic characterization of $>$1300 \kepler planet hosts the California--\kepler Survey \citep[hereafter CKS,][]{Petigura2017, Johnson2017} revealed that the size distribution of close-in ($P<$100~days) small planets is bimodal, with a valley in the completeness-corrected radius distribution between 1.5 and 2 Earth radii \citep{Fulton2017}. The radius valley is widely believed to be a signature of atmospheric loss. This belief is bolstered by determinations of planet densities on either side of the valley: planets below the valley, dubbed super-Earths, have densities consistent with a rocky composition, while planets above the valley, known as sub-Neptunes, require atmospheres of a few percent by mass to explain the low measured densities \citep[e.g.][]{Weiss2014, Rogers2015}. In the atmospheric loss model some fraction of super-Earths are the remnant cores of planets that shed their primordial envelopes, which potentially alleviates the issue of neighboring planets with dissimilar densities \citep[e.g.][]{Lopez2013, OwenMorton2016}. While exploration of the radius valley among planets orbiting low-mass stars has provided support for an alternative hypothesis \citep[formation in a gas-poor disk without the need for atmospheric loss,][]{Cloutier2020}, atmospheric erosion remains the leading theory for planets around Sun-like stars.

%Mechanisms
Atmospheric loss requires energy. Energy deposited into a planet's atmosphere, from an internal or external source, can heat gas to velocities exceeding the planet's escape velocity. External mechanisms of energy deposition include photoevaporation \citep[heating of the planet's thermosphere by X-ray and extreme ultraviolet radiation, e.g.][]{Owen2012} and impacts by planetesimals or planetary embryos \citep{Liu2015, Inamdar2016, Chatterjee2018, Wyatt2020}. Internal energy deposition can be provided by the luminosity of a planet's cooling core \citep[e.g.][]{Ginzburg2016}. Notably, planetary evolution models studying the effect of photoevaporation predicted the existence of a radius valley before it was observed \citep{OwenWu2013, Lopez2013, Jin2014, ChenRogers2016}. However, subsequent studies considering the effects of core-powered mass loss were also able to reproduce the bimodal radius distributon of small planets \citep{Ginzburg2018, Gupta2019, Gupta2020}. Photoevaporation and core-cooling remain the two leading explanations for the radius valley, and both processes may well be important, but to determine the relative importance of the two effects will require a better understanding of the dependence of the valley on other key parameters.

%Gap Width: Atmospheric loss models predict an empty gap.
Determining how empty the radius gap is represents an important step toward understanding its origins. In the initial CKS sample, typical planet radius uncertainties were comparable to the width of the gap so that an intrinsically empty gap would not have been resolved \citep{Fulton2017}. \citet{vanEylen2018} studied planets orbiting a subset of \kepler host stars with precise asteroseismic parameters (including ages ranging from $\sim$2--10~Gyr) and found a gap considerably wider and emptier than that found in the initial CKS sample. Including trigonometric parallaxes from Gaia DR2, \citet{Fulton2018} were able to improve the median \rstar errors by a factor of five in the CKS sample but the gap remained populated. Those authors presented simulations which suggest that the gap is not empty (i.e. solely filled in by noisy data), and that there are real planets in the gap. More recently, however, \citet{Petigura2020} showed that a sizable number of planets in and around the gap have poorly determined radii due to high impact parameters, indicating the gap may be emptier than previously appreciated.

%Orbital period
It also appears that the gap, which is a one-dimensional projection of a higher dimensional manifold, is partially filled in due to a dependence of the gap center on orbital period (or stellar light intensity) and host star mass. The gap center is anti-correlated with orbital period \citep{Fulton2017, vanEylen2018, Martinez2019, MacDonald2019, Loyd2020}, which is considered compatible with both the photoevaporation \citep[e.g.][]{OwenWu2013, OwenWu2017, LopezRice2018, Jin2018} and core-powered mass-loss models \citep[e.g.][]{Gupta2019, Gupta2020}, but incompatible with formation in a gas-poor disk \citep{LopezRice2018}; at larger orbital periods, only the smallest and least massive cores are susceptible to total atmospheric loss, driving the gap to smaller radii. The length of the radius valley, i.e. its outer boundary in either period or insolation, may also provide clues to its origin, though this parameter remains poorly studied. In the photoevaporation model, the radius valley should not extend beyond orbital periods of 30--60~days as the incident XUV flux is believed to be too low to drive substantial mass loss \citep{OwenWu2017}. However, the low completeness of the \kepler data set for small planets at these orbital periods presents a challenge for detecting such a transition point.

%Stellar mass
The gap center is positively correlated with stellar mass \citep{Fulton2018, Wu2019, Cloutier2020, Berger2020b, Hansen2020, vanEylen2021}, \added{although it has been suggested that this trend is due to the relationship between stellar mass and planetary insolation \citep{Loyd2020}}. The measured mass-dependence of the gap has been used to argue support for photoevaporation \citep[e.g.][]{Wu2019}, but requires that the average planet mass scales approximately linearly with host star mass, an assertion that has not been verified for small planets. By comparison, in the core-powered mass-loss model the dependence of the radius gap location on stellar mass is a natural consequence of the dependence of planet equilibrium temperature (which partially determines the mass-loss rate in the Bondi-limited regime) on the stellar mass-luminosity relation \citep[e.g.][]{Gupta2020}.

%Stellar metallicity
As for metallicity, there is tentative evidence for a wider radius valley for metal-rich stars \citep{Owen2018}. Such a dependence could result if the core mass distributions, core bulk densities, or initial atmospheric mass fractions of small planets depend sensitively on the metallicity of the host star, and hence protoplanetary disk. There is evidence that large \kepler planets (2--8~\rearth) are more common around higher metallicity stars \citep{Dong2018, Petigura2018} and that planets at short orbital periods are preferentially larger around higher metallicity stars \citep{Owen2018}. Both findings are compatible with a scenario in which metal-rich stars form more massive cores, on average. It has also been suggested that metal-rich stars host planets with higher atmospheric metallicities, which increases the efficiency of atomic line cooling in photoevaporative flows and decreases mass-loss rates \citep{Owen2018}. In the core-powered mass-loss model, the rate at which sub-Neptunes cool and contract is anti-correlated with the opacity of the envelope, which is assumed to be proportional to the stellar metallicity \citep{Gupta2020}. Thus, in both the photoevaporation and core-cooling models, larger sub-Neptunes and a consequently wider radius valley are expected around more metal-rich stars (for fixed mass and age, and neglecting any potential scaling between metallicity and core mass distributions).

%Timescales
The characteristic timescale for atmospheric loss among close-in exoplanets has been proposed as a key parameter for assessing the relative importance of photoevaporation and core-powered mass loss. Firm observational constraints on that timescale, however, are lacking. Constraining this timescale through exoplanet population studies may provide a means for discerning the relative importance of proposed mass loss mechanisms. Core-powered mass loss is believed to operate over gigayear timescales \citep{Ginzburg2016, Ginzburg2018, Gupta2019, Gupta2020}. By comparison, photoevaporation models predict the majority of mass-loss to occur during the first 0.1~Gyr \citep[e.g.][]{Lopez2012, Lopez2013, Owen2012, OwenWu2013, OwenWu2017}, corresponding roughly to the length of time a Sun-like star spends as a saturated X-ray emitter \citep[e.g.][]{Jackson2012, Tu2015}. However, a more recent study found that the majority of the combined X-ray and extreme UV emission of stars occurs after the saturated phase of high energy emission, implying that XUV irradiation of exoplanet atmospheres continues to be important over gigayear timescales \citep{King2020}. If valid, then observational constraints on exoplanet evolution timescales may not provide a conclusive means for discerning the relative importance of photoevaporation and core-powered mass-loss. \added{Nevertheless, there is evidence that the detected fraction of super-Earths to sub-Neptunes increases over gigayears, suggesting that the sizes of at least some planets evolve on these long timescales \citep{Berger2020b, Sandoval2020}}.

A basic prediction of atmospheric loss models is that the radius gap is wider at younger ages and fills in over time; at a fixed value of high energy incident flux and initial atmospheric mass fraction, photoevaporation models predict that sub-Neptunes with the least massive cores (and smallest core sizes) will cross the gap first, with more massive cores crossing the gap at later times, if at all. As a result, the radius valley is expected to be wider and emptier at early times, progressively filling in with stripped cores of ever larger masses and sizes \citep[e.g.][]{RogersOwen2020}. In the core-powered mass-loss model, \citet{Gupta2020} suggest the average size of sub-Neptunes is expected to decline with age while the average size of super-Earths remains relatively constant, again leading to a wider and emptier radius valley at earlier times. 

While the specific theoretical predictions for the age dependence of the radius valley morphology are uncertain, the fundamental prediction from atmospheric loss models that this feature should weaken with increasing age is a firm conclusion. We aim to investigate this hypothesis using the CKS sample. Here we investigate the time evolution of the exoplanet radius gap. In \S\ref{sec:sample} we describe our sample selection process, including several filters intended to rid our sample of stars or planets with unreliable parameters. Our analysis procedures are discussed in \S\ref{sec:analysis}, and finally we interpret our results and summarize our primary findings in \S\ref{sec:conclusions}.

\section{Sample Selection} \label{sec:sample}
We began with the CKS VII sample published in \citet{Fulton2018}, hereafter F18. The CKS VII sample is a well-characterized subset of all \kepler planet candidates. Stellar characterization for these stars was performed in a homogeneous manner, with spectroscopic \teff, \logg, and \feh derived from high signal-to-noise, high-dispersion Keck/HIRES spectra \citep{Petigura2017, Johnson2017}. F18 derived stellar radii from the Stefan-Boltzmann law using their spectroscopic \teff and bolometric luminosities computed from \gaia DR2 parallaxes \citep{GaiaDR2}, extinction-corrected 2MASS $K_s$ magnitudes \citep{2mass}, and theoretical bolometric corrections from the MESA Isochrones and Stellar Tracks \citep[MIST,][]{Choi2016, Dotter2016}. F18 additionally computed ages for the CKS sample using the \texttt{isoclassify} package \citep{Huber2017}, which also depends on the MIST models. We use the F18 median posterior isochrone ages as one source of age in the analysis that follows.

\begin{figure}
    \centering
    \includegraphics[width=\linewidth]{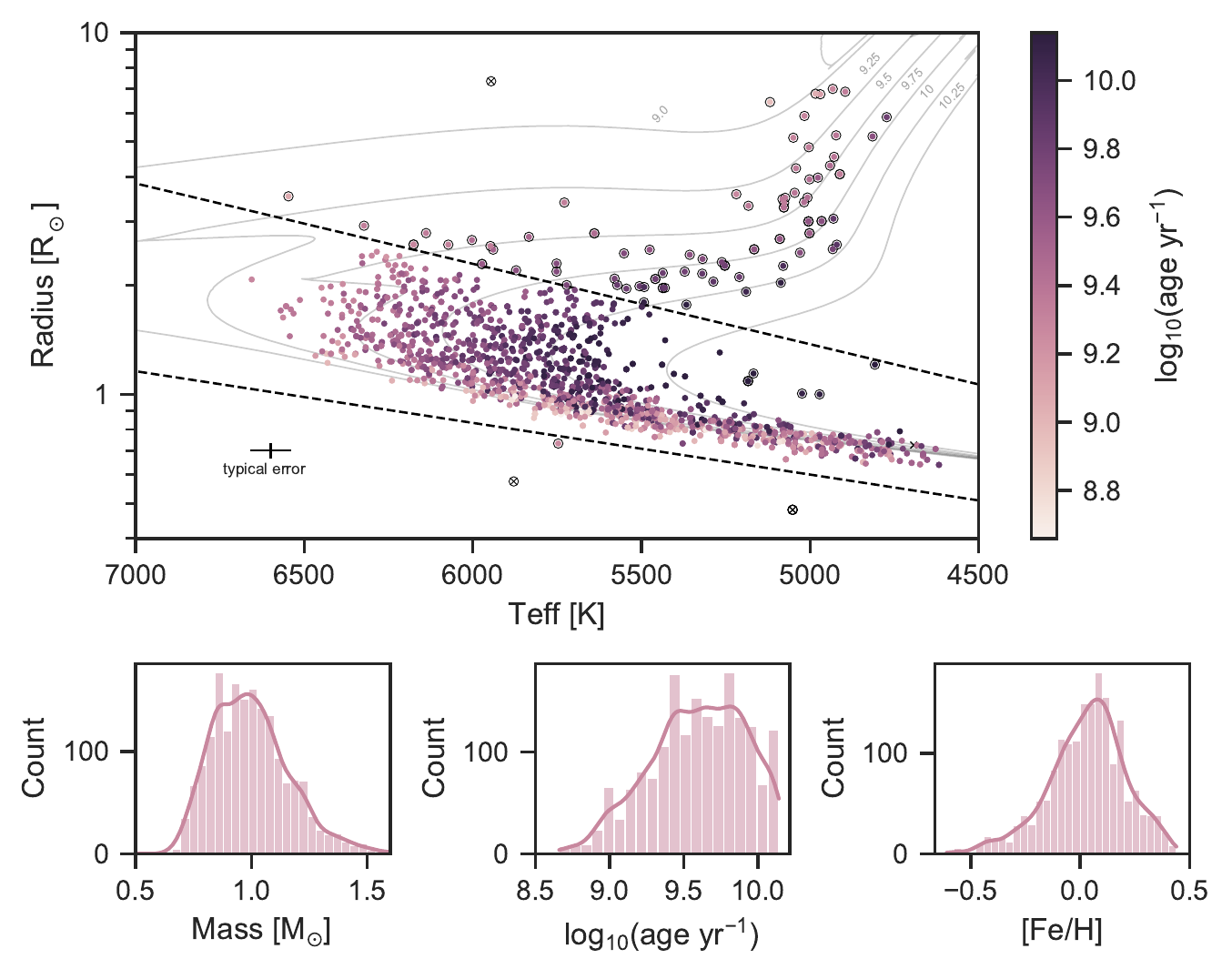}
    \caption{\textit{Top:} H-R diagram for the CKS VII sample \citep{Fulton2018}. Point colors indicate median posterior ages from that work. Dashed lines indicate the dwarf star selection criteria explained in \S\ref{sec:sample}. Points with encircled by black rings were excluded from our analysis, and points indicated by X marks lack ages. Grey curves indicate solar-metallicity, non-rotating MIST v1.1 isochrones \citep{Choi2016, Dotter2016}. \textit{Bottom:} Distributions of host star masses (left), ages (middle), and metallicities (right) after performing the dwarf star cut. Note, stars hosting multiple planets are represented more than once in these distributions. All parameters originate from F18.}
    \label{fig:hrd}
\end{figure}

We constructed several filters, many motivated by the cuts outlined in F18, to refine the sample and select those planets and stars with the most reliable parameters. The filters are enumerated as follows:

\begin{enumerate}
    
    \item \textit{Planet orbital period.} We restricted our analysis to planets with orbital periods $<$100 days. At larger periods \kepler suffers from low completeness, particularly for small planets.
    
    \item \textit{Planet size.} We restricted our analysis to planets with sizes $<$10~\rearth.
    
    \item \textit{Planet radius precision.} We restricted our analysis to planets with fractional radius uncertainties $\sigma_{R_P}/R_P < 20\%$.    
    
    \item \textit{Planet false positive designation.} We excluded planets identified as false positives in Table 4 of the CKS I paper \citep{Petigura2017}, which synthesized dispositions from \citet{Mullally2015}, \citet{Morton2016}, and the NASA Exoplanet Archive \citep[as accessed on Feb. 1, 2017,][]{Akeson2013}.
    
    \item \textit{Stellar radius (dwarf stars).} We restricted our analysis to dwarf stars with the following condition:
    \begin{equation}
        \left ( \frac{T_\mathrm{eff}-5500}{7000} \right ) + 0.15 < \log_{10}\left ( \frac{R_\star}{R_\odot} \right ) <  \left ( \frac{T_\mathrm{eff}-5500}{4500} \right ) + 0.25.
    \end{equation}
    The left hand side of this condition excludes a small number of stars far below the main sequence which may have erroneous parameters. The right hand side excludes stars which have evolved considerably away from the main sequence. This cut is depicted in Figure~\ref{fig:hrd}. We additionally excluded cool stars elevated from the main sequence which would result in unrealistically old ages. This cut was performed by requiring \teff $> T_\mathrm{isoc}$ where $T_\mathrm{isoc}$ is the temperature of a \logage = 10.25, [Fe/H] = +0.25 MIST v1.1 non-rotating isochrone with equivalent evolutionary point (EEP) $<500$.

    \item \textit{Stellar mass.} We wish to isolate the effect of stellar age on the exoplanet radius gap, while minimizing the effects of stellar mass as much as possible. We restricted our sample to stars with masses $0.75 < M_\star/M_\odot < 1.25$, where the masses were derived from stellar evolution models in F18. 
    
    \item \textit{Stellar metallicity.} For the same reason we confined our sample in stellar mass, we restricted our analysis to stars with spectroscopically determined metallicities in the range $-0.3 < $~[Fe/H]~$< +0.3$. 
    
    \item \textit{Isochrone parallax.} We removed stars where the \gaia and F18 ``spectroscopic'' or ``isochrone'' parallaxes differed by more than 4$\sigma$, where the latter quantities were computed in Table 2 of CKS VII. F18 speculated that such discrepancies may be due to flux contamination from unresolved binaries. This cut also removes all stars where the CKS VII isochrone-derived radius, $R_\mathrm{iso}$, differed by more than 10\% from the radius derived from the Stefan-Boltzmann law.  
    
    \item \textit{Stellar dilution (\gaia).} We used the $r_8$ column in Table 2 of CKS VII to exclude stars with closely projected sources detected by \textit{Gaia} that contribute a non-negligible fraction of the optical flux in the \kepler aperture. We excluded stars where additional sources in an 8\arcsec\ radius (two \kepler pixels) contribute more than 10\% of the cumulative $G$-band flux (including the target).
    
    \item \textit{Stellar dilution (Imaging).} As in F18, we excluded KOIs with closely-projected stellar companions bright enough to require corrections to the planetary radii of 5\% or more. Like F18, we use the radius correction factor (RCF) computed by \citet{Furlan2017} based on high-resolution imaging from several authors, accepting planets for which RCF $<$ 1.05.
    
    \item \textit{Unresolved binaries (\gaia).} We excluded stars with \gaia RUWE values $>$1.4, where these values were queried from the \gaia archive.\footnote{https://gea.esac.esa.int/archive/} The renormalized unit weight error, or RUWE, is a goodness-of-fit metric for a single star astrometric model \citep{Lindegren2018}. RUWE is a sensitive indicator of unresolved binaries \citep{Belokurov2020}, which are a concern for planet radius studies due to potential flux dilution or misidentification of the planet host.
    
    \item \textit{Discrepant photometry.} We removed stars with discrepant optical brightnesses, $|G-K_P| > 1$~mag, indicating a potentially erroneous cross-match between the \kepler and \gaia sources. 
    
    \item \textit{Reddening.} We removed stars with reddening estimates of $A_V > \avmax$~mag, where these estimates were sourced from \cite{Lu2020}. Stars with high reddening are more susceptible to erroneously determined stellar parameters. 
    
    \item \textit{Planets with grazing transits.} Due to degeneracies inherent to light curve modeling, planets with grazing transits can have poorly constrained radii. \citet{Petigura2020} showed that there is some level of contamination of the radius valley from planets with grazing transits. Since impact parameters measured from long cadence photometry are unreliable, we follow \citet{Petigura2020} and exclude planets with $R_\tau < 0.6$, where $R_\tau$ is the ratio of the measured transit duration to the duration of a $b=0, e=0$ transit with the same period around the same star. 
    
\end{enumerate}

The overall CKS VII sample contains 1913 planets orbiting 1189 unique stellar hosts. In the analysis that follows, we will refer to the base sample (constructed from the first five cuts enumerated) and the filtered sample (constructed from all of the filters). After applying the filters, the base sample consists of \nplbase planets orbiting \nstbase unique stellar hosts. The filtered sample consists of \nplfilt planets orbiting \nstfilt unique hosts. Later, we find that our analysis is insensitive to many of these restrictions and relax most of them.

\subsection{Rotation period vetting}
\label{subsec:prot}
We supplemented the CKS sample with stellar rotation periods, which we use in \S\ref{subsec:prgyro} to empirically age-rank planet hosts, compiled from the literature. In order to perform an accurate rotation-based selection it is imperative to have reliable rotation periods. To this end we performed visual vetting of the full \kepler light curves for each  star in the CKS sample. For each star, our period vetting procedure consisted of the following steps:

\begin{enumerate}
    \item Retrieve the full \kepler long cadence PDCSAP light curve \citep{Stumpe2012, Smith2012} from MAST\footnote{\url{https://archive.stsci.edu/kepler/search_retrieve.html}}, mask known transits using ephemerides from the KOI cumulative table\footnote{\url{https://exoplanetarchive.ipac.caltech.edu/docs/PurposeOfKOITable.html\#cumulative}}, mask data with nonzero \texttt{PDCSAP\_QUALITY} flags, and median normalize each quarter of data.
    \item Compile published rotation period measurements from four sources in the literature \citep{McQuillan2013, WalkowiczBasri2013, Mazeh2015, Angus2018}.
    \item Perform a Lomb-Scargle (L-S) periodogram analysis of the \kepler PDCSAP light curve using the \texttt{LombScargle} class in the \texttt{astropy.timeseries} package.
    \item Phase-fold the PDCSAP light curve on the L-S peak power period and any published period, as well as on the first harmonic and sub-harmonic of each of the previously mentioned periods.
    \item Generate a vetting sheet including all phase-folded light curves, the L-S periodogram, a 120-day segment of the light curve, and the full light curve.
    \item Visually examine each vetting sheet, recording the preferred period source and assigning a reliability flag to each period determination ($3:$ highly reliable, $2:$ reliable, and $1:$ period could not be unambiguously determined, and $0:$ no periodicity evident).\footnote{An example vetting sheet is presented in Appendix~\ref{sec:appendixa} and all sheets are available through the journal.} 
\end{enumerate}

The results of our rotation period vetting are summarized in Table~\ref{tab:prot}. Of the 1189 unique planet hosts in the CKS VII sample, which are predominantly FGK main-sequence stars, we found that approximately 22\% have highly reliably rotation periods, 23\% have reliable periods, 34\% could not have periods determined unambiguously from the light curve, and 21\% had no clear periodicity evident in the light curve.

\begin{deluxetable*}{ccccccccc}
\tablecaption{Rotation periods of KOIs in CKS VII sample.}
\label{tab:prot}
\tablecolumns{9}
\tablewidth{\linewidth}
\tablehead{\colhead{KOI} & \colhead{KIC} &  \colhead{\prot} & \colhead{\prot Ref.} & \colhead{Flag} &  \colhead{A18 \prot} &  \colhead{M13 \prot} &  \colhead{M15 \prot} &  \colhead{W13 \prot}}
\startdata
          10 &     6922244 &   7.46 &          A18 &         2 &        7.46 &       \nodata &     82.12 &      \nodata \\
          49 &     9527334 &   8.74 &          A18 &         3 &        8.74 &      8.55 &      8.59 &     8.60 \\
          63 &    11554435 &   5.49 &          A18 &         3 &        5.49 &      5.41 &       \nodata &     5.39 \\
\enddata
\tablecomments{Flag meanings are as follows, $3$: highly reliable, $2$: reliable, $1$: true period could not be unambiguously determined, $0$: no periodicity evident. References: A18 \citep{Angus2018}, M13 \citep{McQuillan2013}, M15 \citep{Mazeh2015}, W13 \citep{WalkowiczBasri2013}. Only a portion of this table is shown here to demonstrate its form and content. A machine-readable version of the full table is available.}
\end{deluxetable*}

\section{Analysis} \label{sec:analysis}

\subsection{Evolution of the period-radius diagram: isochrone ages}
\label{subsec:pr}

\begin{figure}
    \centering
    \includegraphics[width=\linewidth]{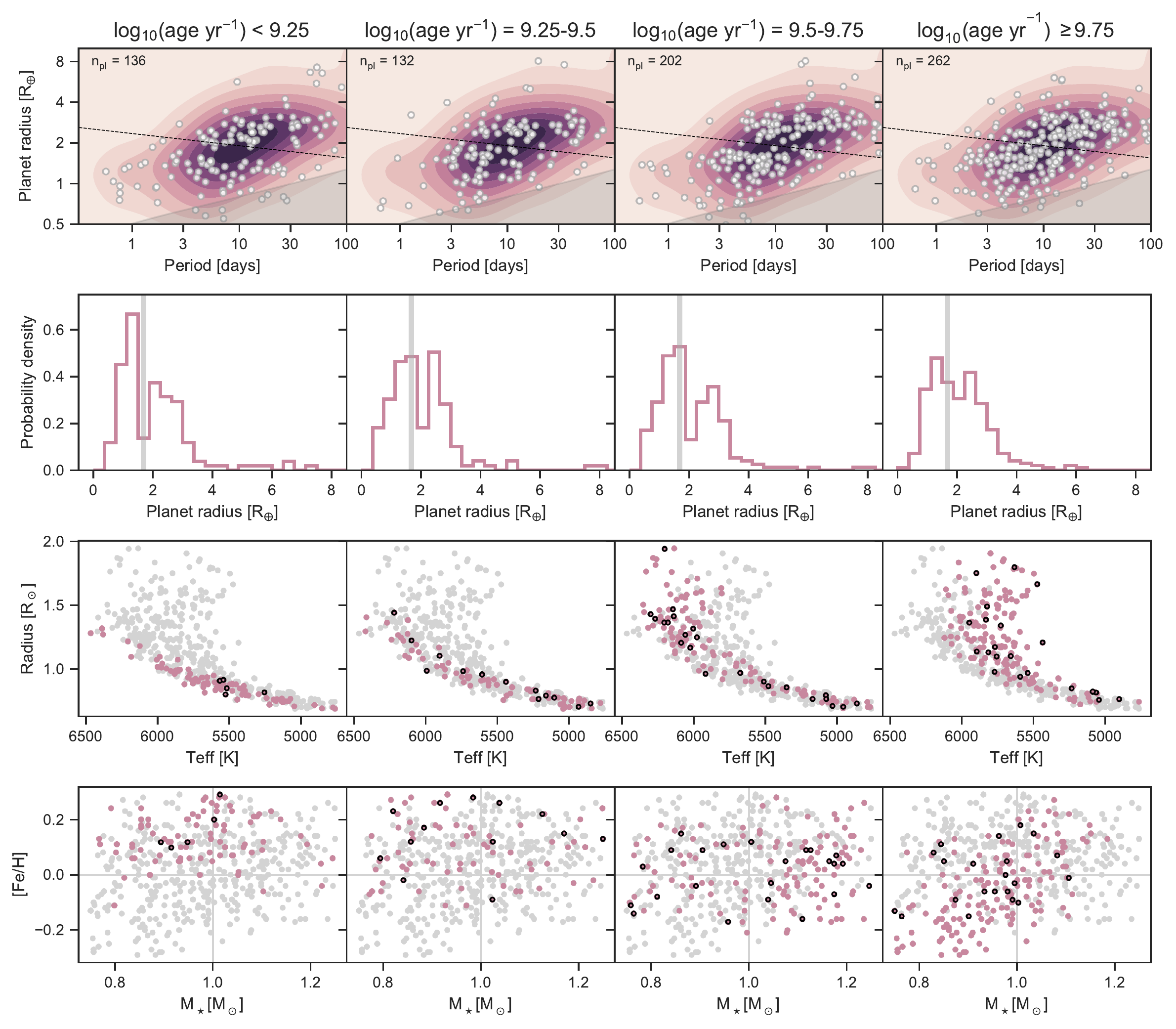}
    \caption{\textit{First row:} Evolution of the \kepler planet population in the period-radius diagram for the filtered CKS sample (age bins indicated above each panel). Contours show Gaussian kernel density estimates of planets in the overall CKS sample. The black dashed line indicates the radius valley derived by V18. The grey shaded region indicates the 25\% pipeline completeness contour calculated from the CKS sample. \textit{Second row:} one-dimensional distributions of planet radii for the samples plotted above in each case. The nominal location of the radius gap from \citet{Fulton2017} is indicated by the vertical grey stripe. \textit{Third row:} our base CKS planet-host sample in the \teff-\rstar plane (grey) and the host stars in the age bins indicated at the top (pink). Stars hosting planets in the radius range 1.6--1.9~\rearth are outlined in black. \textit{Fourth row:} as in the third row, the distribution of planet hosts in the mass-metallicity plane.}
    \label{fig:pr}
\end{figure}

\begin{figure}
    \centering
    \includegraphics[width=\linewidth]{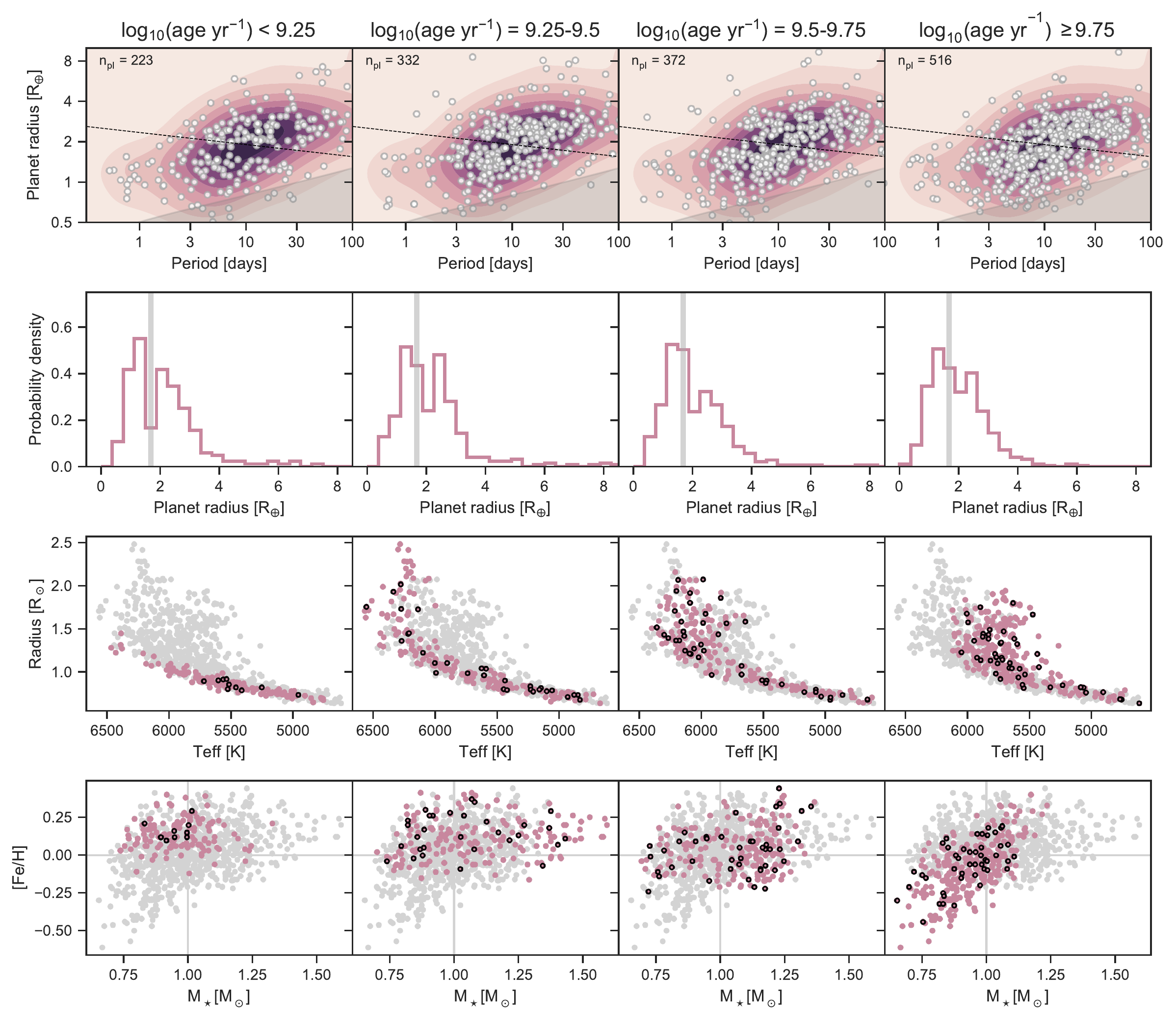}
    \caption{Same as Figure~\ref{fig:pr} for the base sample.}
    \label{fig:pr-base}
\end{figure}

We first plotted the period-radius (P-R) diagram for CKS planets in four bins of \logage: $<$9.25, 9.25--9.5, 9.5--9.75, and $\geq9.75$ for both the filtered and base samples (Figures~\ref{fig:pr} and \ref{fig:pr-base}, respectively).\footnote{The completeness curve in Figures~\ref{fig:pr} and \ref{fig:pr-base} was computed for CKS stars using CKS stellar parameters and the methodology of \citet{Burke2015} as implemented in Python at \url{https://dfm.io/posts/exopop/}.} This binning scheme was in part chosen because there are few planets with \logage $<9$ or $>10$. From these figures we observed a conspicuous void of planets around the radius valley in the youngest age bin ($\lesssim$1.8~Gyr). Moreover, the slope of the void appears to be very close to the slope of the radius valley determined in \citet{vanEylen2018}, hereafter V18, from a subset of planets orbiting stars with precise asteroseismic parameters. However, while the slope of the young planet void appears to be consistent with that of the radius valley, the intercept appears to be different. This is evident from the fact that the lower boundary of young sub-Neptunes straddles the V18 line, while the super-Earths are well-separated from the V18 valley. In other words, there appears to be a dearth of large super-Earths at younger ages, resulting in an apparent shift in the peak of the super-Earth radius distribution to larger radii at older ages. Our interpretation of this shift is discussed in \S\ref{sec:conclusions}. 

While age, mass, and metallicity are correlated in the CKS sample, we show that the distribution of masses in each age bin is not changing drastically. The age-metallicity gradient is stronger, and essentially all stars in the youngest age bin are metal-rich. However, it is also clear that stars hosting planets in the radius valley are not exclusively metal-poor but rather have a wide range of metallicities. Additionally, while the [Fe/H] distributions in the two youngest age bins are broadly similar, the distributions of planets in the P-R diagram are markedly different. While these observations offer some degree of assurance that observed features in the P-R diagram are due to age, rather than mass or metallicity, we explore the effects of stellar mass and metallicity further in \S\ref{subsec:binning} and \S\ref{subsec:distributions}.

\subsection{Evolution of the period-radius diagram: gyrochronology}
\label{subsec:prgyro}
To this point we have only considered isochrone ages from F18 in our analysis. While we present a qualitative validation of the F18 ages in Appendix \ref{sec:appendixb}, there are substantial uncertainties associated with isochrone ages. It is also possible to empirically age-rank the CKS sample with a gyrochronology analysis. Recently, \citet{Curtis2020} presented empirical gyrochrones for several open clusters which enable high-fidelity, model-independent age ranking of solar-type stars with well-determined \teff and rotation periods. 

We separated the CKS sample into ``fast" rotators and ``slow" rotators using the hybrid NGC~6819 + Ruprecht~147 gyrochrone of \citet{Curtis2020}, corresponding to an age of $\sim$2.7~Gyr. To perform this cut we first converted the gyrochrone from a ($B_P$-$R_P$)--\prot relation into a \teff--\prot relation using the color-temperature polynomial relation presented by those authors (valid for the temperatures considered here).\footnote{We opted not to perform the gyrochronology classification in color space because we noted increased scatter in the ($B_P$-$R_P$)--\prot diagram for CKS stars, possibly a result of reddening, metallicity effects, or both.} 

We constructed the young planet sample from the CKS base sample (\S\ref{sec:sample}) by choosing host stars placed between the 0.12~Gyr (Pleiades) and 2.7~Gyr gyrochrones in the \teff-\prot diagram, high reliability \prot flags, RUWE$<1.4$ (to remove unresolved binaries with unreliable rotation periods), and \teff $<6000$~K. The \teff$<6000$~K cut is motivated by the fact that gyrochrones cluster closely for hotter stars, and small temperature uncertainties can translate to large uncertainties in age from a gyrochronology analysis. 

The old rotation-selected planet sample was selected in the same fashion from stars lying \textit{above} the 2.7~Gyr gyrochrone, except for the high reliability \prot flag requirement. We found that we assigned the high reliability flag more often to faster rotators which tend to exhibit higher amplitude and more stable brightness modulations (presumably due to larger, longer-lived spots), while the older, more slowly rotating stars exhibit smaller-amplitude, more sporadic, Sun-like variations potentially due to smaller, short-lived spots. Thus, in the slow-rotator sample we accepted stars with either reliable or highly reliable \prot flags. 

\begin{figure}
    \centering
    \includegraphics[width=\linewidth]{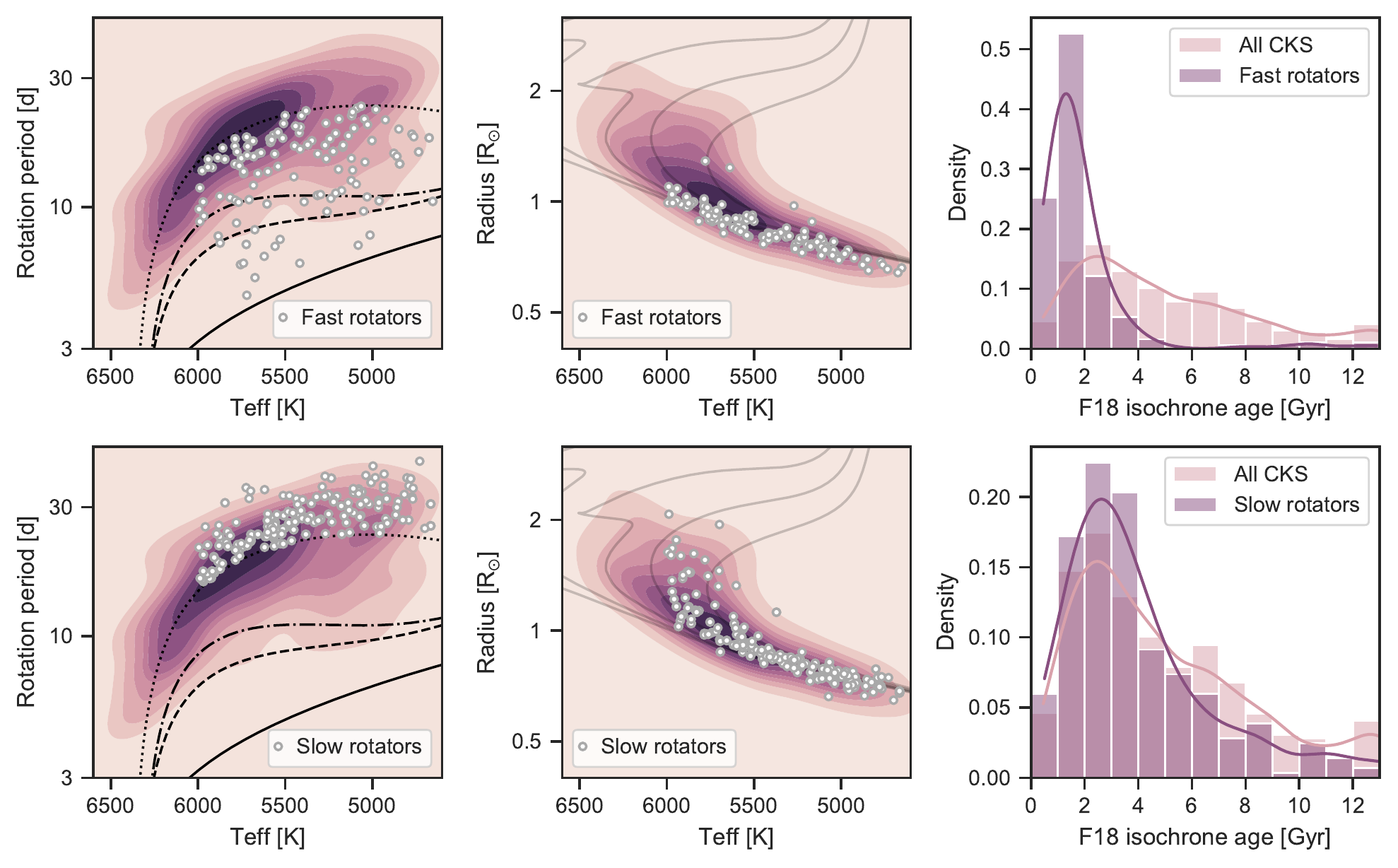}
    \caption{The \teff-\prot diagram (left column), H-R diagram (middle column), and median isochrone age distributions (right column) for the young (top row) and old (bottom row) rotation-selected samples. The shaded contours in the left and middle columns show Gaussian kernel density estimates of the full CKS sample in those respective planes. In the left column, the solid, dashed, dash-dotted, and dotted lines indicate polynomial fits to the empirical gyrochrones of the Pleiades ($\approx$0.12~Gyr), Praesepe ($\approx$0.63~Gyr), NGC~6811 ($\approx$1~Gyr), and NGC~6819 + Ruprecht~147 ($\approx$2.7~Gyr) clusters respectively \citep{Curtis2020}. In the middle panels, the grey curves show solar-metallicity, non-rotating MIST v1.1 isochrones from \logage=9--10 in steps of 0.25 dex.}
    \label{fig:gyro-sample}
\end{figure}

The distributions of the young and old rotation-selected planet hosts in the \teff-\prot and H-R diagrams are shown in Figure~\ref{fig:gyro-sample}, along with the 1-d distributions of their median isochrone ages. We observe that the fast rotating sample does indeed correspond to stars that lie closer to the ZAMS with median isochrone ages strongly skewed towards younger ages (mostly below 3~Gyr) relative to the CKS sample. The more slowly rotating stars show a distribution of median isochrone ages which is practically indistinguishable from the bulk of the CKS sample, but with a significant number of stars in the $\sim$1--3~Gyr range. However, this is likely to be the result of isochrone clustering on the main sequence and not because those stars are actually young. We also note that only 46\% (546/1189) of the stars in our sample have assigned rotation periods from our period vetting procedure. For the remaining stars it was not possible to unambiguously assign a period. Those stars are preferentially more evolved relative to the periodic sample, though they are observed across the H-R diagram. Thus, the modest decrement at old ages in the age distribution of the slow rotators relative to the overall CKS sample may be the result of a finite active lifetime for solar-type stars. 

\begin{figure}
    \centering
    \includegraphics[width=\linewidth]{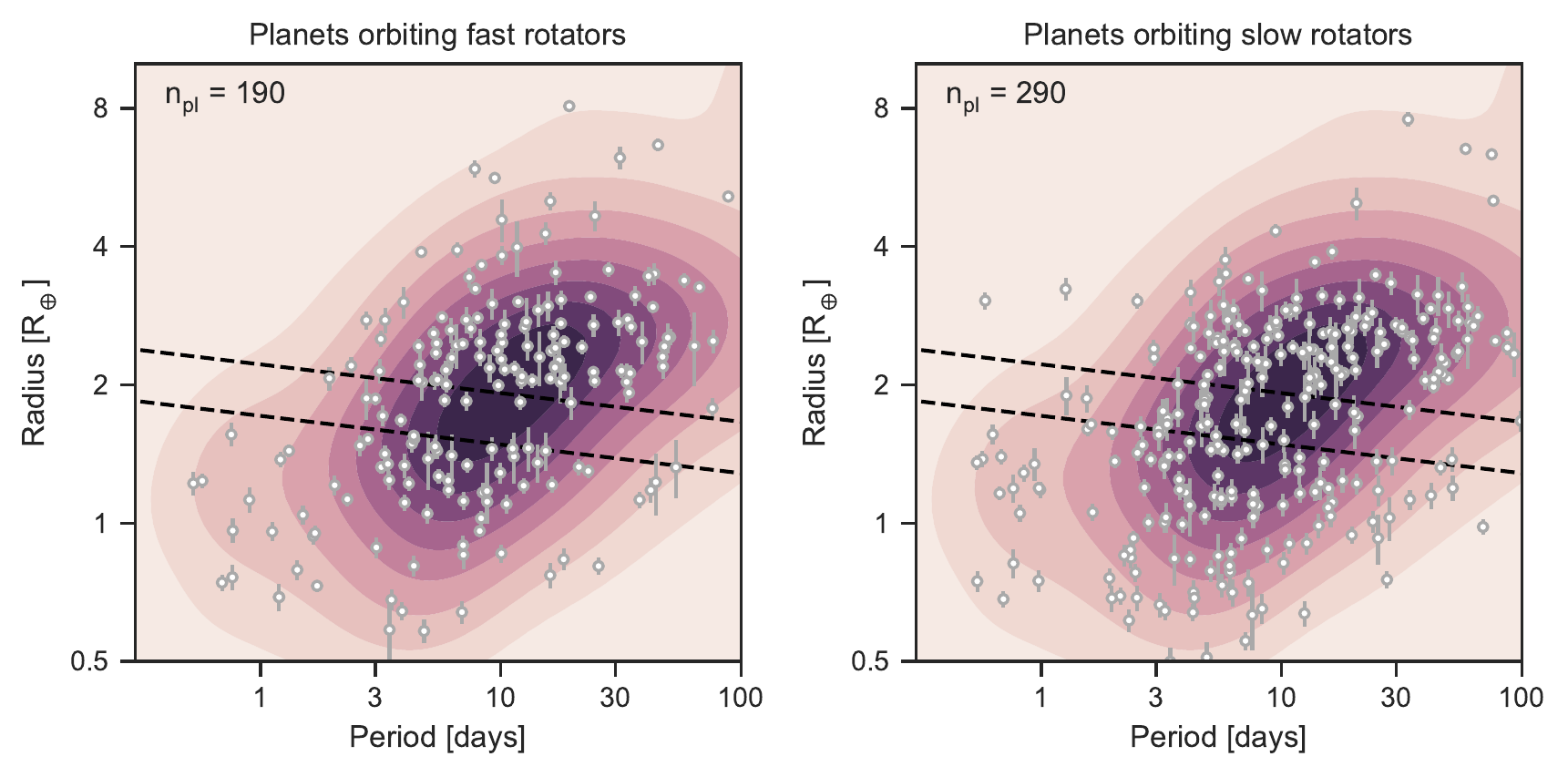}
    \caption{The period-radius diagram for exoplanets orbiting stars rotating more rapidly (left) or more slowly (right) than an empirical 2.7~Gyr gyrochrone. The shaded contours represent a 2D Gaussian kernel density estimation for the overall CKS sample. The black dashed lines indicate the margins of the young planet void \emph{derived from the isochrone-selected sample} in
    \S\ref{subsec:void}.}
    \label{fig:pr-gyro}
\end{figure}

After separating the planet hosts into the fast and slow rotator samples, we then examined the distributions of the corresponding planet populations in the P-R diagram (Figure~\ref{fig:pr-gyro}). We observe qualitatively similar behavior as to what was found when using isochrone ages to perform age cuts. That is, there is a dearth of exoplanets in the radius valley among planets empirically determined to be younger than $\sim$2.7~Gyr from a gyrochronology analysis. Among the planets older than $\sim$2.7~Gyr the radius valley appears more filled in. Moreover, the slope and boundaries of the radius valley in the left panel of Fig.~\ref{fig:pr-gyro} appear to be very close to that derived from the isochrone age-selected sample (as described in \S\ref{subsec:void}). 

\subsection{Measuring the slope of the void}
\label{subsec:void}

\begin{figure}
    \centering
    \includegraphics[width=0.8\linewidth]{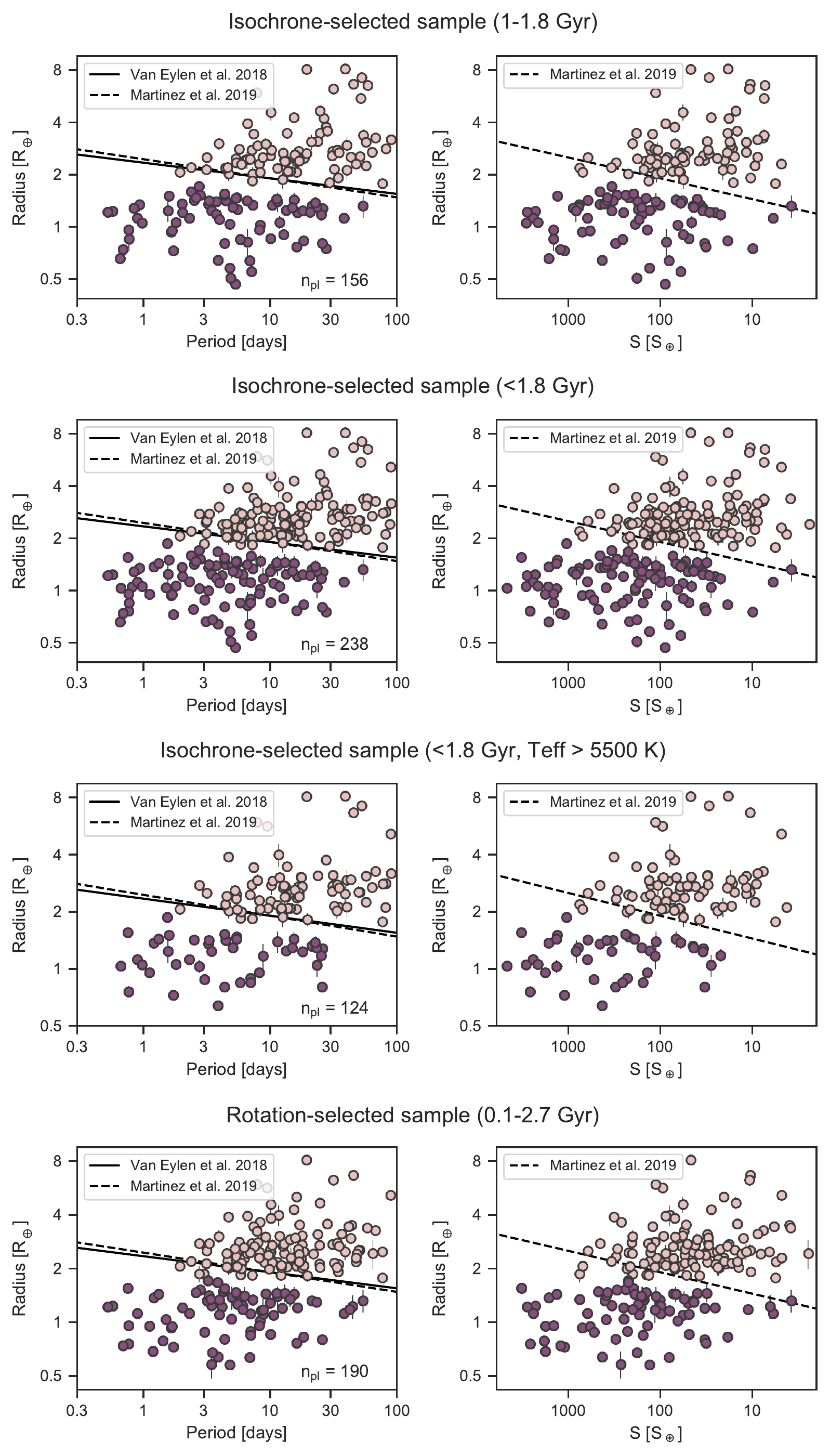}
    \caption{Age-selected samples of planets in the period-radius (left column) and insolation-radius (right column) planes. Age selections are described in \S\ref{sec:sample}. Previously determined equations for the radius valley are shown as the black lines. Point colors indicate the classification used in the SVM analysis (described in \S\ref{subsec:void}).}
    \label{fig:samples}
\end{figure}

As the slope of the radius valley contains information about the mechanism(s) responsible for producing it, we proceeded to characterize the void for four planetary samples described as follows. Each sample is a subset of the base sample, sharing the following cuts: planets orbiting dwarf stars (described in \S\ref{sec:sample}) with $P<100$~d, $R_P<10$~\rearth, and $\sigma_{R_P}/R_P<20$\%. The \samplea sample also employs the age restriction $9 < \logage \leq 9.25$, while the \sampleb sample is produced from the more inclusive criterion $\logage \leq 9.25$. As isochrone ages for hotter stars are more reliable than those of cooler stars, the \samplec sample combines the criteria $\logage \leq 9.25$ and \teff $>5500$~K. Finally, the \sampled sample combines the common cuts with the following criteria: \teff $< 6000$~K, RUWE $< 1.4$, high reliability rotation periods (reliability flag of 3), and positions in the \teff-\prot plane between the empirical Pleiades and NGC~6819+Ruprecht~147 gyrochrones of \citet{Curtis2020}. The distributions of these planetary samples in the P-R and insolation-radius planes are depicted in Figure~\ref{fig:samples}. 

Following the approach of V18 we used support vector machines (SVMs) to find the decision boundary that maximizes the margins between two distinct classes of planets in the period-radius and insolation-radius planes. To label the planets we found that shifting the V18 radius valley equation downwards by \rvshift in \logrp provided an unambiguous separation of planets into two classes for the \samplea sample. Thus, we used the equation, $\log_{10}(R_P/R_\oplus) = -0.09 \log_{10}(P/\text{d}) + 0.3$, to label planets as sub-Neptunes or super-Earths. To implement the SVM classification we used the \texttt{sklearn.svm.SVC} module in Python with a linear kernel \citep{scikit-learn}. 

\begin{figure}
    \centering
    \includegraphics[width=\linewidth]{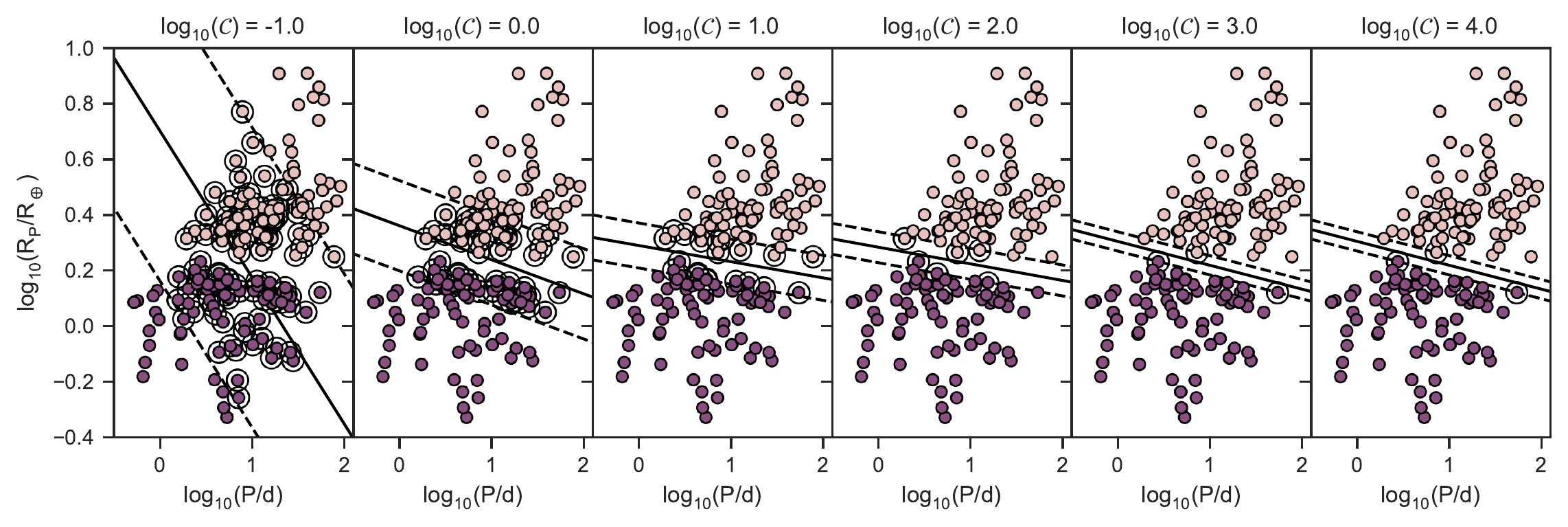}
    \caption{The effect of the regularization parameter, $\mathcal{C}$, on the support vector classification for young CKS planets (the \samplea sample) in the period-radius diagram. In each panel point colors indicate the planet classification provided in the SVM classification analysis. Points encircled by black rings indicate the support vectors. The solid lines indicate the decision surface of maximal separation, while the dashed lines indicate the margins (as discussed in the text).}
    \label{fig:regularization}
\end{figure}

We explored the sensitivity of our results to the regularization parameter, $\mathcal{C}$, finding that for $\mathcal{C}<5$ the SVM misclassifies a large fraction of planets and fails to trace the center of the void which is so readily visible by eye (Figure~\ref{fig:regularization}). In determining the equation of the void we ultimately adopt the slope and intercept derived from the $\mathcal{C}=10$ case, but recommend $\mathcal{C}=1000$ for determining the upper and lower boundaries of the void. To calculate uncertainties on the slope and intercept of the radius valley, we performed $10^3$ bootstrapping simulations, selecting 50 planets (with replacement) randomly from the young planet samples and recording the slope and intercept resulting from the SVM classification for each bootstrapped sample. 

\begin{figure}
    \centering
    \includegraphics[width=\linewidth]{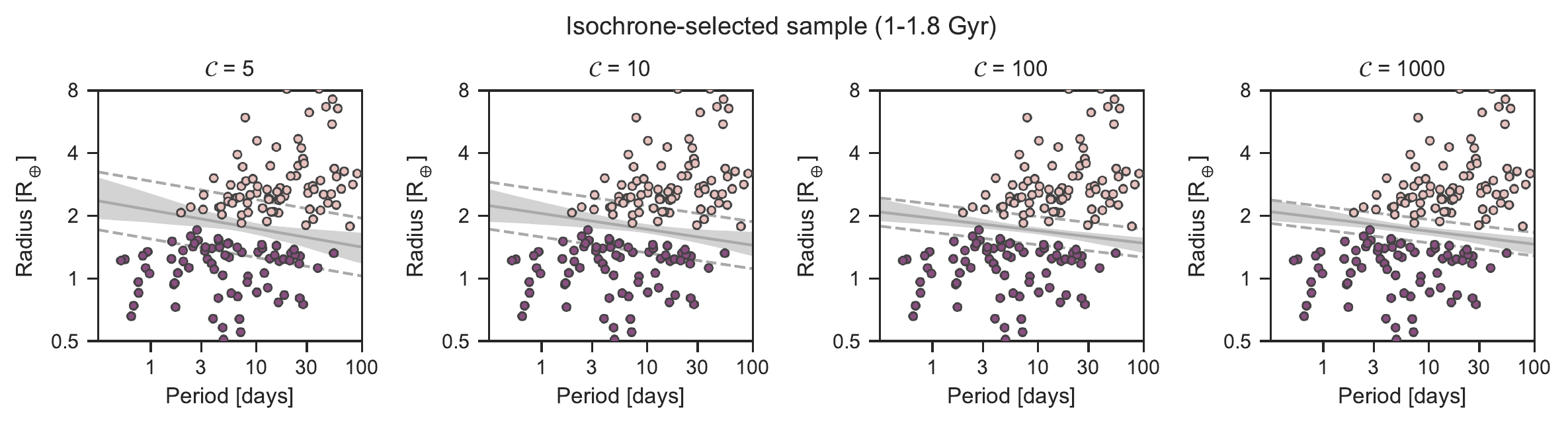}
    \includegraphics[width=\linewidth]{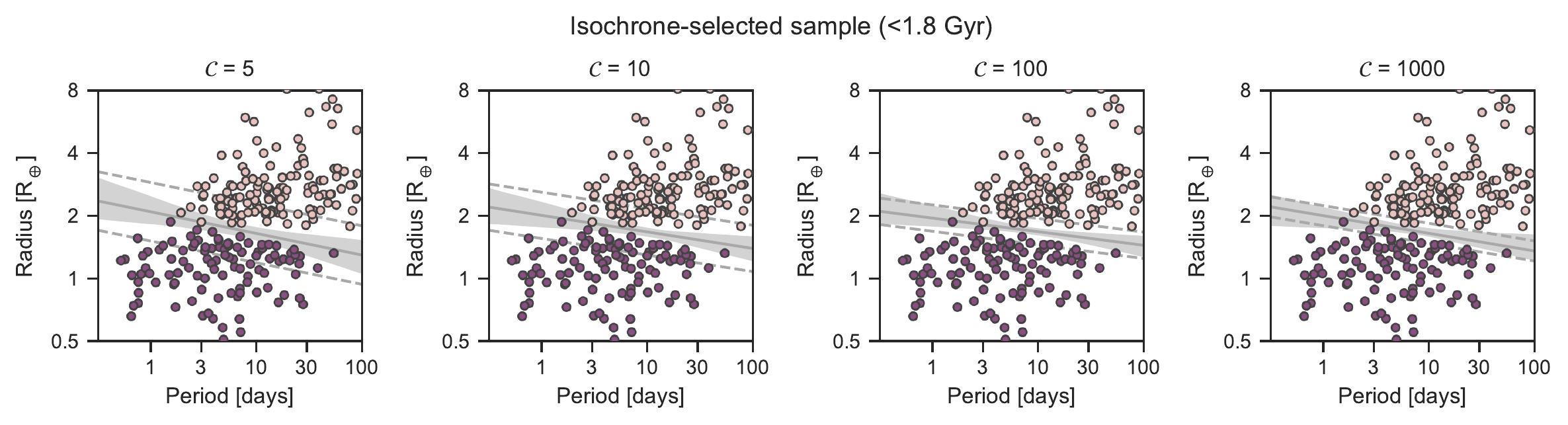}
    \includegraphics[width=\linewidth]{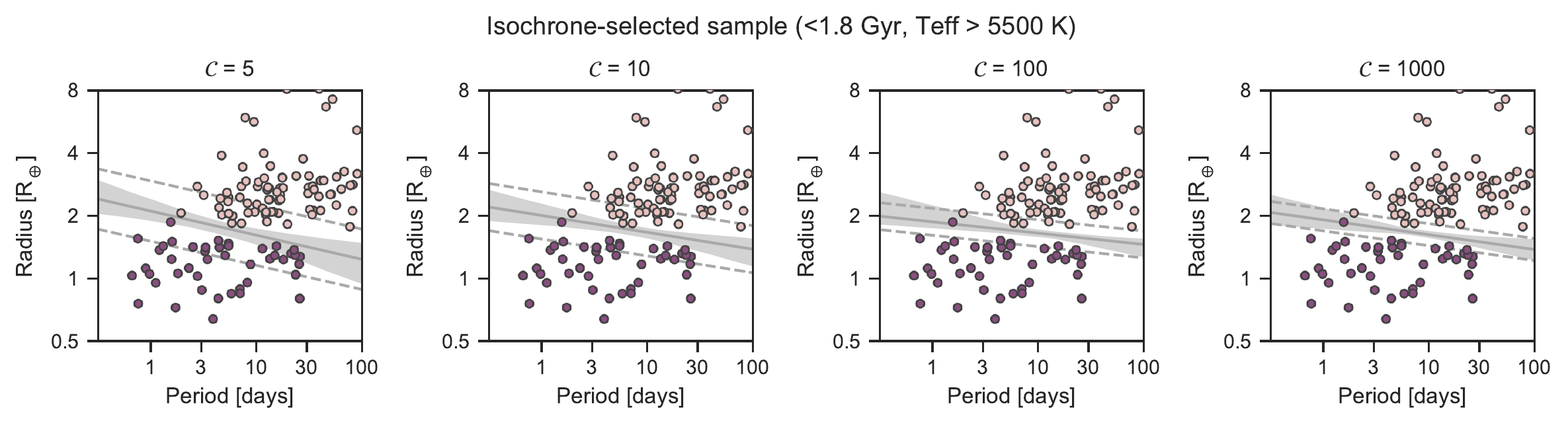}
    \includegraphics[width=\linewidth]{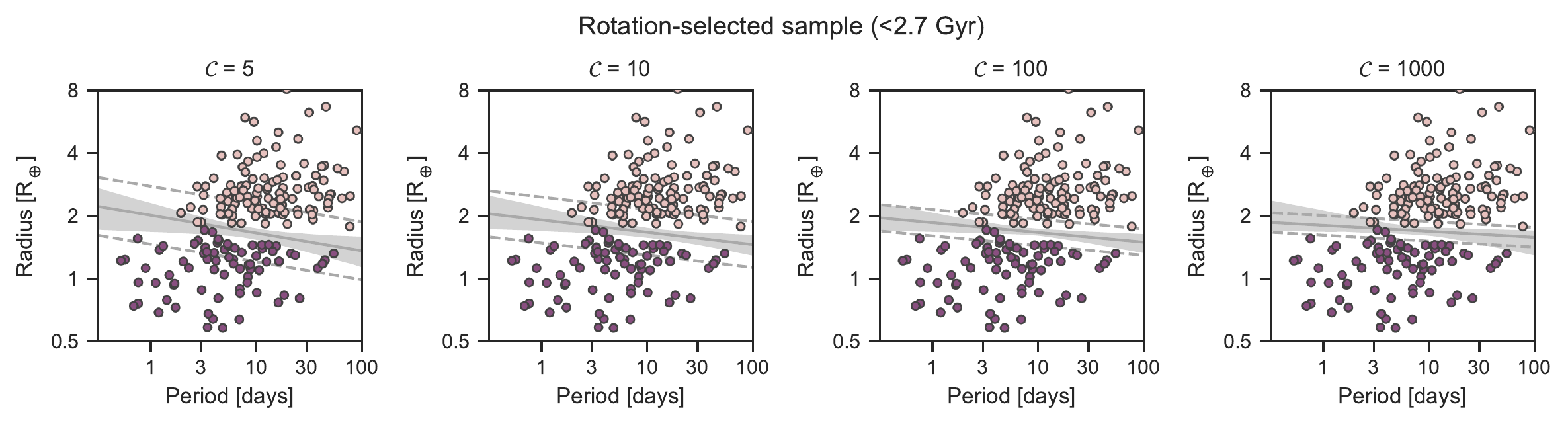}
    \caption{Period-radius diagram for planets in the \samplea (top row), \sampleb (second row), \samplec (third row), and \sampled samples. Point colors indicate the classifications used in the support vector machine (SVM) analysis. The grey line and shaded region show the median and 16th-84th percentile width of the radius valley from the SVM bootstrapping simulations. The dashed lines indicate the median margins from the SVM analysis. The regularization parameter, $\mathcal{C}$, is indicated at the top of each panel.}
    \label{fig:rv}
\end{figure}

\begin{figure}
    \centering
    \includegraphics[width=\linewidth]{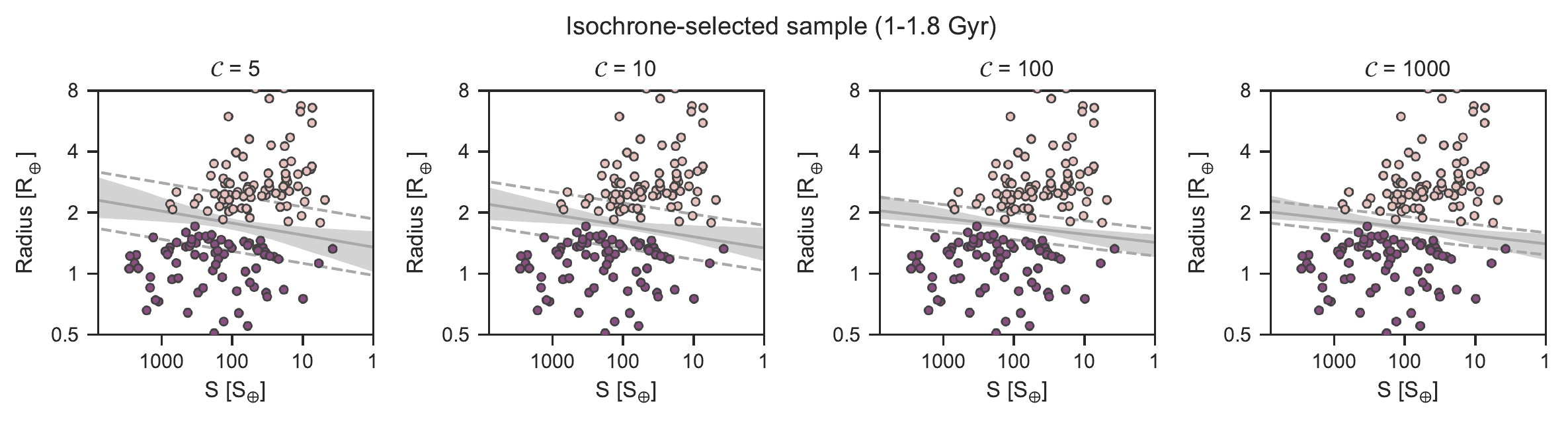}
    \includegraphics[width=\linewidth]{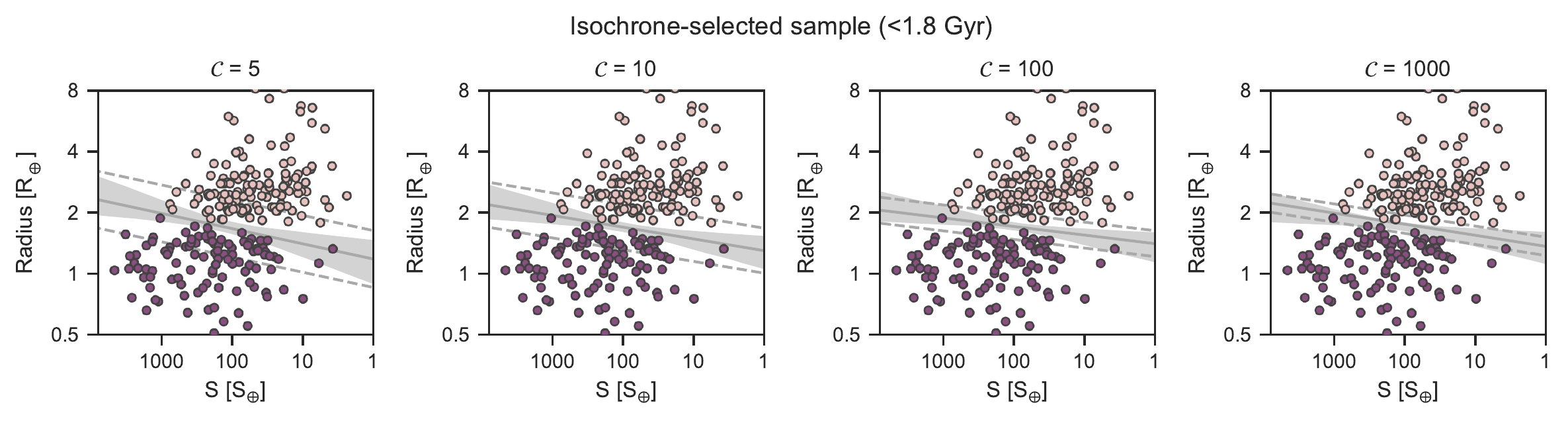}
    \includegraphics[width=\linewidth]{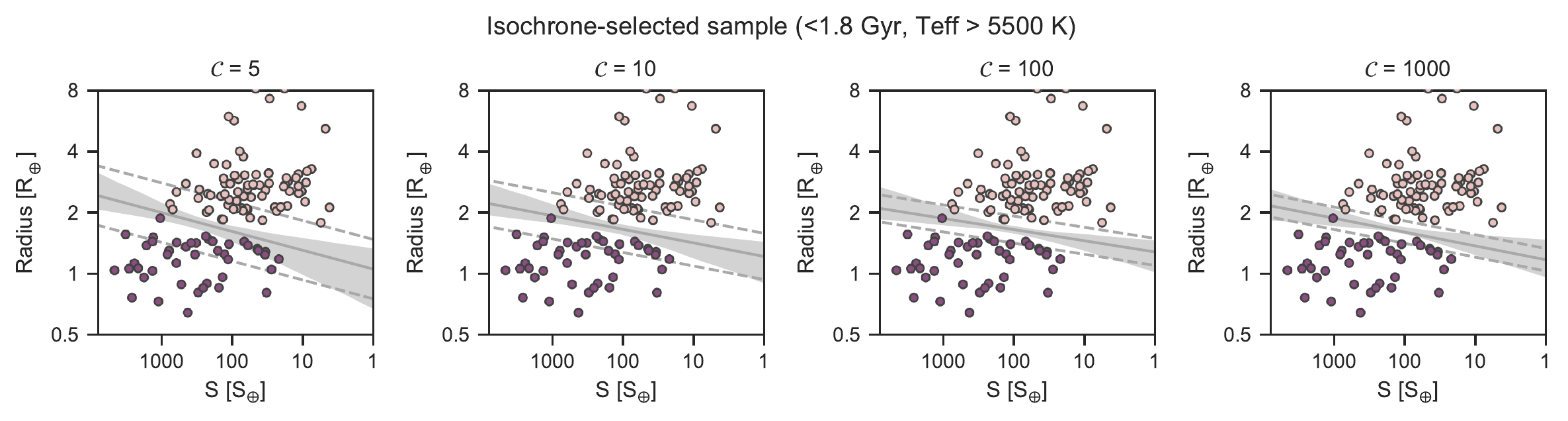}
    \includegraphics[width=\linewidth]{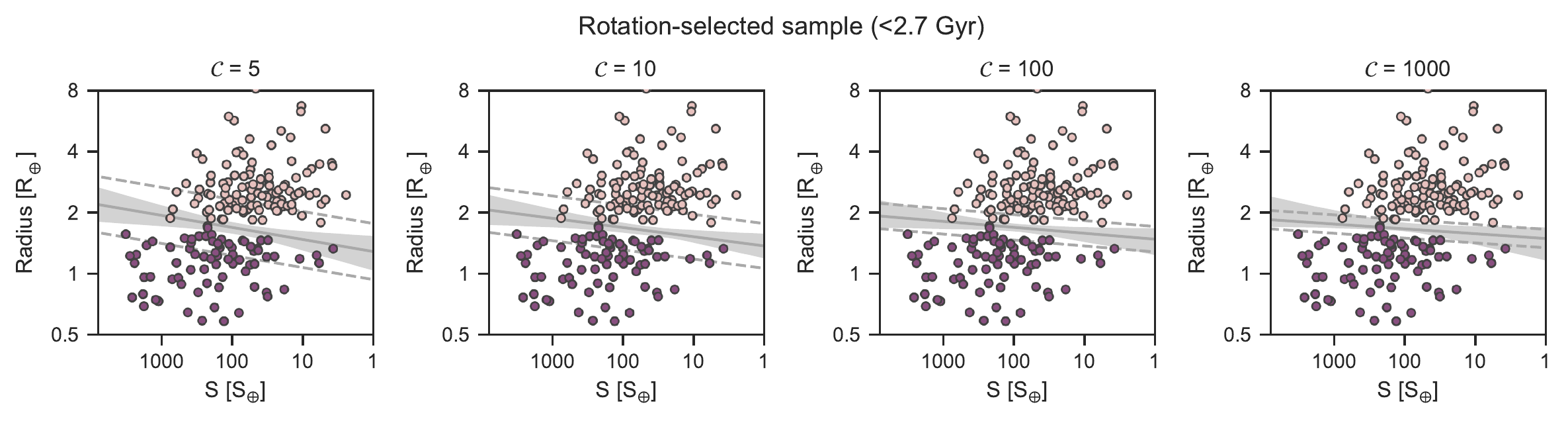}
    \caption{Same as Figure~\ref{fig:rv} for the insolation-radius plane.}
    \label{fig:rvs}
\end{figure}

\begin{figure}
    \centering
    \includegraphics[width=\linewidth]{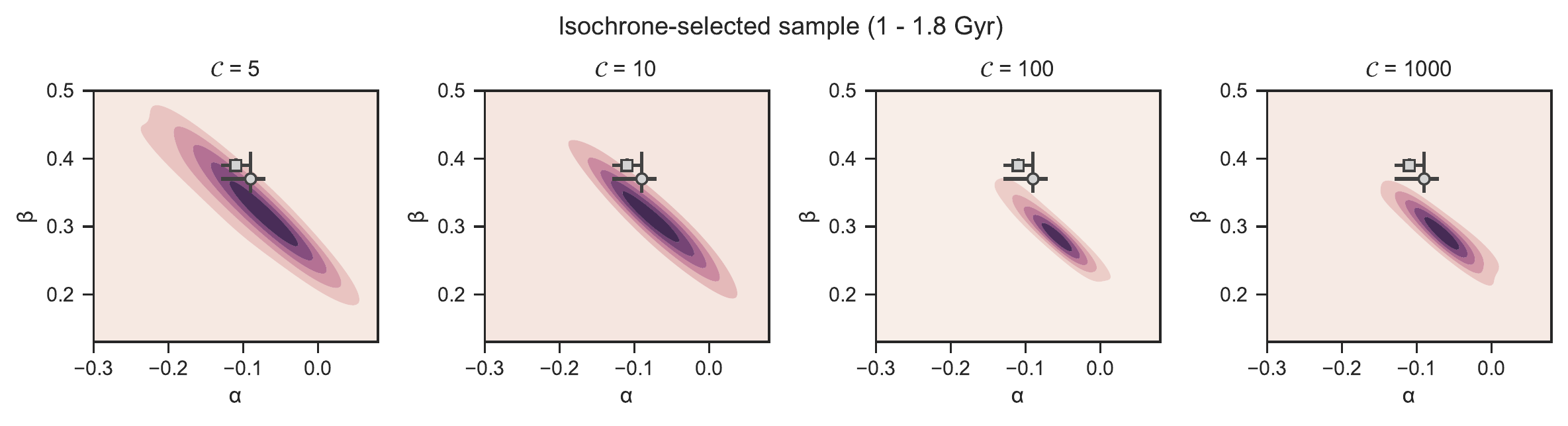}
    \includegraphics[width=\linewidth]{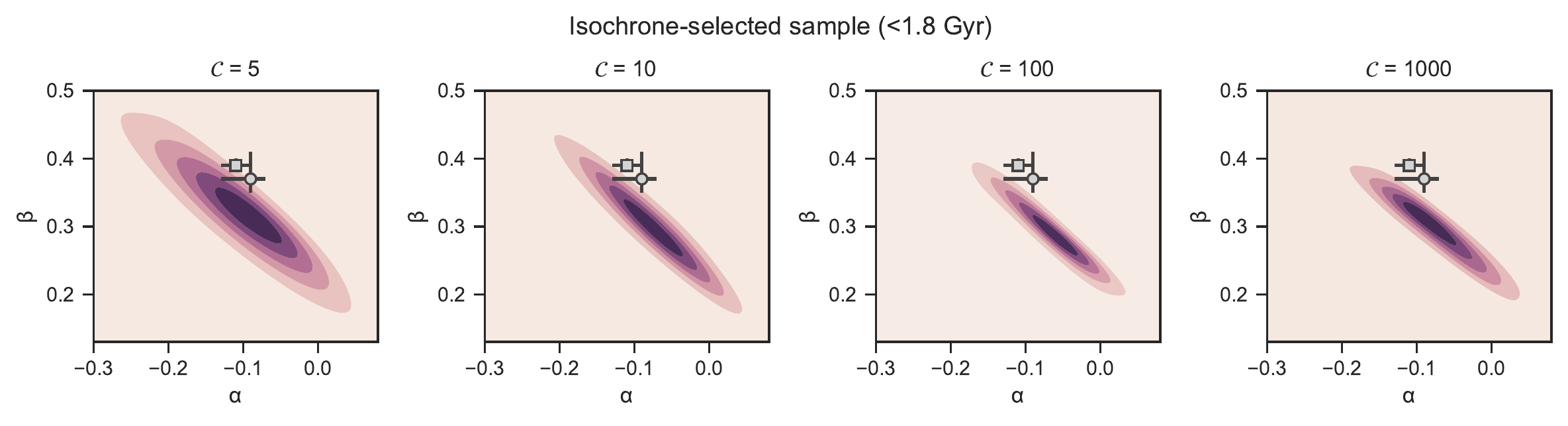}    
    \includegraphics[width=\linewidth]{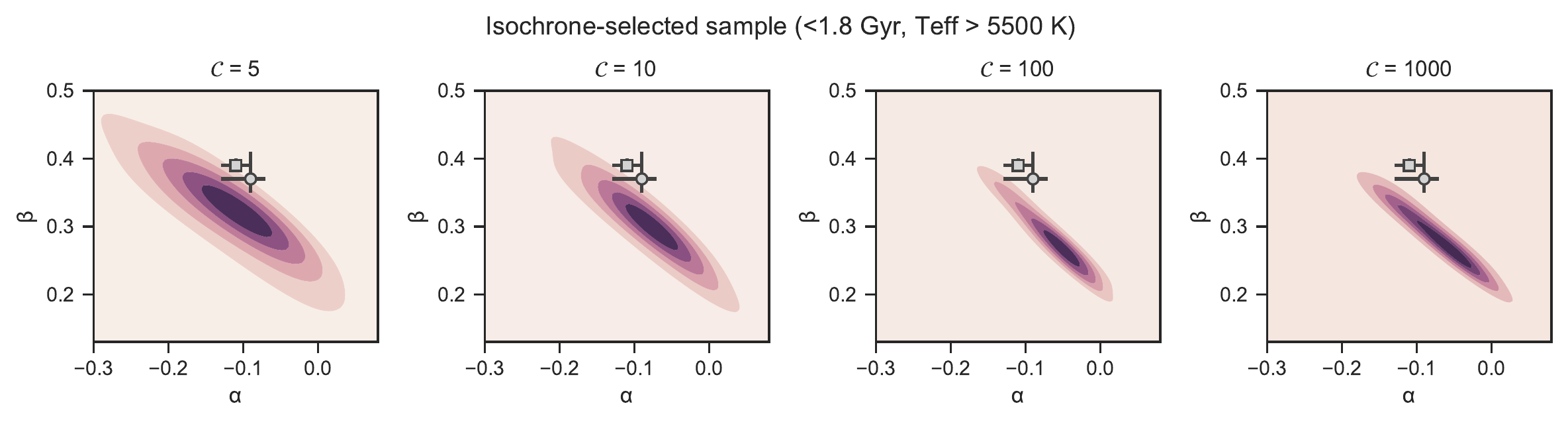}
    \includegraphics[width=\linewidth]{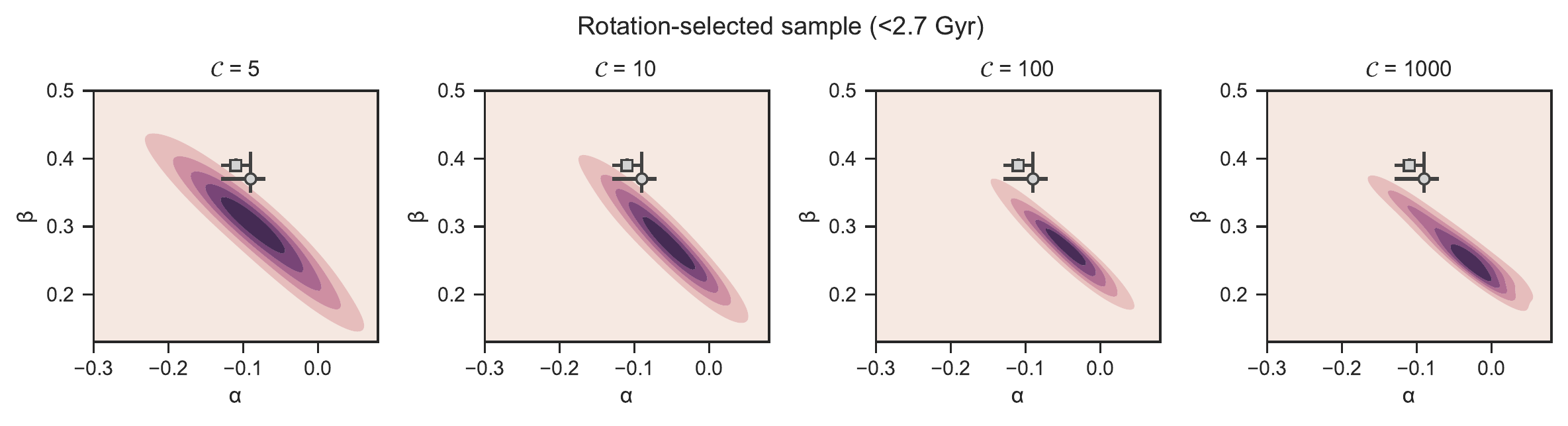}
    \caption{Gaussian kernel density estimation of the distribution of radius valley slopes ($\alpha$) and intercepts ($\beta$) from the SVM bootstrapping simulations with different regularization ($\mathcal{C}$) parameters. The circle with errorbars indicates the values derived in \citet{vanEylen2018} from planets orbiting asteroseismic stars. The square with errorbars indicates the values derived by \citet{Martinez2019} from an independent spectroscopic analysis of the CKS sample.}
    \label{fig:joint}
\end{figure}

Table~\ref{tab:svm} lists the slopes and intercepts for the young planet void inferred from the SVM bootstrapping simulations. Figures~\ref{fig:rv} and \ref{fig:rvs} show the derived radius valley from the bootstrapping simulations and Figure~\ref{fig:joint} shows the distributions of slopes and intercepts from this analysis. For the $\mathcal{C}$ values explored here, the inferred slopes and intercepts of the radius valley are relatively constant. We find in almost all cases that the slope of the valley is consistent with the slope found in V18 at the $\lesssim 1\sigma$ level. However, we find an intercept that is systematically smaller than that found by V18 and \citet{Martinez2019}, by at least 2$\sigma$ and in some cases as much as 10$\sigma$ using the quoted uncertainties from those works. While the statistical significance of this difference is highly dependent on the adopted uncertainty (where ours appear to generally be larger), it is clear from Figure~\ref{fig:samples} that the void we observe is offset from previous determinations of the radius valley. We note that previous works characterized the radius valley among samples with a broader range of ages, while the focus of this analysis is on the younger planets in the CKS sample. In \S\ref{sec:conclusions} we discuss our interpretation of this difference. The level of agreement between the radius valley slopes derived here and in V18 is noteworthy given that the samples we characterize are $\approx$30--100\% larger and selected on the basis of age, rather than radius precision which was the impetus for the V18 sample. 

The upper and lower boundaries of the radius valley are given by the equation:

\begin{equation}
\log_{10}(R_P/R_\oplus)^{\text{upper}}_{\text{lower}} = \alpha \log_{10}(P/\text{d}) + \beta   \pm \gamma \sqrt{1+\alpha^2},    
\end{equation}

in the period-radius plane or,

\begin{equation}
\log_{10}(R_P/R_\oplus)^{\text{upper}}_{\text{lower}} = \delta \log_{10}(S_\mathrm{inc}/S_\oplus) + \epsilon   \pm \zeta \sqrt{1+\delta^2},    
\end{equation}

in the insolation-radius plane. We use the highest $\mathcal{C}$ parameter explored here for determination of the radius valley boundaries as it provides the closest match to the data (i.e. the fewest planets inside those boundaries). 

\begin{deluxetable}{cccccccc}
\tablecaption{Results of SVM bootstrapping simulations.}
\label{tab:svm}
\tablecolumns{8}
\tablewidth{\linewidth}
\tablehead{\colhead{Sample} & \colhead{$\mathcal{C}$} & \colhead{$\alpha$} & \colhead{$\beta$} & \colhead{$\gamma$} & \colhead{$\delta$} & \colhead{$\epsilon$} & \colhead{$\zeta$}}
\startdata
\samplea & 5 & $-0.09^{+0.06}_{-0.06}$ & $0.33^{+0.06}_{-0.07}$ & $0.14^{+0.01}_{-0.01}$ & $0.06^{+0.04}_{-0.05}$ & $0.13^{+0.12}_{-0.06}$ & $0.14^{+0.01}_{-0.01}$\\
\samplea & 10 & $\mathbf{-0.08^{+0.06}_{-0.04}}$ & $\mathbf{0.31^{+0.05}_{-0.05}}$ & $0.11^{+0.01}_{-0.01}$ & $\mathbf{0.06^{+0.03}_{-0.04}}$ & $\mathbf{0.13^{+0.11}_{-0.04}}$ & $0.11^{+0.01}_{-0.01}$ \\
\samplea & 100 & $-0.06^{+0.02}_{-0.04}$ & $0.29^{+0.02}_{-0.04}$ & $0.07^{+0.01}_{-0.01}$ & $0.04^{+0.02}_{-0.03}$ & $0.15^{+0.04}_{-0.06}$ & $0.07^{+0.01}_{-0.01}$\\
\samplea & 1000 & $-0.06^{+0.02}_{-0.04}$ & $0.29^{+0.03}_{-0.03}$ & $0.06^{+0.01}_{-0.01}$ & $0.04^{+0.04}_{-0.01}$ & $0.15^{+0.04}_{-0.08}$ & $0.05^{+0.01}_{-0.01}$\\
\hline
\sampleb & 5 & $-0.1^{+0.07}_{-0.05}$ & $0.32^{+0.06}_{-0.06}$ & $0.14^{+0.01}_{-0.01}$ & $0.08^{+0.04}_{-0.05}$ & $0.07^{+0.1}_{-0.09}$ & $0.14^{+0.01}_{-0.01}$\\
\sampleb & 10 & $-0.08^{+0.05}_{-0.05}$ & $0.3^{+0.04}_{-0.07}$ & $0.11^{+0.01}_{-0.01}$ & $0.06^{+0.04}_{-0.04}$ & $0.11^{+0.08}_{-0.07}$ & $0.11^{+0.01}_{-0.01}$\\
\sampleb & 100 & $-0.07^{+0.04}_{-0.04}$ & $0.29^{+0.04}_{-0.05}$ & $0.06^{+0.01}_{-0.01}$ & $0.04^{+0.04}_{-0.03}$ & $0.15^{+0.06}_{-0.06}$ & $0.06^{+0.01}_{-0.01}$\\
\sampleb & 1000 & $-0.08^{+0.07}_{-0.03}$ & $0.3^{+0.04}_{-0.05}$ & $0.05^{+0.01}_{-0.02}$ & $0.05^{+0.02}_{-0.04}$ & $0.13^{+0.08}_{-0.07}$ & $0.05^{+0.01}_{-0.02}$\\
\hline
\samplec & 5 & $-0.12^{+0.06}_{-0.06}$ & $0.32^{+0.05}_{-0.05}$ & $0.14^{+0.01}_{-0.02}$ & $0.09^{+0.05}_{-0.05}$ & $0.02^{+0.15}_{-0.11}$ & $0.15^{+0.01}_{-0.02}$\\
\samplec & 10 & $-0.08^{+0.05}_{-0.04}$ & $0.3^{+0.05}_{-0.04}$ & $0.11^{+0.01}_{-0.01}$ & $0.07^{+0.03}_{-0.04}$ & $0.08^{+0.11}_{-0.05}$ & $0.11^{+0.01}_{-0.01}$\\
\samplec & 100 & $-0.05^{+0.03}_{-0.05}$ & $0.27^{+0.04}_{-0.04}$ & $0.06^{+0.01}_{-0.01}$ & $0.06^{+0.02}_{-0.04}$ & $0.11^{+0.08}_{-0.05}$ & $0.07^{+0.01}_{-0.01}$\\
\samplec & 1000 & $-0.07^{+0.04}_{-0.04}$ & $0.28^{+0.04}_{-0.04}$ & $0.05^{+0.02}_{-0.01}$ & $0.07^{+0.04}_{-0.03}$ & $0.07^{+0.08}_{-0.09}$ & $0.06^{+0.01}_{-0.01}$\\
\hline
\sampled & 5 & $-0.09^{+0.08}_{-0.04}$ & $0.3^{+0.04}_{-0.08}$ & $0.14^{+0.01}_{-0.01}$ & $0.06^{+0.04}_{-0.04}$ & $0.11^{+0.09}_{-0.07}$ & $0.14^{+0.01}_{-0.01}$\\
\sampled & 10 & $-0.06^{+0.06}_{-0.04}$ & $0.28^{+0.05}_{-0.05}$ & $0.11^{+0.01}_{-0.01}$ & $0.05^{+0.03}_{-0.03}$ & $0.14^{+0.06}_{-0.05}$ & $0.11^{+0.01}_{-0.01}$\\
\sampled & 100 & $-0.05^{+0.04}_{-0.03}$ & $0.27^{+0.03}_{-0.05}$ & $0.06^{+0.01}_{-0.01}$ & $0.03^{+0.02}_{-0.04}$ & $0.17^{+0.06}_{-0.05}$ & $0.06^{+0.01}_{-0.01}$\\
\sampled & 1000 & $-0.03^{+0.02}_{-0.06}$ & $0.26^{+0.04}_{-0.04}$ & $0.05^{+0.02}_{-0.01}$ & $0.02^{+0.05}_{-0.02}$ & $0.17^{+0.07}_{-0.07}$ & $0.05^{+0.01}_{-0.02}$\\
\enddata
\tablecomments{Equation for the radius valley in the period-radius diagram is of the form $\log_{10}(R_P/R_\oplus) = \alpha \log_{10}(P/\text{d}) + \beta$. In the insolation-radius diagram it is $\log_{10}(R_P/R_\oplus) = \delta \log_{10}(S_\mathrm{inc}/S_{\oplus}) + \epsilon$. Adopted values in bold.}
\end{deluxetable}

\subsection{Calculation of false alarm probability}
\label{subsec:chance}

\begin{figure}
    \centering
    \includegraphics[width=\linewidth]{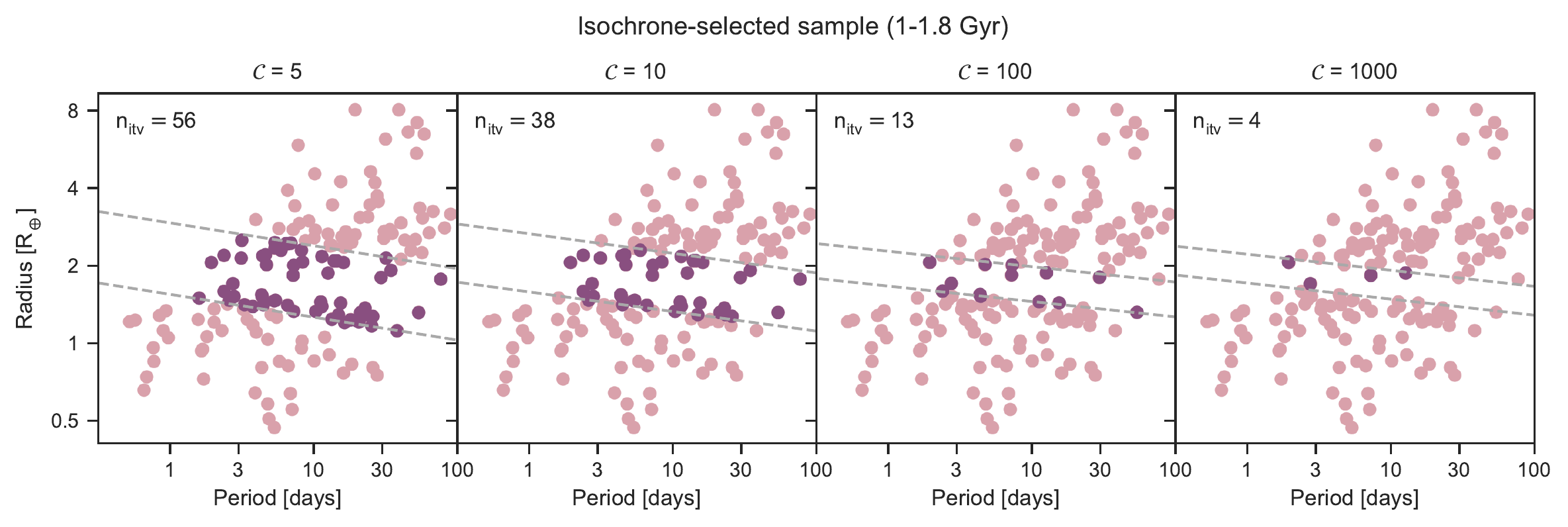}
    \caption{The effect of regularization parameter, $\mathcal{C}$, on the boundaries of the radius valley (indicated by the dashed lines) and hence the number of planets in the valley (dark points). The \samplea sample is shown for illustration.}
    \label{fig:margins}
\end{figure}

To determine whether the void we observe in the period-radius diagram could be due to chance, we performed simulations to determine the probability of finding $n_\mathrm{itv}$ or fewer planets in the void from $N$ planets selected at random (without replacement) from the CKS sample. Here $N$ is the total number of planets in each of the samples described in \S\ref{sec:sample}, i.e. \npla, \nplb, \nplc, and \npld for the \samplea, \sampleb, \samplec, and \sampled samples, respectively. The definition of the radius valley boundaries and hence the true number of planets in the valley, $n_\mathrm{itv,true}$, are sensitive to the regularization parameter, $\mathcal{C}$, and the specific sample used in the SVM bootstrapping analysis (see Figure~\ref{fig:margins}). For each sample and each $\mathcal{C}$ value we performed $10^4$ simulations, selecting $N$ planets randomly (without replacement) from the overall CKS sample, modeling the planet period and radius uncertainties with normal distributions, and recording the number of planets in the valley, $n_\mathrm{itv,sim}$. The false alarm probability was then computed as the fraction of total trials which satisfied the condition $n_\mathrm{itv,sim} \leq n_\mathrm{itv,true}$. The results of these simulations are tabulated in Table~\ref{tab:fap}. For $\mathcal{C}=5$, we find false alarm probabilities in the range of 18--30\%, but from Figure~\ref{fig:margins} it is clear that the SVM margins in this case are so wide as to not provide an accurate description of the void boundaries. The same may be argued for the $\mathcal{C}=10$ case, but even then we find false alarm probabilities $<$10\%. Finally, for the $\mathcal{C}=100,1000$ cases, for which the SVM margins precisely trace the boundaries of the void, we found false alarm probabilities $\ll$1\%. 

We also performed Monte Carlo simulations to estimate the probability that the void could be produced from random selection among those planets orbiting hosts rotating more slowly than the empirical 2.7~Gyr gyrochrone (see Fig.\ref{fig:gyro-sample}). We started by performing the same generic cuts on the overall CKS sample as were performed on the rotation-selected sample (main-sequence stars, no false positives, RUWE $<1.4$, \teff $< 6000$~K, and planets with $P<100$~d, $R_P<10$~\rearth, and radius precision $<20$\%). Then, in $10^4$ simulations we modeled the uncertainties in the planet periods and radii using normal distributions, selected \npld planets at random (without replacement) from the slow rotator sample, and computed the number of planets in the void. Here, \npld is the total number of planets in the fast rotator sample. We found that the probability of finding a comparably empty void from the slow rotator sample is $<$2\% when using the \sampled margins and $\mathcal{C}=100,1000$. Using the same $\mathcal{C}$ values but margins derived from the isochrone samples raises the false alarm probability, but only to $\sim$2--6\% at most. Taken together, we conclude that the emptiness of the observed void is not due to chance. 

\begin{deluxetable*}{rrrrr}
\tablecaption{False alarm probabilities.}
\label{tab:fap}
\tablecolumns{5}
\tablewidth{\linewidth}
\tablehead{\colhead{Sample} & \colhead{$\mathcal{C}$=5} &  \colhead{$\mathcal{C}$=10} & \colhead{$\mathcal{C}$=100} & \colhead{$\mathcal{C}$=1000}}
\startdata
\samplea & 30\% & 3.0\% & 0.09\% & $<0.01$\% \\
\sampleb & 18\% & 6.5\% & 0.08\% & 0.01\% \\
\samplec & 15\% & 14\% & 0.2\% & 0.1\%\\
\sampled & 22\% & 4.5\% & 0.03\% & 0.08\% \\
\enddata
\end{deluxetable*}

\subsection{Effects of stellar mass and metallicity}
\label{subsec:binning}
Our analysis separates the data set into age bins in order to understand the evolution of planets on a population level. Since age, mass, and metallicity are correlated in the CKS sample, it is difficult to entirely disentangle the effects of each parameter on the distributon of planets in the P-R diagram. We explored how sensitive our analysis is to specific binning schemes by recording the fractional number of planets in the valley over a two-dimensional grid of bin centers and bin widths in age, mass, and metallicity. We used the definition of the radius valley boundaries expressed in \S\ref{subsec:void} for this purpose (specifically, the margins given by the $\mathcal{C}=1000$, \samplea sample case). 

\begin{figure}
    \centering
    \includegraphics[width=\linewidth]{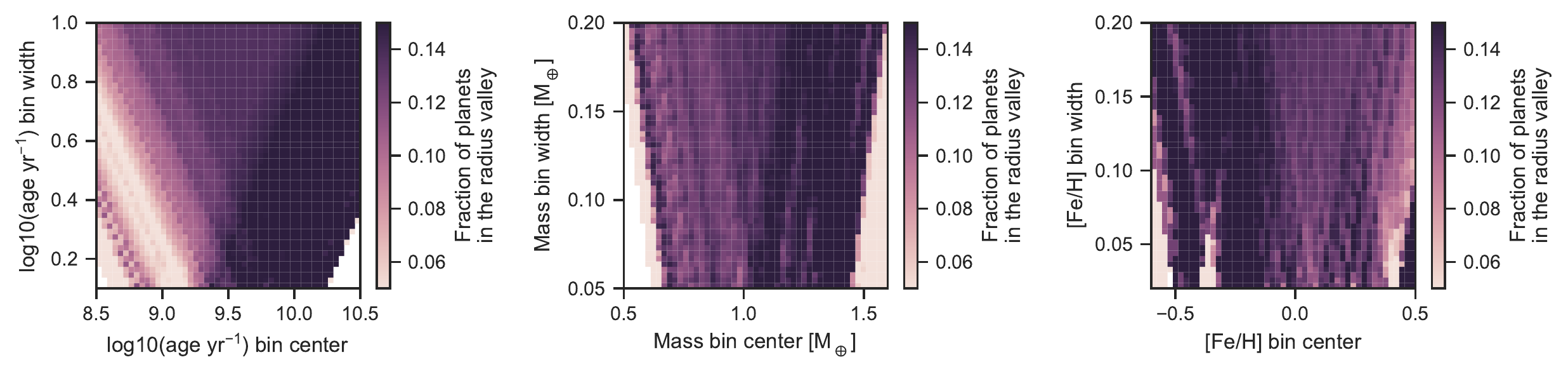}
    \caption{Effects of binning schemes (in age, mass, and metallicity from left to right) on the fractional occupancy of the radius valley. Note the color scaling is the same for each panel.}
    \label{fig:binning}
\end{figure}

Figure~\ref{fig:binning} shows the results of this exercise; the young planet void is apparent as the light, broad diagonal stripe in the left panel. For small bin widths, the minimum density of the radius valley is achieved for a bin center between $9 <$~\logage~$< 9.25$. This is unsurprising, as the boundaries of the radius valley were identified in and derived for just such a binning strategy. However, as the bin width in \logage increases, the minimum density of the valley shifts systematically towards younger bin centers. This suggests that the filling of the radius valley is due to preferentially older planets. Fig.~\ref{fig:binning} also reveals that there is no binning strategy in mass or metallicity that can produce a comparably empty void (in a fractional sense) except in finely tuned regions of parameter space where sample sizes are small. However, while not as pronounced, we do note the radius valley (as defined in this exercise) appears emptier for lower-mass stars and metal-rich stars. The latter observation is consistent with the finding of an apparently wider radius valley for metal-rich hosts within the CKS sample \citep{Owen2018}, though we note that age and [Fe/H] are anti-correlated. 

\begin{figure}
    \centering
    \includegraphics[width=\linewidth]{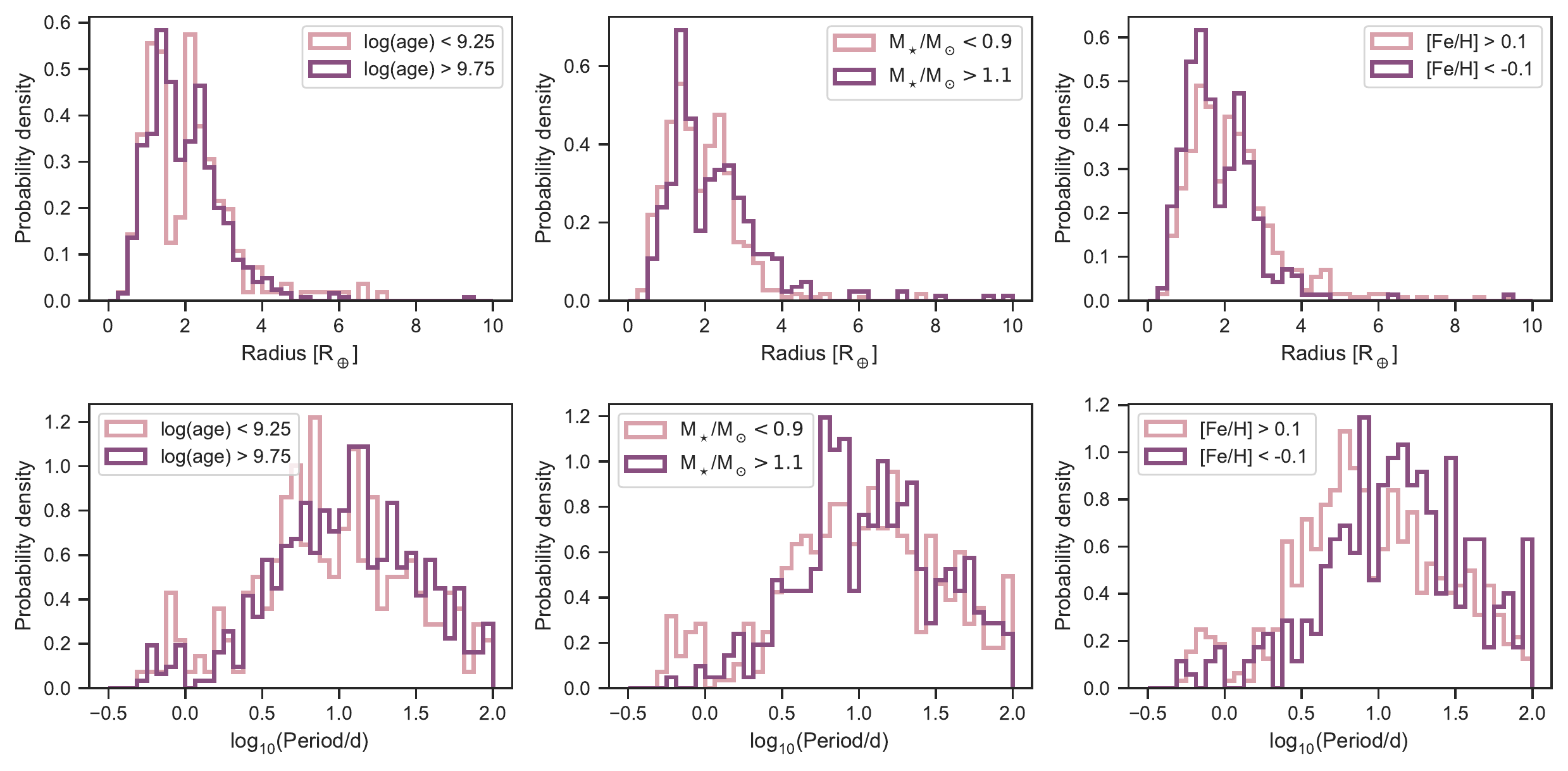}
    \caption{Planet radius (top row) and period (bottom row) distributions among our CKS base sample for host stars of different age (left column), mass (middle column), and metallicity (right column).}
    \label{fig:hist}
\end{figure}

We also examined the one-dimensional radius and period distributions for CKS planets in the extremes of the stellar age, mass, and metallicity axes (Figure~\ref{fig:hist}). The purpose of this exercise was to highlight any exoplanet demographic trends as a function of these key stellar parameters. Interestingly, even without completeness corrections or occurrence rate calculations, several of the now well-established trends in the \kepler planet population are evident from Figure~\ref{fig:hist}: larger sub-Neptunes around more massive stars \citep{Fulton2018, Wu2019, Cloutier2020} and more metal-rich stars \citep{Petigura2018}, the rising occurrence of ultra-short period ($P<1$~d) planets with decreasing stellar mass \citep{SanchisOjeda2014}, and the rising occurrence of short-period planets ($1<P<10$~d) of all sizes with increasing metallicity \citep{Petigura2018}. The other trend that is apparent is the dearth of planets in the radius valley for young stars. The trend of a wider radius valley around more metal-rich stars found by \citet{Owen2018} is not immediately obvious, but relative to that study we use updated planetary radii from F18, do not include completeness corrections, and perform slightly different cuts.

\subsection{Accounting for age uncertainties}
In \S\ref{subsec:chance} we assessed the probability that the observed void was due to chance selection of planets from the overall CKS data set, and in \S\ref{subsec:binning} we explored the sensitivity of the void occupancy to binning schemes in mass, age, and metallicity. Here we attempt to account for stellar age uncertainties in examining the void occupancy as a function of age. As discussed in the Appendix, there may be substantial uncertainties in stellar ages, particularly if those ages originate from isochrones. As a result, when binning in stellar age there is considerable uncertainty in the degree of contamination by stars with inaccurate ages.

\begin{figure}
    \centering
    \includegraphics[width=\linewidth]{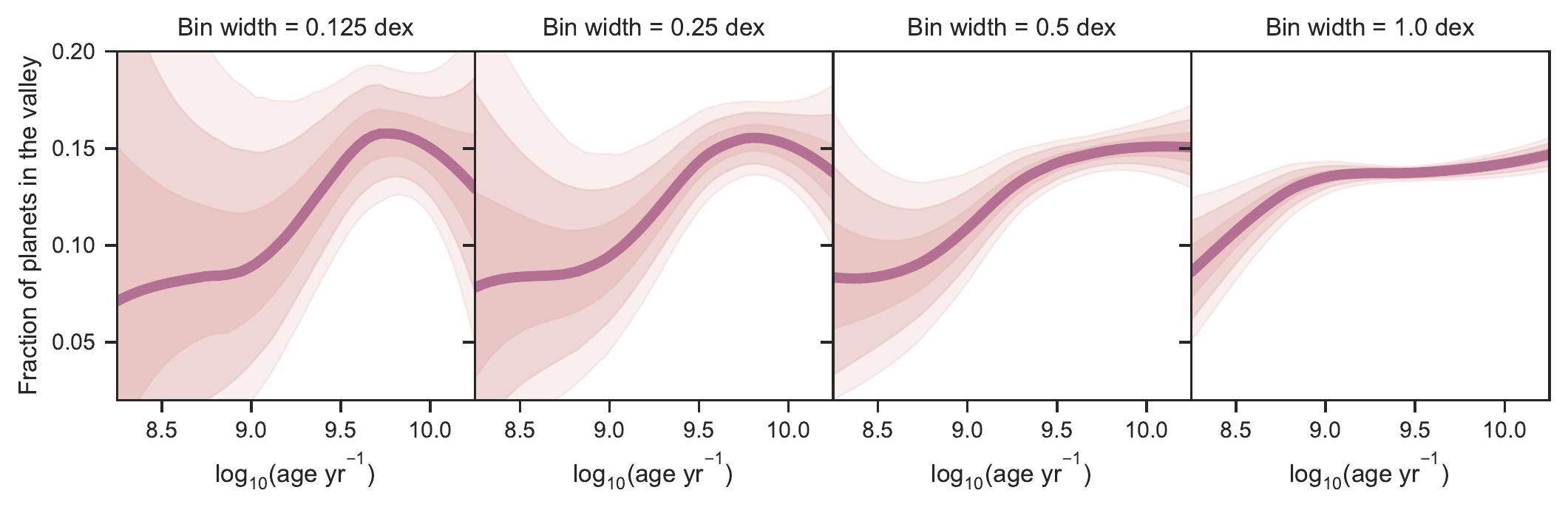}
    \caption{Results of Monte Carlo simulations exploring the occupancy of the radius valley as a function of age, accounting for uncertainties in stellar ages. The dark line shows the median trend resulting from the simulations while the shaded contours show the 68.3, 95.4, and 99.7 percentile ranges. The curves have been smoothed with a Savitzky-Golay filter for clarity.}
    \label{fig:trend}
\end{figure}

To mitigate the effect of stars with spurious ages, we performed Monte Carlo simulations in which the ages were modeled as normal distributions in \logage centered on the median values published in F18 with widths taken as the maximum of the lower and upper age uncertainties for each star. While this is not the same as drawing from the empirical posterior probability density functions in age (which are not available) it is a crude proxy. For 50 bin centers in \logage from 8.25--10.25 we then performed $10^3$ Monte Carlo simulations in each bin, randomly generating ages as described above, to measure the fraction of planets in the valley as a function of age. We measured the fraction of planets in the valley relative to the total number of planets in each age bin for bin widths of 0.125, 0.25, 0.5, and 1.0 dex. The results of this analysis are shown in Figure~\ref{fig:trend}. The scarcity of planets with \logage~$<9$ and \logage~$>10$ leads to large uncertainties in the trend at both extremes, in addition to the larger age uncertainties at younger ages. However, we observe a marginally significant increase in the fraction of planets located in the radius valley between $8.75<$\logage$<9.75$. This is in agreement with Figures~\ref{fig:pr} and \ref{fig:pr-base} which show that the radius valley appears weaker among the oldest planets in the CKS sample. Computing planet occurrence rates in the valley as a function of age might lead to a more robust conclusion on the trend noted here, but is outside the scope of the present work. 

\subsection{In what ways are planets in the valley different?}
\label{subsec:distributions}
In an effort to quantify the parameters that are most important in contributing to the filling of the radius valley, we performed a k-sample Anderson-Darling (A-D) test \citep{Scholz1987} with \texttt{scipy.stats.anderson\_ksamp} to test the null hypothesis that the distribution of a given variable for stars hosting planets in the valley was drawn from the same distribution of that parameter among our base CKS sample. 

\begin{figure}
    \centering
    \includegraphics[width=0.49\linewidth]{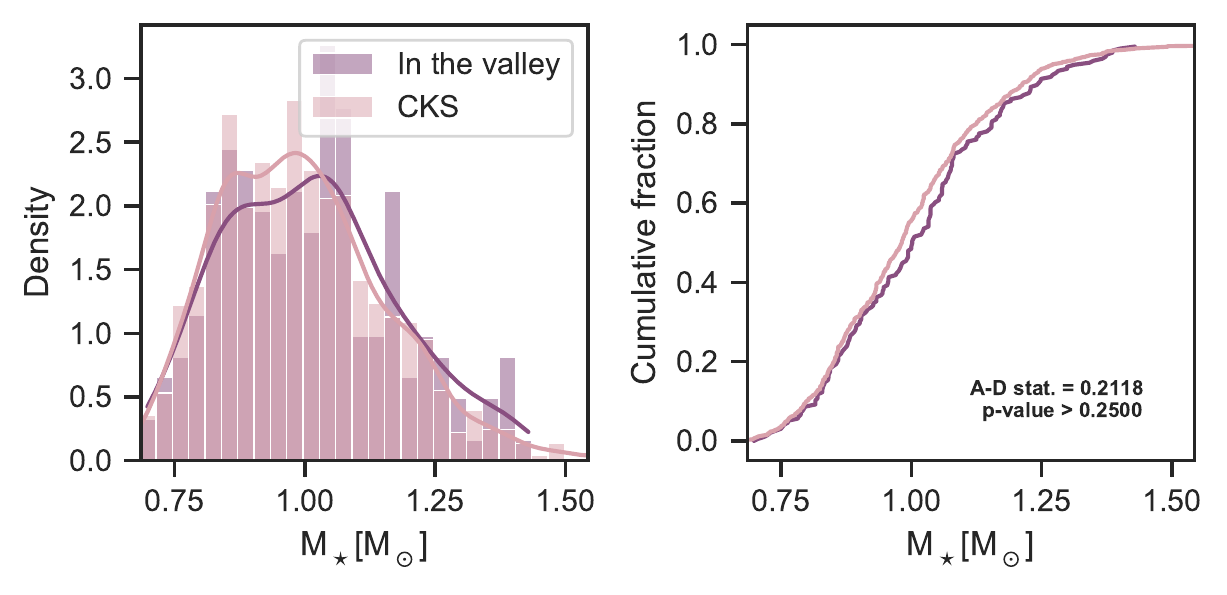}
    \includegraphics[width=0.49\linewidth]{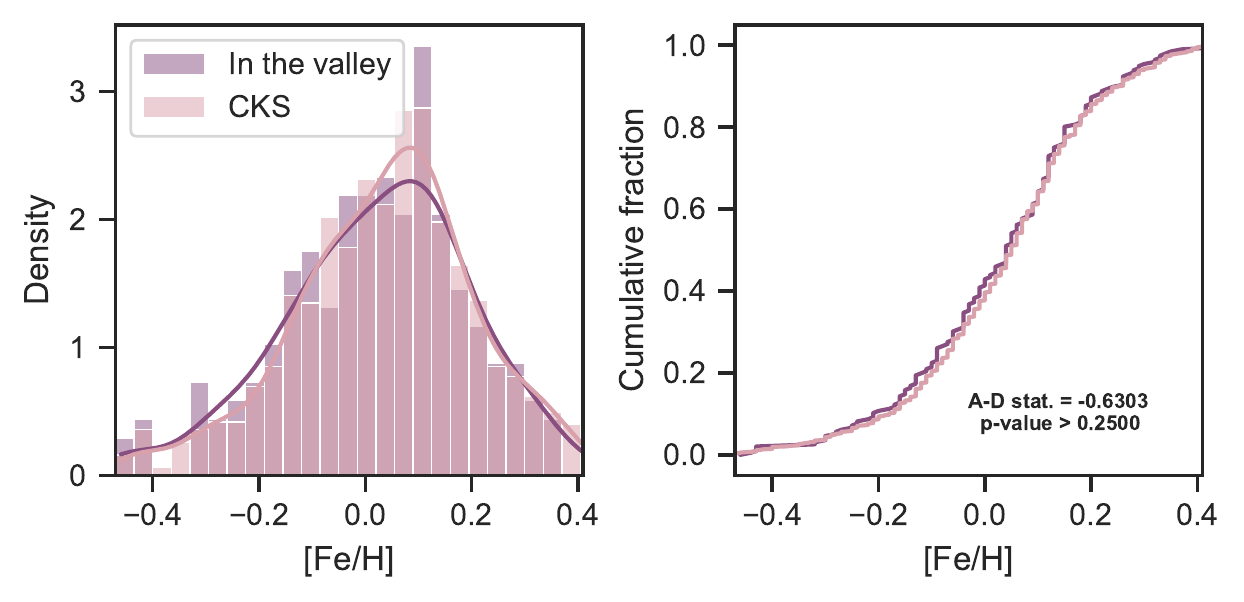}
    \includegraphics[width=0.49\linewidth]{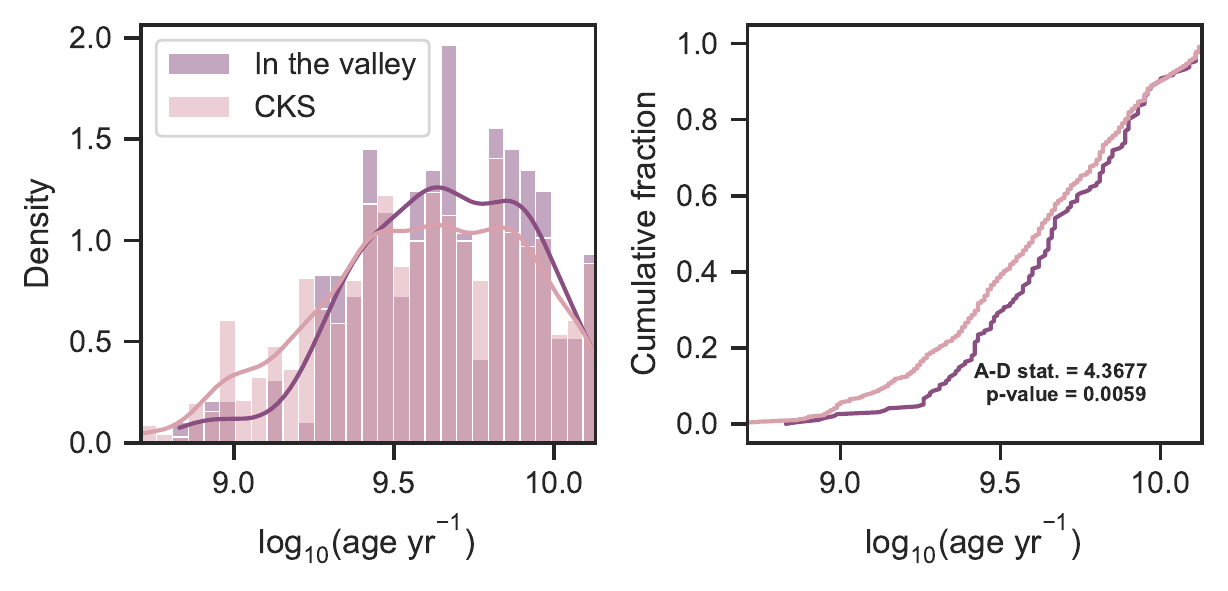}
    \includegraphics[width=0.49\linewidth]{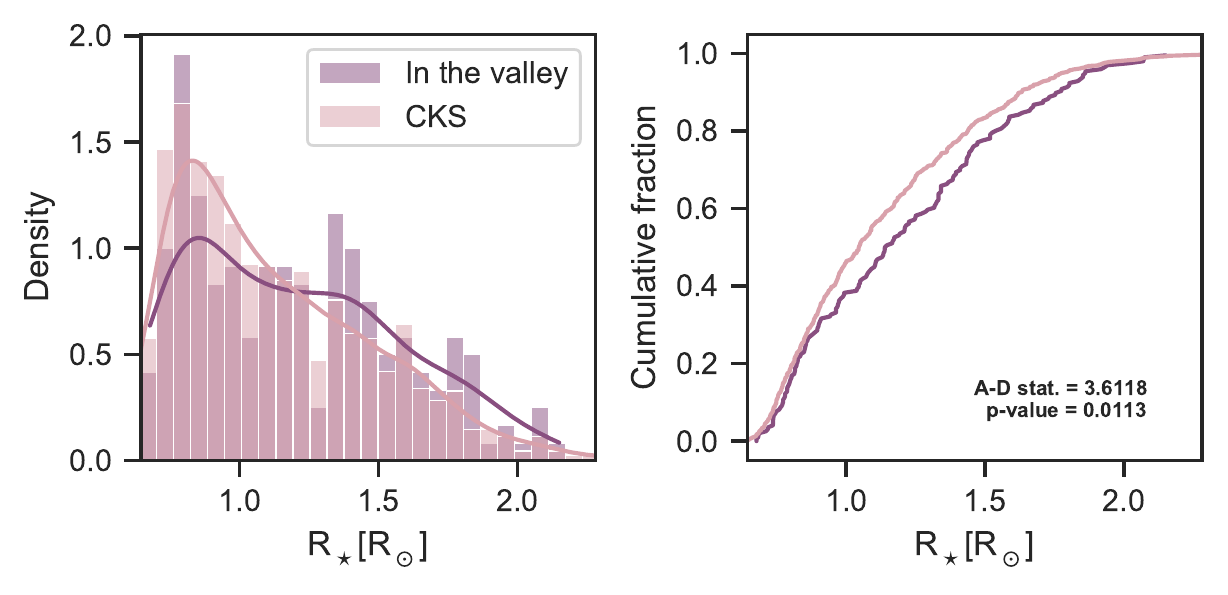}
    \includegraphics[width=0.49\linewidth]{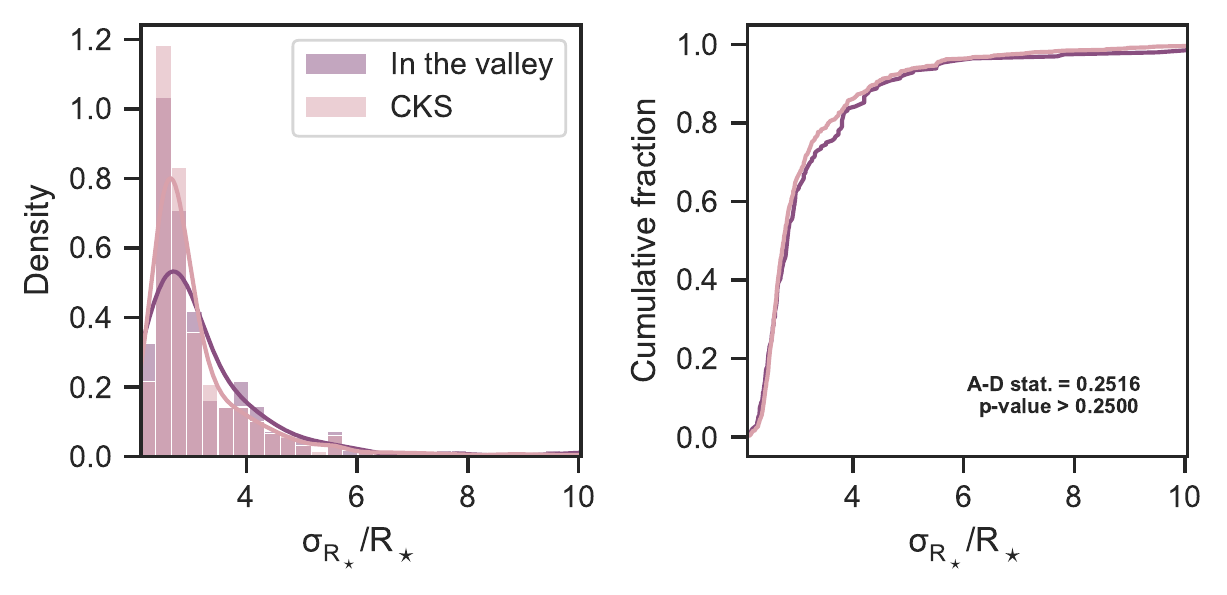}
    \includegraphics[width=0.49\linewidth]{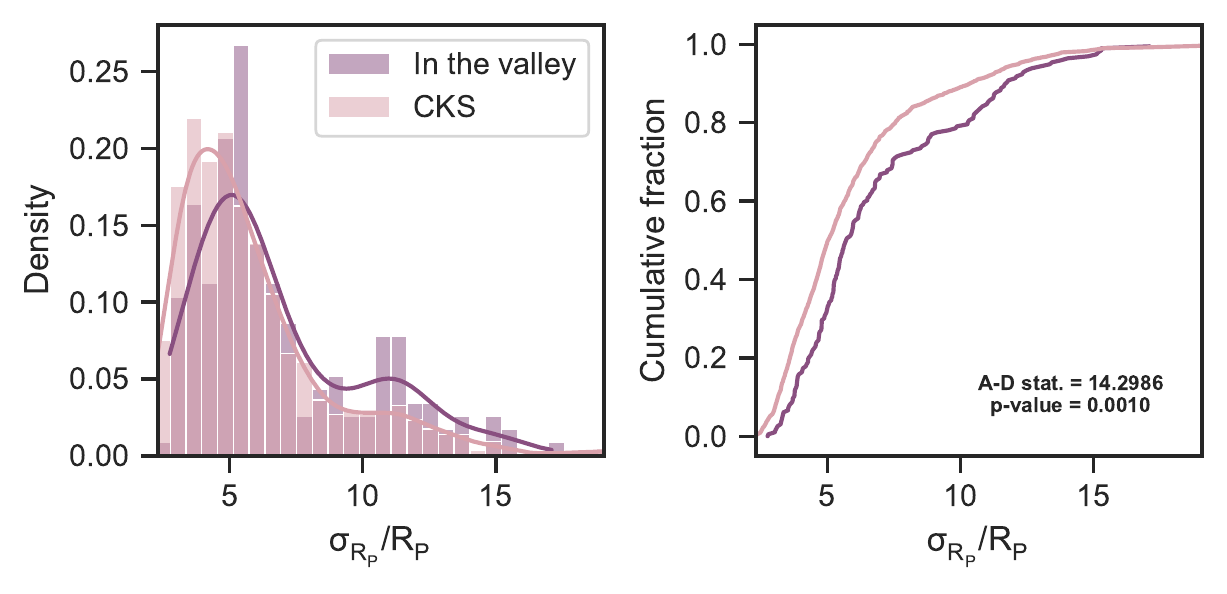}
    \caption{Planet radius (top row) and period (bottom row) distributions for host stars of different age (left column), mass (middle column), and metallicity (right column). Distributions shown are for our CKS base sample.}
    \label{fig:dist}
\end{figure}

The results of this exercise are summarized in Table~\ref{tab:ad} and select parameter distributions are shown in Figure~\ref{fig:dist}. In this exercise we have assumed the equation for the radius valley and its boundaries are given by the fourth row of Table~\ref{tab:svm}. This choice is motivated by the fact that higher regularization parameters correspond to tighter boundaries of the valley, offering a cleaner separation between planets in, above, or below the valley. Additionally, for each parameter we restrict our analysis to those CKS stars/planets for which that parameter is defined (i.e. we exclude targets missing data for a given variable). We also apply the common cuts described in \S\ref{subsec:void} before performing the A-D tests. In the case of \prot, we also restricted the sample to \teff $<6000~K$, where rotation periods are more reliable indicators of age.

Of the nine parameters with the highest normalized k-sample A-D test statistics (and $p$-values $<$0.05), all but two pertain directly or indirectly to the star's evolutionary state: \prot from various sources, \prot flag, \rvar (a measure of the photometric variability amplitude), the median posterior age from isochrones, and \rstar. The other two parameters are fractional \rp precision and $r_8$ (a measure of flux dilution). Thus, of the parameters investigated, those which contribute most to the filling of the radius valley either relate to stellar age or may be associated with erroneous measurements of the planetary radii. Inspection of the parameter distributions (like those shown in Fig.~\ref{fig:dist}) reveals that planets in the valley tend to orbit stars which are older, larger, less likely to have a securely detected rotation period, rotating more slowly, and photometrically quieter. Planets in the valley also have lower $r_8$ values relative to the CKS base sample. Na\"ively, one might expect higher $r_8$ values among planets in the radius valley as flux dilution can lead to erroneous planet radius measurements. However, planets in the radius valley are by definition small, so it is perhaps not surprising that there is a preference for stars not affected by crowding. 

We also found that the stellar mass and metallicity distributions for stars hosting planets in the radius valley are statistically indistinguishable from those of CKS base sample. This lends further support to the notion that the feature identified in the CKS data set from age selections is not due to correlations between stellar age, mass, and metallicity.

The parameter which appears to be most important in contributing to the filling of the radius valley is planet radius precision. This suggests that the radius valley may be emptier than is suggested by current data. The fractional stellar radius precision does not, however, contribute to the filling of the valley. This is not surprising as the typical error budget for a planet's radius in the CKS sample is dominated by the \rprstar uncertainty from light curve fitting rather than the stellar radius uncertainty \citep{Petigura2020}. In \S\ref{subsec:confounding} we examine the possibility that a correlation between planet radius precision and age could conspire to produce the observed void.

\begin{deluxetable*}{ccccc}
\tablecaption{Results of k-sample Anderson-Darling tests.}
\tabletypesize{\scriptsize}
\label{tab:ad}
\tablecolumns{5}
\tablewidth{\linewidth}
\tablehead{\colhead{Parameter} & \colhead{Ref.} & \colhead{A-D test stat.} &  \colhead{A-D p-value} & \colhead{Sample size (valley/control)}}
\startdata
$\sigma_{R_P}/R_P$ & F18 & 14.30 & 0.0010 & 196/1443 \\
\prot flag & D21 & 6.08 & 0.0015 & 196/1443 \\
\prot & M15 & 4.82 & 0.0040 & 135/1055 \\
\rvar & M15 & 4.67 & 0.0045 & 180/1334 \\
\logage & F18 & 4.37 & 0.0059 & 196/1443 \\
\rstar & F18 & 3.61 & 0.011 & 196/1443 \\
$r_8$ & F18 & 2.53 & 0.030 & 196/1443 \\
\prot & M13 & 2.17 & 0.042 & 36/371 \\
\prot & D21 & 1.95 & 0.051 & 55/592 \\
\prot & A18 & 0.67 & 0.17 & 109/873 \\
SNR$_1$ & D21 & 0.65 & 0.18 & 190/1420 \\
\teff & F18 & 0.62 & 0.18 & 196/1443 \\
\rtau & P20 & 0.30 & $>$0.25 & 190/1415 \\
$\sigma_{R_\star}/R_\star$ & F18 & 0.25 & $>$0.25 & 196/1443 \\
\mstar & F18 & 0.21 & $>$0.25 & 196/1443 \\
Parallax & F18 & 0.096 & $>$0.25 & 196/1443 \\
CDPP3 & D21 & -0.16 & $>$0.25 & 190/1420 \\
\prot & W13 & -0.19 & $>$0.25 & 34/335 \\
$A_V$ & B20 & -0.50 & $>$0.25 & 188/1390 \\
$G$ mag & DR2 & -0.56 & $>$0.25 & 190/1420 \\
$A_V$ & L21 & -0.57 & $>$0.25 & 187/1382 \\
\feh & F18 & -0.63 & $>$0.25 & 196/1443 \\
RUWE & D21 & -0.64 & $>$0.25 & 190/1418 \\
RCF & F18 & -0.94 & $>$0.25 & 55/423 \\
\enddata
\tablecomments{References: A18 \citep{Angus2018}; D21 (this work); DR2 \citep{GaiaDR2}; F18 \citep{Fulton2018};  L21 \citep{Lu2020}; M15 \citep{Mazeh2015}; M13 \citep{McQuillan2013}; P20 \citep{Petigura2020}; W13 \citep{WalkowiczBasri2013}}
\end{deluxetable*}

\subsection{Confounding scenarios}
\label{subsec:confounding}
A possibility not yet explored is that the radius valley is inherently empty but, for some reason, planets orbiting stars with younger assigned ages in the CKS sample have more precise radii. In \S\ref{subsec:distributions} we established that the planet radius precision is the most important parameter contributing to the filling of the radius valley. We quantified the correlation between fractional planet radius precision and log(age) by computing the Spearman rank correlation coefficient for planets with $P<100$~d, $R_P<10$~\rearth, non-false positive dispositions, and main-sequence host stars (filters 1,2,4, and 5 from \S\ref{sec:sample}). We used the \texttt{scipy.stats.spearmanr} function for this purpose and found a small $p$-value ($2\times10^{-4}$) but a very weak correlation coefficient ($\rho < 0.1$). 

\begin{figure}
    \centering
    \includegraphics{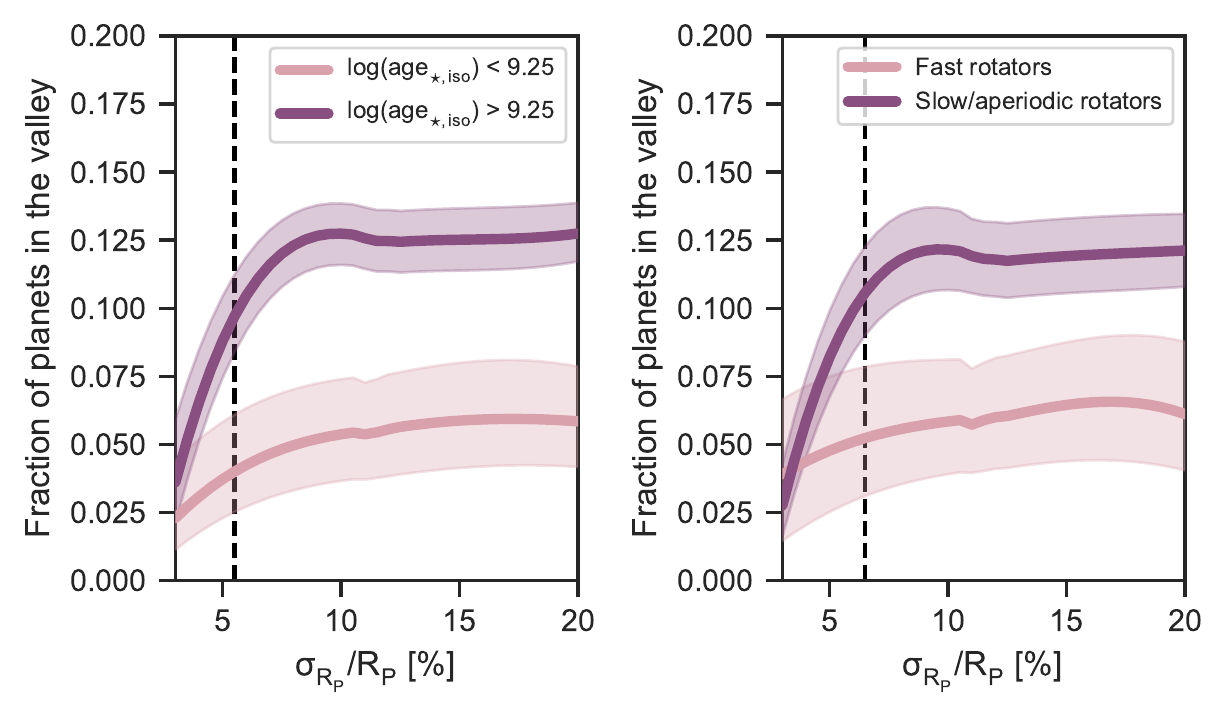}
    \caption{Fraction of planets in the radius valley as a function of maximum fractional radius uncertainty. For each age-selected sample shown, the fraction of planets in the valley is computed for the subsample of planets with fractional radius uncertainties lower than the value on the ordinate. Solid lines and shaded bands show the median and 16th--84th percentile range from bootstrapping simulations, respectively. In each panel the fiducial dashed line shows the value of fractional radius uncertainty for which both young and old samples contain more than 150 planets (corresponding to 21$\pm$4 expected planets in the valley if selected at random from the CKS base sample).}
    \label{fig:precision}
\end{figure}

To further investigate the impact of radius precision we computed the fraction of planets in the valley for young and old samples as a function of fractional radius precision allowed. For a given sample of planets and over a grid of radius precision thresholds, we selected the planets with fractional radius uncertainties smaller the threshold value and computed the ratio of planets in the valley to the total number of planets meeting the radius precision requirement. We performed $10^3$ bootstrapping simulations (including modeling of the planet radii as normal distributions) to determine the uncertainties on these trends, which are shown in Figure~\ref{fig:precision}. We found that the radius valley is comparably empty for young and old planets if the fractional radius precision is required to be better than $\sim$5\%. However, this is not unexpected as our CKS base sample size diminishes steeply below fractional radius uncertainties of 7\%. For reference, the median fractional radius uncertainties for the CKS base sample, young isochrone age-selected sample (\sampleb), and old isochrone age-selected sample are 5.0\%, 4.5\%, and 5.2\%, respectively. 

\begin{figure}
    \centering
    \includegraphics{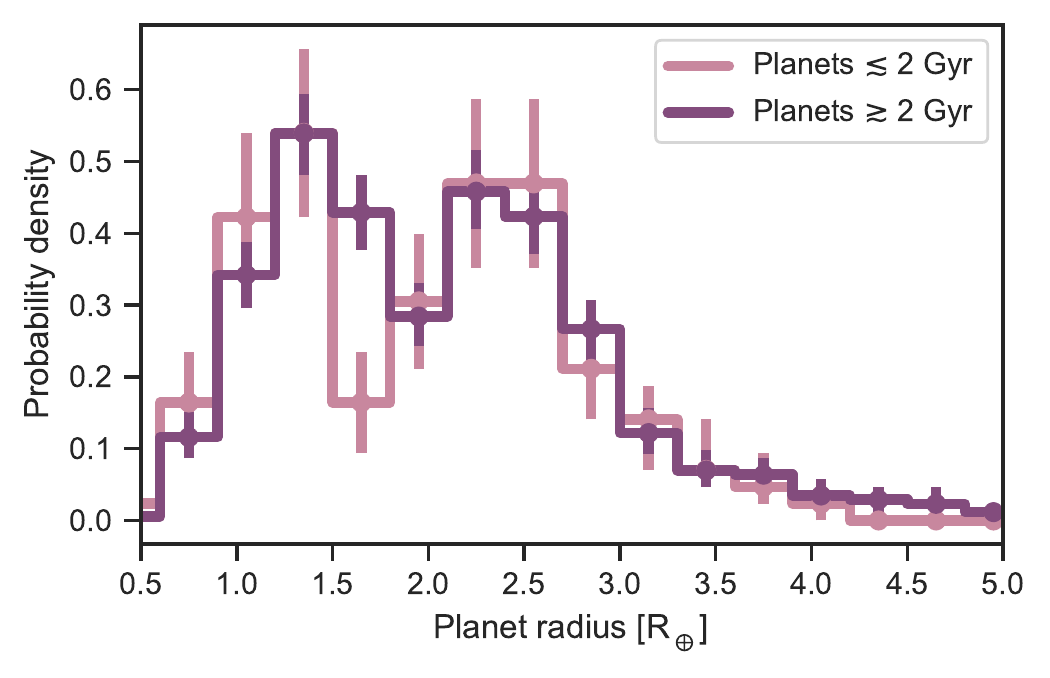}
    \caption{Small planet size distributions among CKS planets with fractional radius uncertainties better than 5\%. Uncertainties (the 16th and 84th percentiles) determined from bootstrapping simulations with planet radii modeled as normal distributions.}
    \label{fig:fivepct}
\end{figure}

Finally, we examined the one-dimensional radius distributions for the young and old isochrone age-selected samples with planet radii known to better than 5\%. We performed $10^3$ bootstrapping simulations (again modeling the planet radii as normal distributions) to determine uncertainties on these distributions. The results are shown in Figure~\ref{fig:fivepct}. While radius precision is clearly an important parameter in determining the occupancy of the radius valley, we conclude that it is unlikely to explain the entire deficit observed for the young planet sample.

\subsection{Is the radius gap empty?}
\label{subsec:empty}

Figures~\ref{fig:pr}, \ref{fig:pr-base}, and \ref{fig:pr-gyro} give the impression that the radius valley progressively fills in over time, becoming weaker or disappearing entirely among older planet populations. This interpretation appears to be at odds with the results of V18 who observed a clean gap in the period-radius distributon of planets orbiting asteroseismic host stars, which are preferentially older than the stars in our young sample.\footnote{We note that cross-matching the V18 sample with \citet{SilvaAguirre2015} and F18 reveals that the asteroseismic sample contains host stars with a broad range of ages, from $\approx$2--12.5~Gyr.} Notably, planets with ages $\gtrsim$3~Gyr in the CKS base sample have a median radius precision of 5.3\%, while planets in the V18 sample have a median radius precision of 3.3\%. Similarly, in \S\ref{subsec:confounding} we found that age-dependent differences in the radius valley filling factor can be resolved at least partially by restricting analysis to planets with the most precise radii.

\begin{figure}
    \centering
    \includegraphics[width=\linewidth]{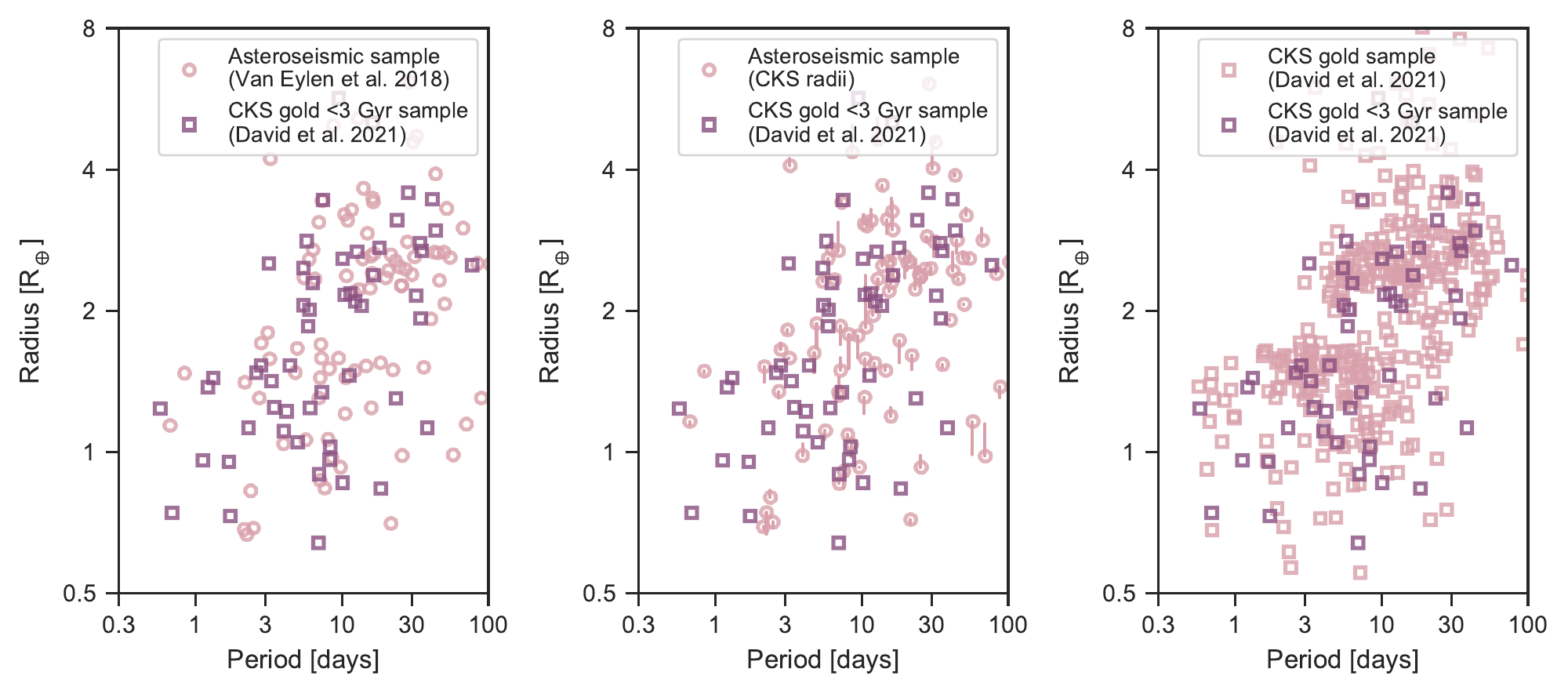}
    \caption{Planet distributions in the period-radius diagram for the samples described in \S\ref{subsec:empty}. \added{In the middle panel planets in the V18 asteroseismic sample are plotted using the F18 radii. Lines connect each planet in that sample to its radius as determined by V18.}}
    \label{fig:gold}
\end{figure}

To further investigate this issue we constructed a new sample, the \gold sample, which implements several reliability cuts. In addition to the cuts of the base sample, the \gold sample is restricted to planets with fractional radius precision $<6\%$, fractional $R_P/R_*$ precision $<6\%$, non-grazing transits (\rtau$>0.6$), RCF$< 1.05$, $A_V < 0.5$~mag, RUWE~$< 1.1$, agreement between the F18 isochrone-derived and trigonometric parallaxes, and a KOI reliability score $>0.99$ from the Q1-Q17 DR25 catalog. We then split this sample into young, \goldyng, and old, \goldold, samples. The young sample includes the restriction that the stellar age inferred from both isochrones and gyrochronology is $<3$~Gyr, while the old sample requires a planet host has a median isochrone age $>3$~Gyr and does not have a rotation period consistent with an age $<3$~Gyr. For both the \goldyng and \goldold samples we confirmed that the corresponding distributions in $R_P$ precision, $R_P/R_*$ precision, and single transit SNR (defined as $(R_P/R_*)^2/\text{CDPP3}$) were not statistically different either from each other or from the overall distributions in the \gold sample, yielding $p$-values $>0.25$ in each case from k-sample A-D tests. The \goldyng sample in comparison to the V18 and \gold samples in the period-radius diagram are shown in Fig.~\ref{fig:gold}. From that figure it appears that the reliability cuts have a significant impact on how well-defined the super-Earth and sub-Neptune distributions are, as well as how empty the gap appears at all ages, though it is not entirely devoid of planets. Furthermore, it is clear from Fig.~\ref{fig:gold} that the gap in the \goldyng sample is offset from the gap in both the V18 and \gold samples, indicating that the difference is unlikely to be due to systematic differences in planet radii between the two studies. \added{For reference, we also show the period-radius distribution of V18 planets using the CKS radii, which highlights the importance of the precise ($R_P/R_\star$) values used by V18 in resolving the gap \citep[for a detailed discussion see][]{Petigura2020}.}

\begin{figure}
    \centering
    \includegraphics[width=\linewidth]{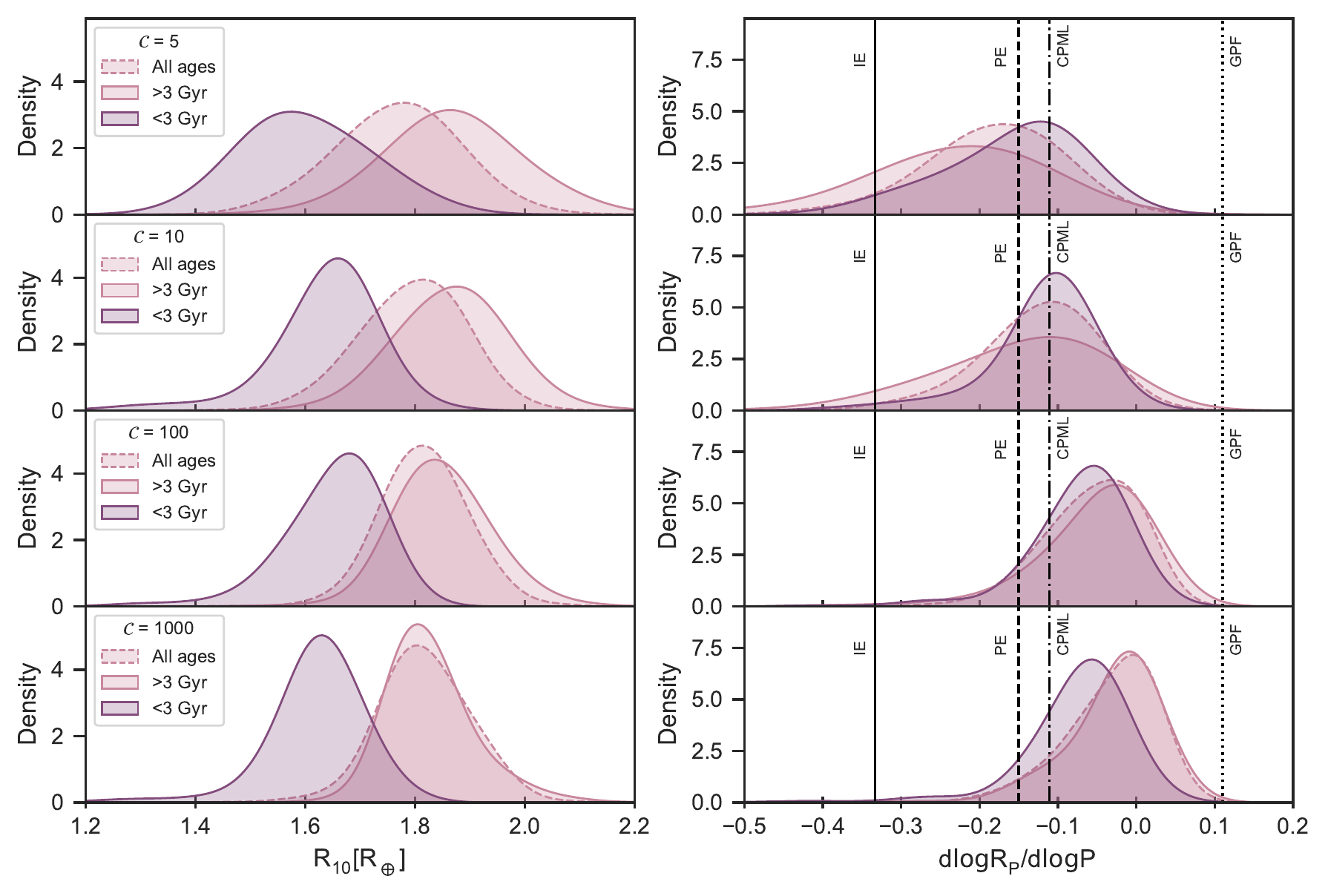}
    \caption{Smoothed Gaussian kernel density estimates of the distributions of the intercepts (left) and slopes (right) resulting from the SVM bootstrapping analysis of the CKS \texttt{gold} samples presented in \S\ref{subsec:empty}. At left, the $R_{10}$ parameter indicates the center of the radius gap at an orbital period of 10 days. Each row corresponds to a different regularization parameter, $\mathcal{C}$, indicated in the figure legend. At right, vertical lines indicate predictions from the impact erosion \citep[IE,][]{Wyatt2020}, photoevaporation \citep[PE,][]{LopezRice2018}, core-powered mass-loss \citep[CPML,][]{Gupta2019}, and gas-poor formation \citep[GPF,][]{LopezRice2018} theories.}
    \label{fig:gold-results}
\end{figure}

We proceeded to perform the same SVM analysis as was presented in \S\ref{subsec:void} with one difference: classification of the samples into super-Earths and sub-Neptunes was performed using the threshold $R_P=1.8$~\rearth rather than using a period-dependent classification scheme. The reason for this choice is because this scheme clearly works well for the \goldyng sample and allows us to test the sensitivity of the our analysis to the classification step. The results of our analysis are presented in Fig.~\ref{fig:gold-results} and Table~\ref{tab:svm-gold}. We find that despite the simplified classification scheme, a negative slope in the period-radius diagram is still preferred (though the data are also consistent with no orbital period dependence). Furthermore, we observe that, independent of regularization parameter, there is a persistent offset in the center of the gap for the \goldyng sample, compared to both the \gold and \goldold samples.  

Related to this last point, we emphasize that the gap identified in this work is primarily due to a lack of large super-Earths at young ages, as opposed to a difference in the sub-Neptune size distribution or some combination of the two. This is most apparent in Fig.~\ref{fig:fivepct} and Fig.~\ref{fig:gold}. We note that the young planet samples are always smaller than the control samples, and the dearth of large super-Earths at young ages could be due in part to small number statistics. To assess the probability of this scenario we performed 10$^4$ simulations and measured the fraction of outcomes in which the number of $>1.5$~\rearth planets in a control sample was equal to or fewer than the number of $>1.5$~\rearth planets in the young sample. In each simulation, we selected 40 planets at random from the control sample, corresponding to the size of our young super-Earth sample. We accounted for planet radius uncertainties in both the young and control samples by modeling the radii as normal distributions given their published uncertainties. For the control samples we used the CKS gold and V18 samples, where we used the V18 radius valley equation to select only the super-Earths in each. In both cases we found $\lesssim$2\% of the simulations resulted in outcomes where the number of $>1.5$~\rearth super-Earths was greater in the young sample than in the control sample. We also compared the young and control super-Earth size distributions with a k-sample A-D test, finding the null hypothesis can be rejected at the 1\% level.

In conclusion, we propose a solution to resolve the apparent tension described in the beginning of this section and to explain all of the observations to date: the radius gap is intrinsically empty, or at least emptier than previously appreciated, but its precise location shifts with the age of the planetary population. Since the radius gap appears to have an orbital period dependence, a gap that is intrinsically empty in the period-radius plane will always appear filled in when projected along the radius axis, even if radii are known perfectly. Similarly, if the location of the gap also depends on host star mass, age, or metallicity, as has been suggested, then the gap will only appear empty in sufficiently narrow projections of parameter space.
\added{While this proposed solution would help explain some of our observations, we emphasize that we have not conclusively shown it to be the case. Confirming or rejecting this hypothesis may be possible with (1) a larger sample providing sufficient coverage of the period-radius plane across the variables of interest, and/or (2) a thorough, multivariate investigation of the radius gap in order to find the projection of the data resulting in the emptiest gap. We leave such an investigation to future works and emphasize that planet radius uncertainties (resulting from inaccurate light curve fits, stellar radius uncertainties, or more pernicious sources such as flux dilution from unresolved binaries) remain an obstacle to our understanding of the radius gap.}

%\added{The interpretation of an empty radius gap may seem to be at odds with simulations presented in F18 which attempted to quantify the astrophysical spread in the sizes of super-Earths and sub-Neptunes. Those authors used the true periods of planets in the CKS sample and drew planet sizes from two uniform distributions centered at 1.2~\rearth and 2.4~\rearth, optimizing the shared fractional width of these size distributions to match the number of true detections in pre-defined regions of super-Earth and sub-Neptune parameter space.}

\begin{deluxetable}{cccccccc}
\tablecaption{Results of SVM bootstrapping simulations for the CKS gold samples.}
\label{tab:svm-gold}
\tablecolumns{8}
\tablewidth{\linewidth}
\tablehead{\colhead{Sample} & \colhead{$\mathcal{C}$} & \colhead{$\alpha$} & \colhead{$\beta$} & \colhead{$\gamma$} & \colhead{$\delta$} & \colhead{$\epsilon$} & \colhead{$\zeta$}}
\startdata
\gold & 5 & $-0.18^{+0.09}_{-0.07}$ & $0.42^{+0.07}_{-0.07}$ & $0.15^{+0.01}_{-0.02}$ & $0.11^{+0.05}_{-0.05}$ & $0.02^{+0.13}_{-0.07}$ & $0.15^{+0.01}_{-0.02}$\\
\gold & 10 & $\mathbf{-0.12^{+0.07}_{-0.06}}$ & $\mathbf{0.37^{+0.05}_{-0.06}}$ & $0.11^{+0.01}_{-0.01}$ & $\mathbf{0.07^{+0.04}_{-0.04}}$ & $\mathbf{0.11^{+0.11}_{-0.08}}$ & $0.11^{+0.01}_{-0.01}$\\
\gold & 100 & $-0.05^{+0.06}_{-0.04}$ & $0.31^{+0.04}_{-0.05}$ & $0.06^{+0.01}_{-0.01}$ & $0.03^{+0.03}_{-0.04}$ & $0.19^{+0.1}_{-0.05}$ & $0.06^{+0.01}_{-0.01}$\\
\gold & 1000 & $-0.02^{+0.05}_{-0.05}$ & $0.28^{+0.04}_{-0.06}$ & $0.03^{+0.01}_{-0.01}$ & $0.01^{+0.02}_{-0.04}$ & $0.24^{+0.05}_{-0.07}$ & $0.03^{+0.01}_{-0.01}$\\
\hline
\goldold & 5 & $-0.22^{+0.09}_{-0.11}$ & $0.49^{+0.1}_{-0.1}$ & $0.16^{+0.02}_{-0.02}$ & $0.15^{+0.04}_{-0.06}$ & $-0.04^{+0.12}_{-0.11}$ & $0.16^{+0.02}_{-0.02}$\\
\goldold & 10 & $-0.13^{+0.09}_{-0.08}$ & $0.4^{+0.07}_{-0.09}$ & $0.12^{+0.01}_{-0.02}$ & $0.09^{+0.03}_{-0.06}$ & $0.06^{+0.13}_{-0.08}$ & $0.12^{+0.01}_{-0.02}$\\
\goldold & 100 & $-0.03^{+0.05}_{-0.04}$ & $0.29^{+0.04}_{-0.05}$ & $0.06^{+0.01}_{-0.01}$ & $0.03^{+0.03}_{-0.03}$ & $0.2^{+0.08}_{-0.06}$ & $0.06^{+0.01}_{-0.01}$\\
\goldold & 1000 & $-0.01^{+0.04}_{-0.03}$ & $0.27^{+0.04}_{-0.04}$ & $0.04^{+0.01}_{-0.01}$ & $0.01^{+0.03}_{-0.02}$ & $0.24^{+0.04}_{-0.06}$ & $0.04^{+0.01}_{-0.01}$\\
\hline
\goldyng & 5 & $-0.14^{+0.06}_{-0.07}$ & $0.34^{+0.06}_{-0.05}$ & $0.15^{+0.01}_{-0.02}$ & $0.07^{+0.03}_{-0.04}$ & $0.08^{+0.11}_{-0.05}$ & $0.14^{+0.01}_{-0.02}$\\
\goldyng & 10 & $-0.1^{+0.03}_{-0.04}$ & $0.33^{+0.03}_{-0.05}$ & $0.11^{+0.01}_{-0.01}$ & $0.06^{+0.02}_{-0.02}$ & $0.11^{+0.04}_{-0.04}$ & $0.11^{+0.01}_{-0.01}$\\
\goldyng & 100 & $-0.05^{+0.02}_{-0.05}$ & $0.28^{+0.04}_{-0.03}$ & $0.07^{+0.01}_{-0.01}$ & $0.03^{+0.06}_{-0.01}$ & $0.18^{+0.02}_{-0.13}$ & $0.07^{+0.01}_{-0.01}$\\
\goldyng & 1000 & $-0.05^{+0.02}_{-0.04}$ & $0.27^{+0.02}_{-0.03}$ & $0.06^{+0.01}_{-0.01}$ & $0.03^{+0.06}_{-0.01}$ & $0.16^{+0.01}_{-0.13}$ & $0.06^{+0.01}_{-0.01}$\\
\hline
\enddata
\tablecomments{Equation for the radius valley in the period-radius diagram is of the form $\log_{10}(R_P/R_\oplus) = \alpha \log_{10}(P/\text{d}) + \beta$. In the insolation-radius diagram it is $\log_{10}(R_P/R_\oplus) = \delta \log_{10}(S_\mathrm{inc}/S_{\oplus}) + \epsilon$. Adopted values in bold.}
\end{deluxetable}

\section{Discussion \& Conclusions}
\label{sec:conclusions}

\begin{figure}
    \centering
    \includegraphics[width=\linewidth]{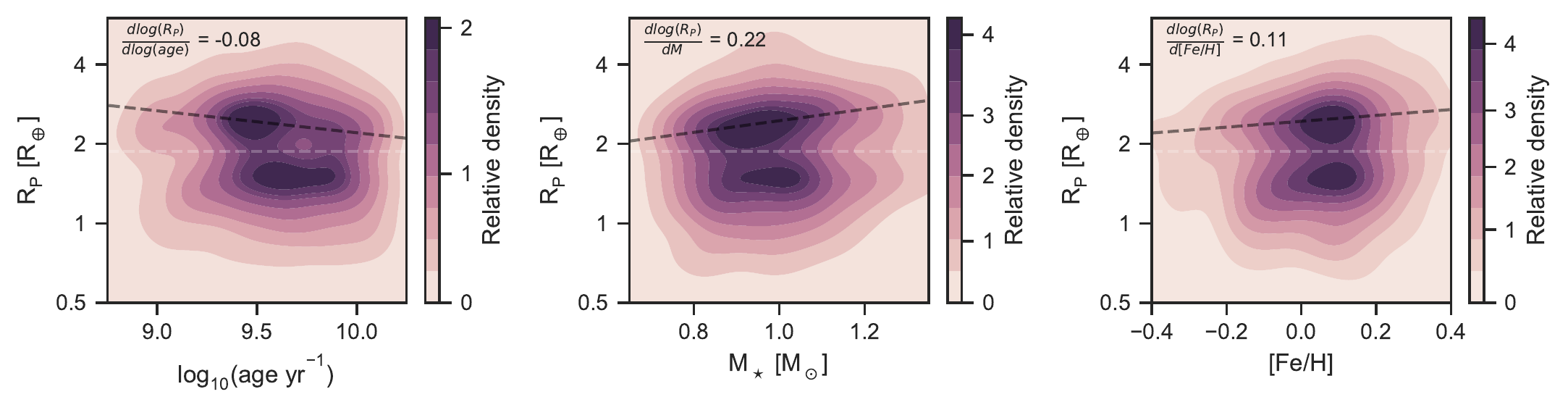}
    \caption{2D Gaussian kernel density estimates of the distribution of CKS planets in the age-radius (left), stellar mass-radius (middle), and metallicity-radius (right) planes. Our base sample is shown with the additional requirement of fractional radius uncertainties $<6\%$. The light dashed line indicates the nominal location of the radius valley. The dark dashed lines are not fits but are drawn as a visual guide, with their slopes indicated in the upper left of each panel.}
    \label{fig:triptych}
\end{figure}

We observe a nearly empty void in the period-radius plane for close-in ($P < 100$~d) exoplanets orbiting stars younger than $\sim$2--3~Gyr. The void was first identified among a sample of planets with median posterior isochrone ages $<$1.8~Gyr, but is also present among planets with stars rotating faster than an empirical 2.7~Gyr gyrochrone. The difference between these two timescales could conceivably be due to systematic offsets between the CKS isochrone ages and ages implied from a gyrochronology analysis. Because the gyrochrone used to perform our sample selection is calibrated to open clusters with main-sequence turnoff ages, the longer timescale may more accurately reflect the lifetime of this feature in the period-radius diagram.

We derived equations for the center of this void, which we refer to as the young planet gap, as a function of orbital period, $P$, and insolation, $S_\mathrm{inc}$: 

    \begin{align}
    \log_{10}(R_P/R_\oplus)_\mathrm{ypg} &= -0.08(^{+0.06}_{-0.04}) \log_{10}(P/\text{d}) + 0.31 (\pm 0.05), \\
    \log_{10}(R_P/R_\oplus)_\mathrm{ypg} &= 0.06(^{+0.03}_{-0.04}) \log_{10}(S_\mathrm{inc}/S_\oplus) + 0.13(^{+0.04}_{-0.11}).
    \end{align}

For periods in the range of 3--30 days, describing the bulk of our sample, this places the center of the radius valley at 1.87--1.56~\rearth. Over this same period range, the lower boundary of the void is in the range of $\sim$1.6--1.4~\rearth while the upper boundary is at $\sim$2.1--1.8~\rearth.

\added{From a subset of the CKS sample created using reliability and precision cuts we similarly derived equations for the radius valley valid for all ages:}

    \begin{align}
    \log_{10}(R_P/R_\oplus) &= -0.12(^{+0.07}_{-0.06}) \log_{10}(P/\text{d}) + 0.37(^{+0.05}_{-0.06}), \\
    \log_{10}(R_P/R_\oplus) &= 0.07(\pm 0.04) \log_{10}(S_\mathrm{inc}/S_\oplus) + 0.11(^{+0.11}_{-0.08}).
    \end{align}

The slope of the void in the P-R diagram is consistent at the 1$\sigma$ level with the slope of the radius valley measured from the asteroseismic sample in V18, but with an intercept that is smaller by $\sim3\sigma$, using the uncertainty reported by those authors. The smaller intercept among the ``young" planet sample corresponds to a shift in the radius valley towards smaller radii, and would be compatible with a prolonged mass-loss timescale for the sub-Neptune progenitors of the largest observed super-Earths. \added{An alternative explanation could be the late-time formation of secondary or ``revived'' atmospheres through endogenous or exogenous processes \citep[e.g.][]{KiteBarnett2020, KiteSchaefer2021}. Differentiating between these two hypotheses might be achieved with detailed composition modeling or atmospheric studies of the largest super-Earths.}
    
The shallow, negative slope of the void is compatible with models of atmospheric loss through photoevaporation \citep[e.g.][]{OwenWu2013, OwenWu2017, LopezRice2018, Jin2018} or core-cooling \citep{Gupta2019, Gupta2020}, but incompatible with the steeper negative slope implied for one model of impact-driven atmospheric erosion \citep{Wyatt2020}. The negative slope we find is also incompatible with the positive slope predicted by models of late stage formation in a gas-poor disk \citep{LopezRice2018}. However, we note that the void is only marginally inconsistent with being flat when adopting a more conservative regularization parameter in the SVM analysis. The slope of this void in the insolation-radius plane is shallower (by 2--3$\sigma$) than the slope found by \citet{Martinez2019}.

Both rotation-selected and isochrone-selected planet samples show the same qualitative trend: an absence of large super-Earths at young ages. We estimate that the probability of this feature being due to chance is $<$1\%, for both the isochrone-selected and rotation-selected samples. We also showed that this feature is relatively insensitive to various data reliability filters and is unlikely to be the result of correlations between stellar age, mass, and metallicity. Simulations accounting for age and planet radius uncertainties show an increasing fraction of planets residing in this gap as a function of age (see Figure~\ref{fig:trend}). The occupancy of the radius valley is also clearly dependent on the precision of planetary radii and we note that the differences between the young and old planetary samples diminish with more stringent precision requirements (Figure~\ref{fig:precision}). However, resolving the discrepancies entirely requires discarding more than half of the CKS sample. A larger sample size and higher precision planetary radii for the entire CKS sample would help to more securely determine how much of the discrepancy between young and old planet populations is astrophysical and how much is due to noise.  

While a more detailed study of planet radius demographics as a function of age, mass, and metallicity is left for future works, we note that our findings are broadly consistent with expectations from both the core-powered mass-loss and photoevaporation theories. This is most evident from Figure~\ref{fig:triptych}, where the sub-Neptune size trends with age, mass, and metallicity among those CKS planets with the most precise radii are shown in relation to scalings which approximately, though not exactly, mimic those presented in \citet{Gupta2020}. The slope in the stellar mass versus planet radius plane is shallower than predicted in core-cooling models, but more similar to that predicted by photoevaporation models provided that planet mass scales approximately linearly with stellar mass \citep{Wu2019}. The gigayear timescale we find for evolution of the radius valley is more compatible with core-powered mass-loss models than the canonical timescale of 0.1~Gyr from photoevaporation. However, although photoevaporation models predict the radius gap to emerge on a timescale of 0.1~Gyr, some small fraction of planets are expected to cross the gap on timescales of $\sim$1~Gyr or more \citep{RogersOwen2020}, which is compatible with our observations. 

\begin{figure}
    \centering
    \includegraphics[width=\linewidth]{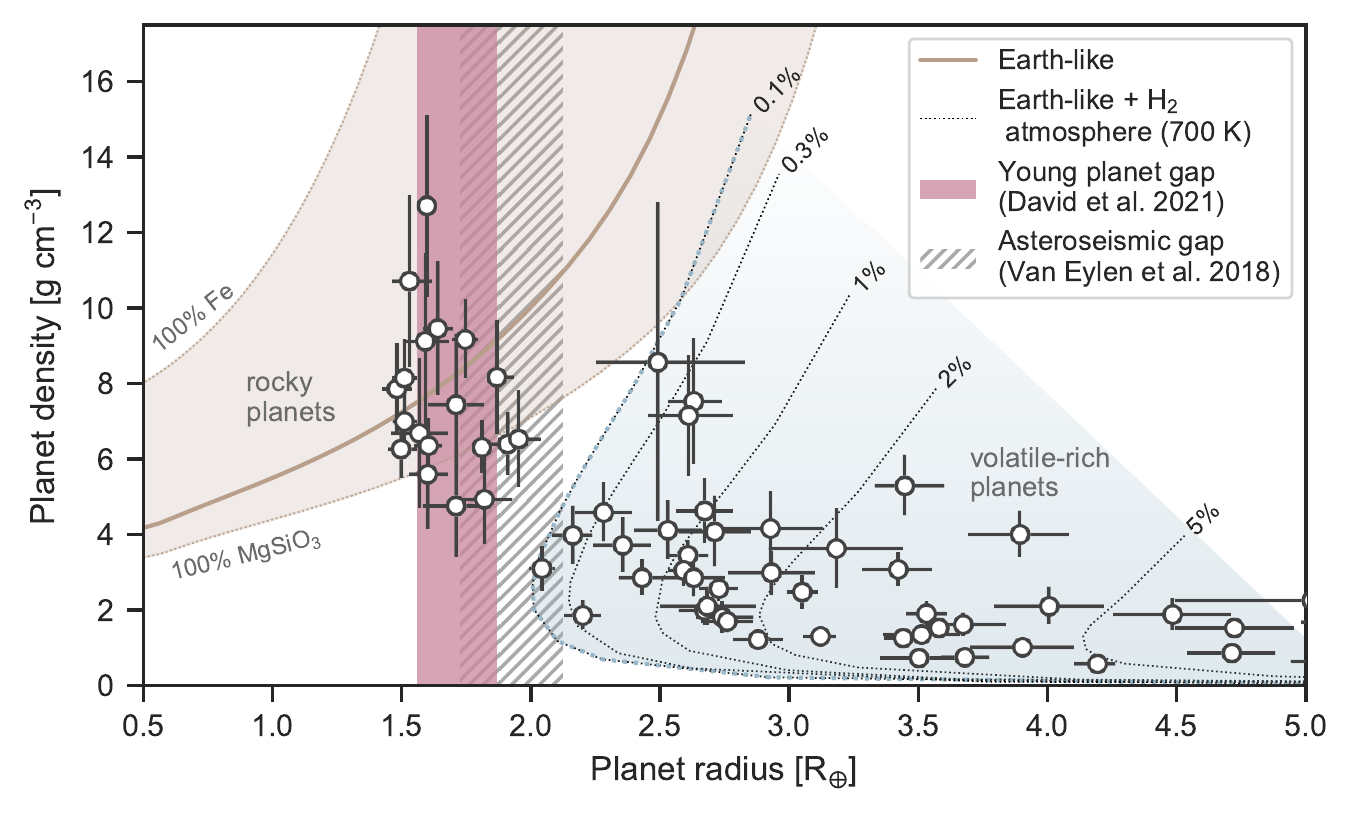}
    \caption{Well-characterized exoplanets in the radius-density plane. Planet composition curves from \citet{Zeng2019} are shown. The beige line indicates an Earth-like rocky composition (32.5\% Fe + 67.5\% MgSiO$_3$), and the similarly shaded swath is bounded by the pure Fe (upper bound) and pure MgSiO$_3$ curves (lower bound). Dotted lines indicate composition curves for Earth-like cores with H$_2$ atmospheres at 700~K for different atmospheric mass fractions, which are indicated above each curve. The vertical pink band indicates the range of planetary radii at the center of the young planet gap identified in this work, for orbital periods in the 3--30~d range. The hatched band indicates the equivalent radius range, for the same orbital periods, of the gap identified in \citet{vanEylen2018}.}
    \label{fig:composition}
\end{figure}

The difference in the radius distributions between young and old planets is primarily driven by an absence of large super-Earths (1.5--1.8~\rearth) at young ages, rather than an absence of small sub-Neptunes (see Figure~\ref{fig:fivepct}). As a result, the precise location of the radius valley is shifted to larger planet sizes at older ages. To better understand the compositions of planets missing from the young planet radius distribution, we compiled data for well-characterized, confirmed exoplanets from the NASA Exoplanet Archive \citep{Akeson2013}. We selected planets with masses known to 25\% precision or better, radii with 10\% precision or better, orbital periods $<100$~days, and host stars with $4500~K < \teff < 6500~K$ to match the CKS sample. We computed the bulk densities of these planets and compared the distribution of planets in the radius-density plane to composition curves from \citet{Zeng2019}. Among these well-characterized planets, we observe a clean separation in the radius-density plane \citep[also observed by][]{Sinukoff2018} between planets which are consistent with rocky compositions and those which require a significant volatile component (such as a H$_2$-He atmosphere, H$_2$O-dominated ices/fluids, or some combination of the two) to explain their bulk compositions (Figure~\ref{fig:composition}). We also observed that the radius valley identified in \citet{vanEylen2018} bridges the gap between planets in these two composition regimes. Meanwhile, the young planet gap identified in this work appears to correspond only to planets in the rocky composition regime. Thus, assuming the atmospheric loss hypothesis is correct, the planets which eventually fill the young planet gap may correspond to the large end of the size distribution of stripped cores.

This is an important point in the context of disentangling correlations between stellar mass, age, and metallicity in the CKS sample. The evidence for a wider radius valley among metal-rich rich stars is driven mostly by larger sub-Neptunes on average, with one explanation being the decreased cooling efficiency of planets with higher metallicity envelopes \citep{Owen2018}. By comparison, we observe a sub-Neptune size distribution which is relatively constant below 3~Gyr while the average size of super-Earths appears to increase over this same time frame (Figure~\ref{fig:triptych}). These observations are not easily explained by the anti-correlation between age and [Fe/H] in the CKS sample or the na\"ive expectation of more massive cores around metal-rich stars from core accretion models. Given that a planet's size is correlated with its mass, one physical interpretation for this observation is that the largest, most massive cores lose their atmospheres at later times. 

It is also worth noting that a prolonged mass-loss timescale for some super-Earths might help to explain the rising occurrence of long-period super-Earths with decreasing metallicity observed by \citet{Owen2018}. Those authors noted that such planets are difficult to explain in the photoevaporation model and might have instead formed after the protoplanetary disk dispersed, akin to the canonical view of terrestrial planet formation in the Solar System. However, we note that metallicity and age are correlated in the CKS sample with the median age of the metal-poor sample in \citet{Owen2018} being approximately 0.4~dex older than the metal-rich sample. If mass-loss, regardless of the mechanism, proceeds over gigayear timescales then one might expect a rising occurrence of super-Earths with increasing age (and hence decreasing metallicity).

In a companion paper, \citet{Sandoval2020} found tentative evidence that the fraction of super-Earths to sub-Neptunes rises with system age from $\sim$1--10~Gyr. That work accounted for uncertainties in stellar ages, planetary radii, and the equation for the radius valley itself. The result is in agreement with a previous finding by \citet{Berger2020b} who found, among planets orbiting stars more massive than the Sun, the fraction of super-Earths to sub-Neptunes is higher among older stars ($>$1~Gyr) than it is for younger stars ($<$1~Gyr). Collectively, the present work and the studies mentioned above provide evidence for the evolution of small planet radii over gigayear timescales.

The code and data tables required to reproduce the analyses and figures presented in this paper are made publicly available.\footnote{\url{https://github.com/trevordavid/radius-gap}}.

\appendix
\section{Rotation period vetting sheets}
\label{sec:appendixa}
\begin{figure}
    \centering
    \includegraphics[width=\linewidth]{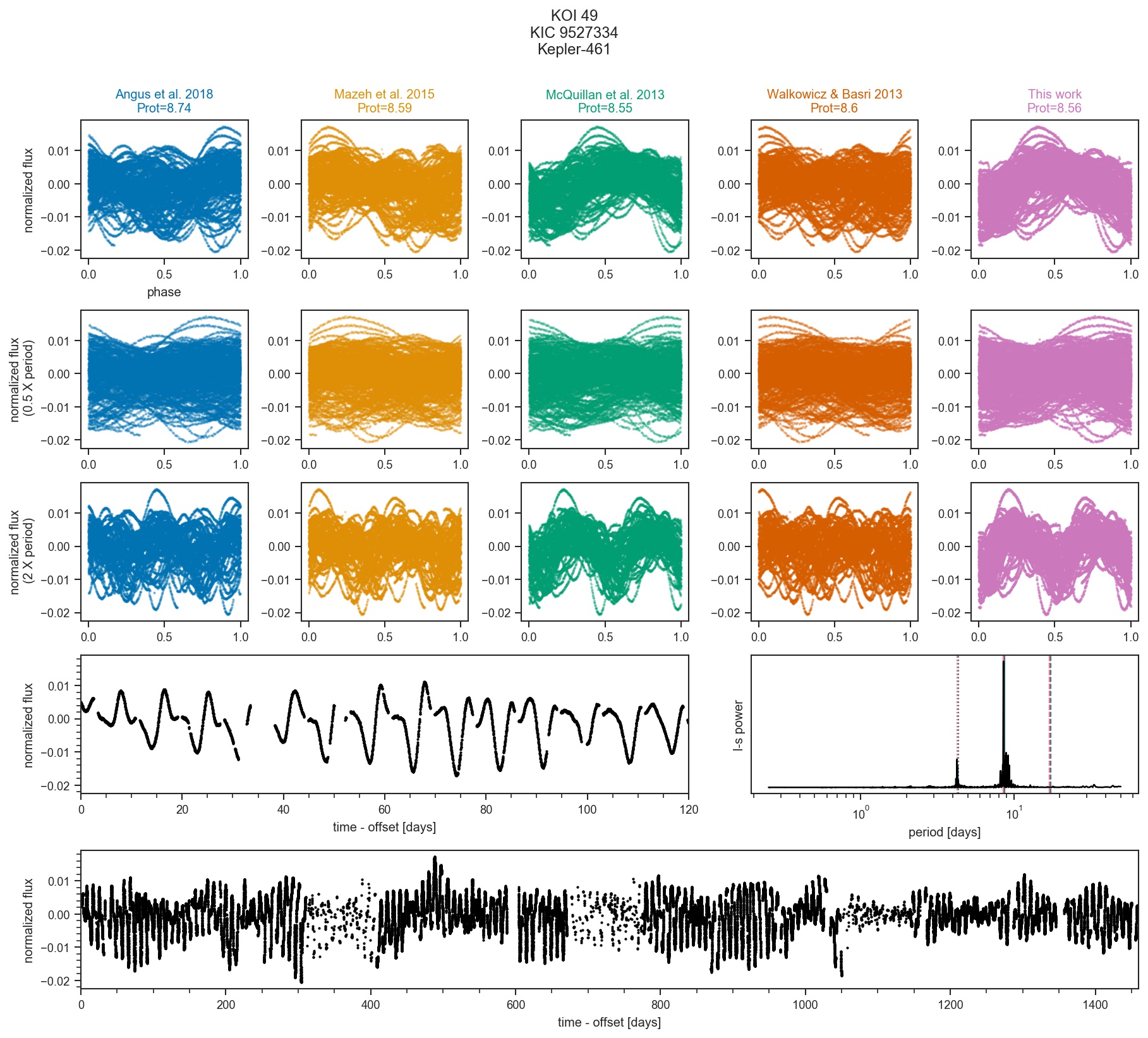}
    \caption{Example rotation period vetting sheet. The \kepler light curve is phase-folded on periods determined from the literature (first row), as well as the first harmonic (second row) and sub-harmonic (third row). The period determinations of different authors (indicated at top) are presented in a column-wise fashion and color-coded for convenience. In the fourth row the first 120 days of the light curve (left) and a Lomb-Scargle periodogram (right) are shown. In the bottom row the full \kepler light curve is shown.}
    \label{fig:protvetting}
\end{figure}

We provide rotation period vetting sheets (as described in \S\ref{subsec:prot}) as a figure set here. An example sheet is shown in Fig.~\ref{fig:protvetting}.

\section{Stellar age validation}
\label{sec:appendixb}
As we are concerned with the time evolution of the exoplanet radius gap, our study hinges on the accuracy of the stellar ages. Main-sequence stars, which constitute the majority of \kepler planet hosts, typically have large age uncertainties; this is because the changes in a star's observable properties over its main sequence lifetime are small relative to typical measurement uncertainties in those properties. However, the high precision of \gaia parallaxes and photometry has enabled the determination of relatively \textit{precise} stellar ages from isochrones. The median lower and upper uncertainties on \logage for stars in the CKS sample are 0.12 and 0.14 dex, respectively. 

A true assessment of the \textit{accuracy} of stellar ages is not possible; essentially all methods for stellar age determination are model-dependent and benchmarks to calibrate these methods are lacking \citep{Soderblom2010}. However, because the \kepler field is so well-studied, it is at least possible to determine the degree of agreement between isochrone ages published by different authors. It is also possible to determine the agreement between ages determined from isochrones versus those determined from gyrochronology or asteroseismology.

To validate the ages used in this study, we compared the isochrone age estimates from F18 with those published in the \gaia-\kepler Stellar Properties Catalog \citep[GKSPC, hereafter B20,][]{Berger2020b, Berger2020a}, asteroseismic ages determined in \citet{SilvaAguirre2015}, and with gyrochronology ages determined here. We note that while GKSPC ages exist for a far larger portion of the \kepler sample, we restrict our analysis here to only those stars which overlap with CKS VII. 

\begin{figure}
    \centering
    \includegraphics[width=\linewidth]{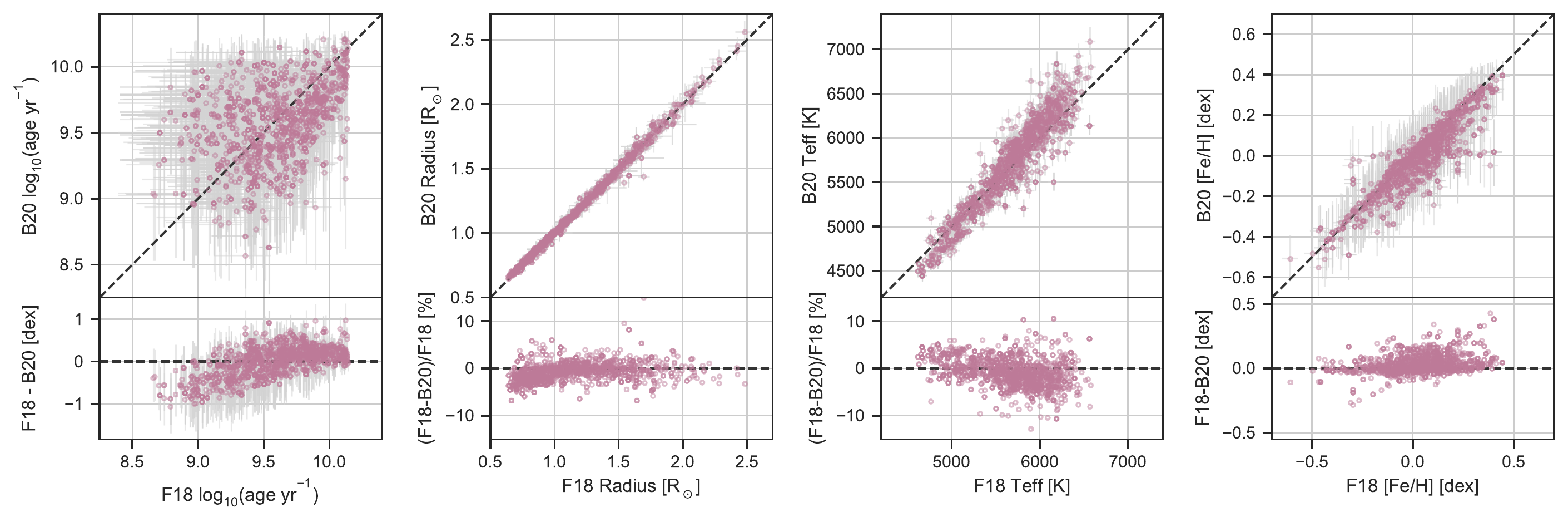}
    \caption{Comparison of F18 and B20 isochrone ages, stellar radii, \teff, and \feh (from left to right). Residuals are show in the bottom row for each panel.}
    \label{fig:iso-age-compare}
\end{figure}

Figure~\ref{fig:iso-age-compare} shows the comparison of ages and other parameters from F18 and B20. For 80\% of the stars with age estimates in both catalogs, the age estimates agree to within $\sim$0.4~dex. The median offset in ages is 0.075 dex, with F18 ages being systematically older, but this shift is smaller than typical age uncertainties from either catalog. The age discrepancies should not be due to differences in the adopted stellar models; both F18 and B20 use the \texttt{isoclassify} package \citep{Huber2017} to compute ages from MIST v1.1 models \citep{Choi2016, Dotter2016}.

\begin{figure}
    \centering
    \includegraphics[width=\linewidth]{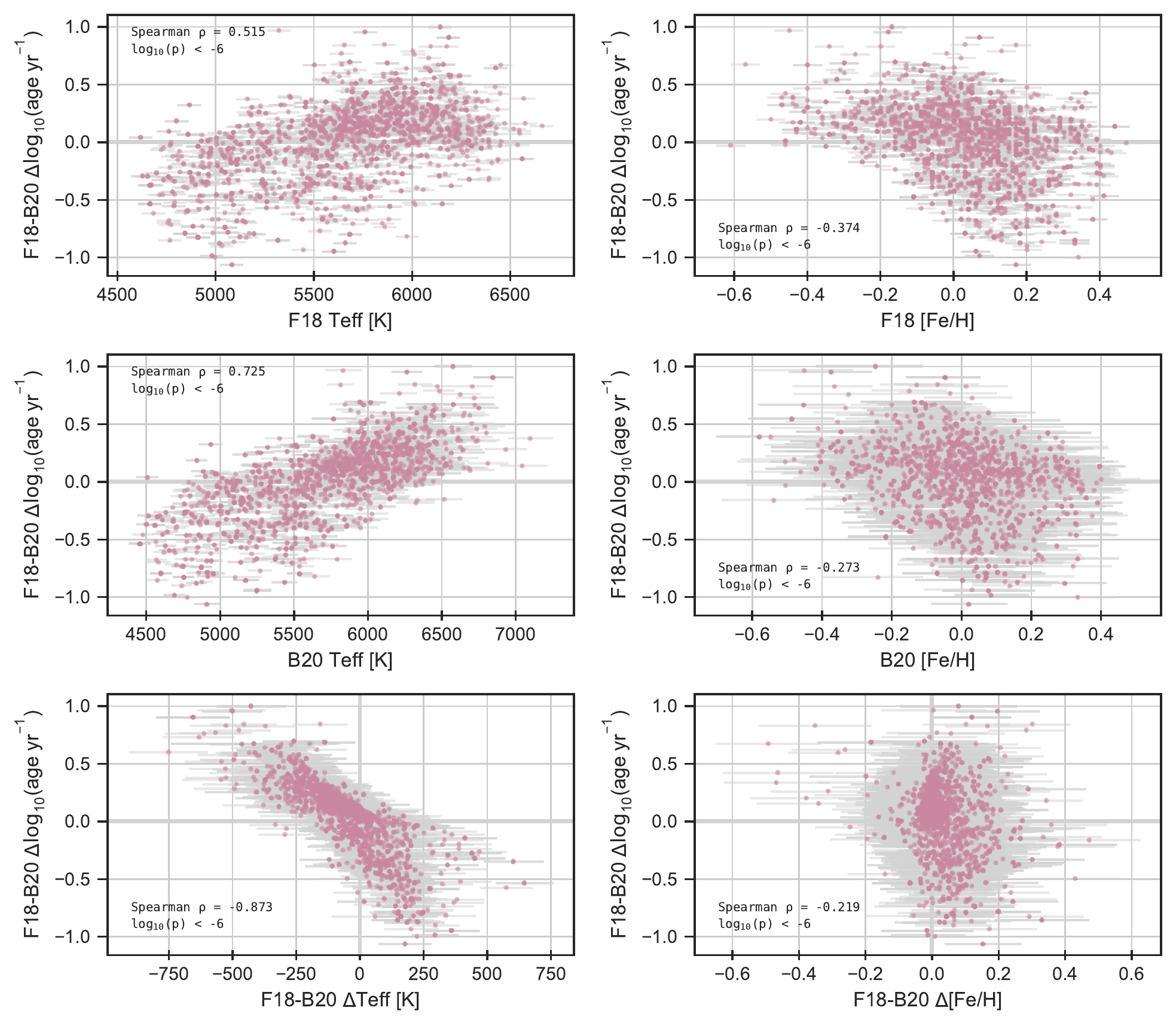}
    \caption{Trends in age discrepancy between the CKS (F18) and GKSPC (B20) catalogs. Errors on the age differences are omitted for clarity.}
    \label{fig:agediff-trends}
\end{figure}

To better understand the origins of the age discrepancies between the two studies, we searched for correlations between \deltaage and other parameters in the dataset. We found that \deltaage is most strongly correlated with \deltamstar and \deltateff (see Figure~\ref{fig:agediff-trends}). As both F18 and B20 determine mass and age simultaneously from stellar models, and these two parameters are intrinsically related, the \deltamstar--\deltaage correlation is unsurprising. The correlation with \deltateff, however, is more informative. F18 derived \teff from high-resolution spectroscopy, while B20 derived \teff from isochrones using photometry (specifically Sloan $g$ and 2MASS $K_s$) and parallaxes as input.

We examined the dependence of \teff-color relations on [Fe/H] and $A_V$ and found, when using the CKS spectroscopic parameters, that [Fe/H] can explain most of the dispersion in the \teff-color relations. By contrast, when using the photometric \teff and \feh from B20, there is no clear metallicity gradient in the \teff-color relations. 

We find that \deltateff is more strongly correlated with the reddening values (sourced either from B20, L21, or \gaia) than it is with any of the metallicity parameters. While reddening might help to explain temperature and age differences for some sources, we note that differences in photometric and spectroscopic temperature scales persist independent of reddening corrections \citep{Pinsonneault2012}.

\begin{figure}
    \centering
    \includegraphics[width=\linewidth]{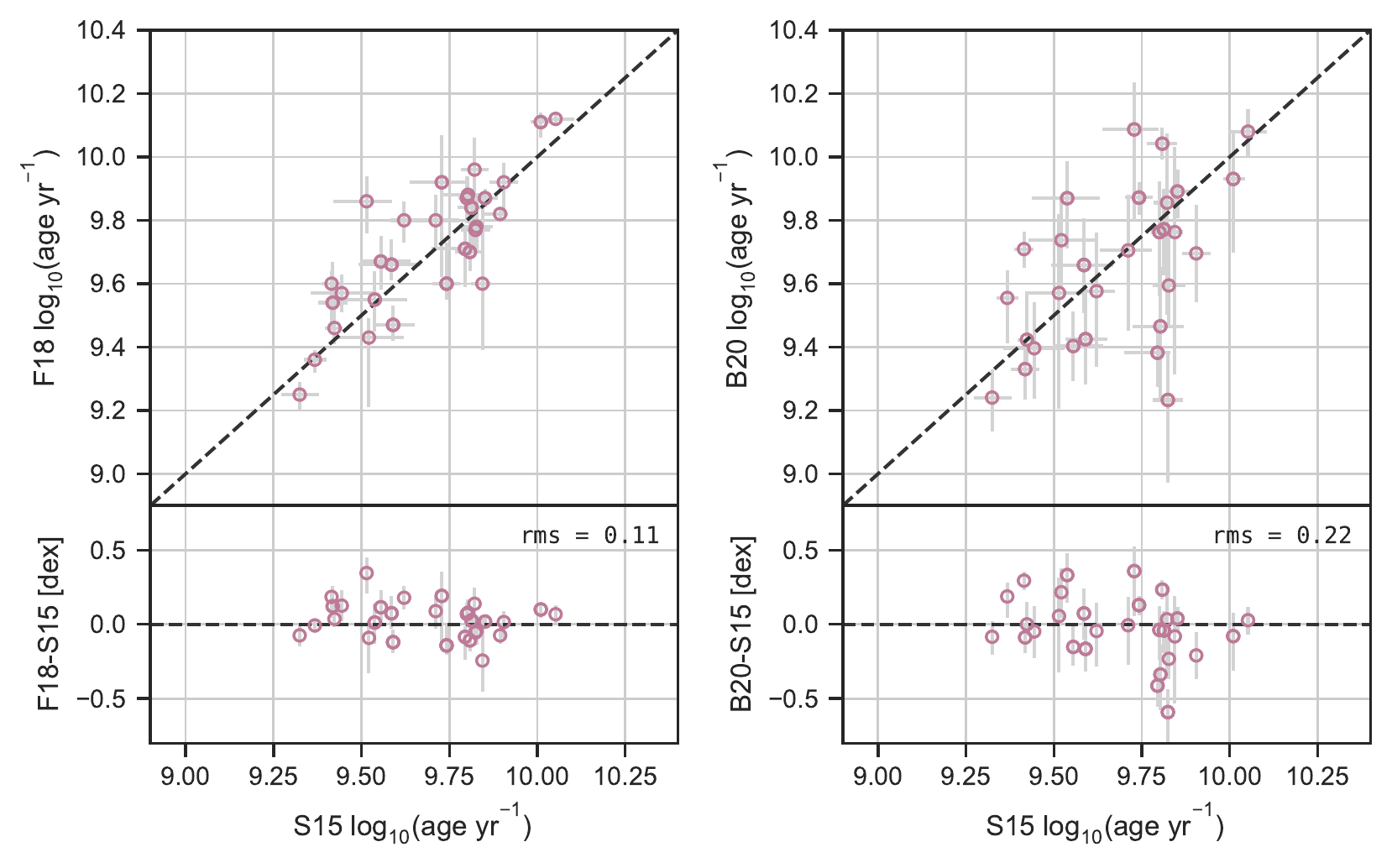}
    \includegraphics[width=\linewidth]{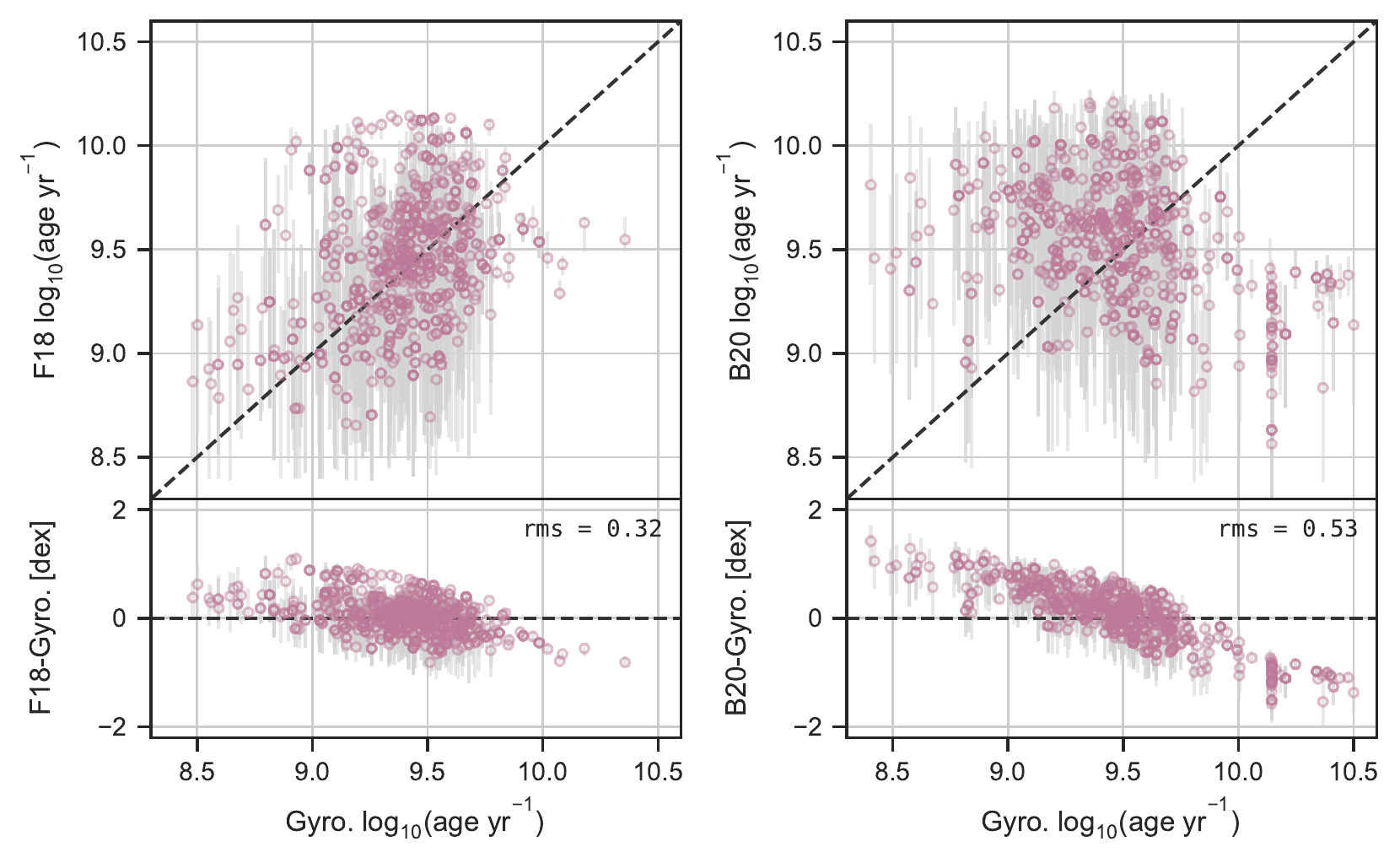}
    \caption{Comparison of F18 and B20 ages with asteroseismic ages from \citet{SilvaAguirre2015} (top panels) and gyrochronology ages from this work.}
    \label{fig:astero}
\end{figure}

We next compared the F18 and B20 ages with those determined from precise asteroseismic parameters. \citet{SilvaAguirre2015} determined ages for a sample of 33 \kepler planet candidate host stars with high signal-to-noise asteroseismic observations, achieving a median statistical uncertainty of 14\% on age. We compare the ages from F18 and B20 with the asteroseismic ages in Figure~\ref{fig:astero}. We find reasonably good agreement with the asteroseismic ages for both F18 and B20. The residual scatter between the isochrone ages and asteroseismic ages is 0.11 dex for F18 and 0.22 dex for B20. In both cases, the residuals are comparable to the median age uncertainties from those catalogs.

Gyrochronology ages were computed using the \stardate software package \citep{stardate, Angus2019}.\footnote{\url{https://github.com/RuthAngus/stardate}} The \stardate ages were computed in the gyrochronology mode alone rather than in the combined isochrone-fitting and gyrochronology mode. The \stardate gyrochronology relations are calibrated in \gaia color space. Using the rotation periods we vetted in \S\ref{sec:sample}, we noticed increased scatter in the ($B_P-R_P$)--\prot plane compared to the \teff-\prot plane, where \teff is the CKS spectroscopic temperature from F18. As such, rather than using the star's actual \gaia colors which are susceptible to reddening, we converted the F18 spectroscopic \teff and B20 photometric \teff to predicted \gaia colors using the relation in \citet{Curtis2020} which was calibrated for stars with negligible reddening. Using the vetted \prot and predicted ($B_P-R_P$) colors we then computed the gyrochronology ages (without uncertainties). Our comparison of the isochrone ages and gyrochronology ages is shown in Figure~\ref{fig:astero}. We note that there is better agreement between F18 and the gyrochronology ages at young ages ($<$1~Gyr). 

\begin{figure}
    \centering
    \includegraphics[width=\linewidth]{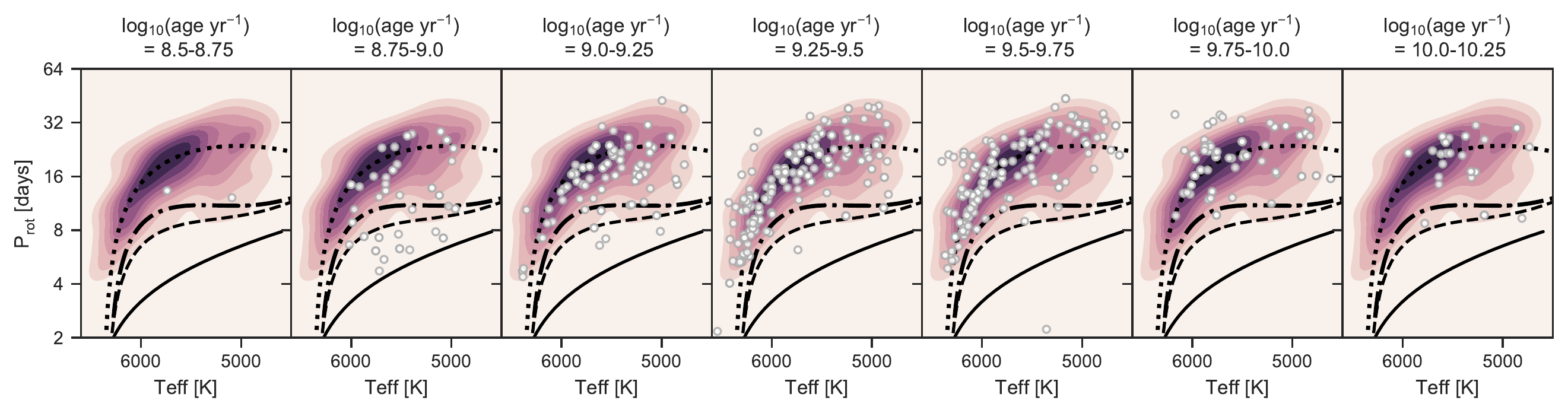}
    \caption{
    In each panel, contours show the Gaussian kernel density estimate of CKS planet hosts in the \teff-\prot plane. Points depict stars with F18 isochrone ages indicated in the titles of each panel.
    The solid, dashed, dash-dotted, and dotted lines indicate polynomial fits to the empirical gyrochrones of the Pleiades (\logage $\approx 8.1$), Praesepe (\logage $\approx 8.8$), NGC~6811 (\logage $\approx 9$), and NGC~6819 + Ruprecht~147 (\logage $\approx 9.4$) clusters respectively \citep{Curtis2020}.
    }
    \label{fig:cks-gyro}
\end{figure}

We note that the \stardate model has not been updated to include recently-determined open cluster rotation period sequences in its calibration. As such, we can compare the CKS sample to empirical gyrochrones from \citet{Curtis2020}. This comparison is shown in Figure~\ref{fig:cks-gyro}, which shows that the F18 isochrone ages do not always map predictably onto the \teff-\prot plane. For example, in the F18 log(age) bin of 8.75-9~dex ($\approx$0.6--1~Gyr), approximately half of the stars fall below the 1~Gyr gyrochrone and half lie above it. Similarly, in the F18 log(age) bin of 9.5-9.75~dex ($\approx$3.2--5.6~Gyr), a non-negligible number of stars fall below the log(age)$\approx$9.4 ($\approx$2.5~Gyr) gyrochrone. However, we note that the vast majority of stars with F18 isochrone ages of log(age)$<9.25$ fall below the log(age)$\approx$9.4 gyrochrone. This is in agreement with the comparison made to the \stardate gyrochronology ages, in the sense that the majority of stars with F18 isochrone ages $\lesssim$1.8~Gyr appear to be younger than $\lesssim$2.7~Gyr from a gyrochronology analysis.

\begin{figure}
    \centering
    \includegraphics[width=\linewidth]{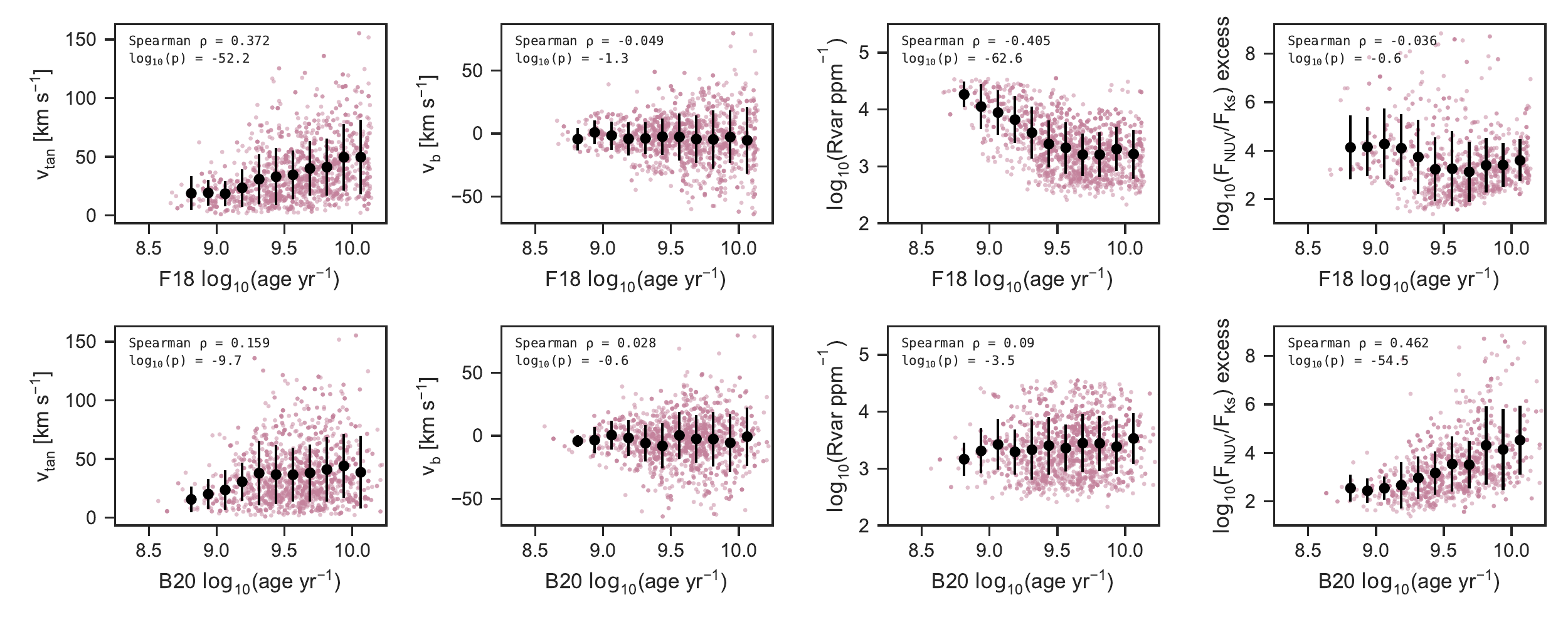}
    \caption{Validation of isochronal age estimates. The $v_\mathrm{tan}$ and $v_b$ velocities (left and middle columns) and \rvar as a function of isochronal ages from F18 (top row) and B20 (bottom row). Spearman rank correlation coefficients ($\rho$) and p-values are printed in the top left corner of each panel. Black points with errorbars indicate the mean and standard deviation of the data binned by 0.125 dex in log(age).}
    \label{fig:observables}
\end{figure}

Finally, we also examined the evolution of other physical parameters known to correlate with age, such as the variability amplitude \rvar, NUV excess, and the velocity dispersion. We tracked velocity dispersion using \vtan, the velocity tangential to the celestial sphere, and \vb, the velocity in the direction of the galactic latitude, sourced from \citet{Lu2020}. \galex \textit{NUV} magnitudes were obtained from \citet{Olmedo2015}. For a crude approximation of the NUV excess, we performed a quadratic fit to the full \kepler Q1-Q17 DR25 sample in the ($G-G_{RP}$) vs. ($NUV-K_s$) color-color diagram. The NUV excess was then defined as a star's ($NUV-K_s$) color minus the quadratic color-color trend. No de-reddening was performed. Figure~\ref{fig:observables} shows the evolution of these parameters as a function of age. For both the F18 and B20 isochrone ages, we observe increasing dispersion in \vtan and \vb with age, as expected. The strongest expected correlation is observed for \rvar \citep[sourced from][]{Lu2020} and F18 age, with \rvar declining for the first $\sim$3~Gyr before plateauing. The average NUV excess appears to decline over a similar timescale when using the F18 ages, though that trend is less significant and there may be residual systematics from the manner in which we computed the excess. Both the \rvar and NUV excess trends are expected, as starspot coverage, variability amplitudes, and chromospheric activity are known to decline with age. By comparison, when using the B20 ages the behavior of \rvar and NUV excess with age is not as expected. 
We conclude by noting that, while substantial uncertainties remain for isochronal ages, there is qualitative agreement between the CKS ages and ages (or age indicators) derived from independent methods. In some of the comparisons above the CKS and GKSPC ages perform comparably well, though it is at the youngest ages ($\lesssim$3~Gyr) where the GKSPC ages do not reproduce some expected trends. As the evolution of small planets at early times is a primary focus of this work, we adopt the CKS ages.

\acknowledgments This paper is dedicated to the memory of John Stauffer, a valued mentor whose energy and determination were an inspiration to many. We thank the anonymous referee for a thorough and insightful review, as well as Eric Ford, Christina Hedges, David W. Hogg, and Josh Winn for helpful discussions. TJD is especially grateful to Chelsea Yarnell for her irreplaceable support throughout the COVID-19 pandemic. This paper includes data collected by the {\em Kepler} mission, funded by the NASA Science Mission directorate. This work presents results from the European Space Agency (ESA) space mission Gaia. Gaia data are being processed by the Gaia Data Processing and Analysis Consortium (DPAC). Funding for the DPAC is provided by national institutions, in particular the institutions participating in the Gaia MultiLateral Agreement (MLA). The Gaia mission website is \url{https://www.cosmos.esa.int/gaia}. The Gaia archive website is \url{https://archives.esac.esa.int/gaia}. This work made use of the gaia-kepler.fun crossmatch database created by Megan Bedell.
\vspace{5mm}
\facilities{Kepler; Gaia; GALEX}

\software{\texttt{astropy} \citep{astropy13, astropy18},
          \texttt{jupyter} \citep{jupyter},
          \texttt{matplotlib} \citep{matplotlib},
          \texttt{numpy} \citep{numpy},
          \texttt{pandas} \citep{pandas-soft, pandas-proc},
          \texttt{seaborn} \citep{seaborn},
          \texttt{scikit-learn} \citep{scikit-learn}
          \texttt{scipy} \citep{scipy}
          }

%\bibliography{main}{}
%\bibliographystyle{aasjournal}

\listofchanges

\end{document}